\documentclass[prd,preprintnumbers,twocolumn,eqsecnum,floatfix,a4paper,nofootinbib,superscriptaddress,amsmath,amssymb]{revtex4-2}

\usepackage{color}
\usepackage{acronym}
\usepackage{amsmath,amssymb,graphicx,mathtools}
\usepackage{bm,lipsum}
\usepackage{comment}
\usepackage{times}
\usepackage{microtype}
\usepackage[utf8]{inputenc}
\usepackage{booktabs}
\usepackage[normalem]{ulem}
\usepackage[varg]{txfonts}
\usepackage{bm,graphics,graphicx,epsfig,xcolor,ulem}
\usepackage[colorlinks, pdfborder={0 0 0}]{hyperref}
\usepackage{multirow}
\usepackage{pifont}
\usepackage{array,threeparttable}
\usepackage{diagbox}
\usepackage{boldline}
\usepackage{gensymb}
\usepackage{url}

\definecolor{LinkColor}{rgb}{0.75, 0, 0}
\definecolor{CiteColor}{rgb}{0, 0.5, 0.5}
\definecolor{UrlColor}{rgb}{0, 0, 0.75}
\hypersetup{linkcolor=LinkColor}
\hypersetup{citecolor=CiteColor}
\hypersetup{urlcolor=UrlColor}
\allowdisplaybreaks
\definecolor{nicegreen}{rgb}{0.1,0.5,0.1}

\begin{document}
\title{Prospects for Direct Detection of Black Hole Formation in Neutron Star Mergers with Next-Generation Gravitational-Wave Detectors}

\newcommand{\psuigc}{\affiliation{Institute for Gravitation and the Cosmos, The Pennsylvania State University, University Park, PA, 16802, USA}}
\newcommand{\psuphys}{\affiliation{Department of Physics, The Pennsylvania State University, University Park, PA, 16802, USA}}
\newcommand{\psuastro}{\affiliation{Department of Astronomy \& Astrophysics, The Pennsylvania State University, University Park, PA, 16802, USA}}
\newcommand{\aei}{\affiliation{Max Planck Institute for Gravitational Physics (Albert Einstein Institute), Am Mühlenberg 1, Potsdam 14476, Germany}}
\newcommand{\cardiff}{\affiliation{School of Physics and Astronomy, Cardiff University, Cardiff, UK, CF24 3AA}}
\newcommand{\ukyphys}{\affiliation{Department of Physics and Astronomy, University of Kentucky, Lexington, KY, 40506-0055, USA}}

\author{Arnab Dhani}
\email{arnab.dhani@aei.mpg.de}
\aei
\psuigc
\psuphys

\author{David Radice}
\thanks{Alfred P. Sloan fellow}
\psuigc
\psuphys
\psuastro

\author{Jan Sch\"{u}tte-Engel}
\affiliation{Department of Physics, University of Illinois at Urbana-Champaign, Urbana, IL 61801, USA}
\affiliation{Illinois Center for Advanced Studies of the Universe, University of Illinois at Urbana-Champaign, Urbana, IL 61801, USA}

\author{Susan Gardner}
\ukyphys

\author{Bangalore Sathyaprakash}
\psuigc
\psuphys
\psuastro
\cardiff

\author{Domenico Logoteta}
\affiliation{Dipartimento di Fisica, Universit\`{a} di Pisa, Largo B.  Pontecorvo, 3 I-56127 Pisa, Italy}
\affiliation{INFN, Sezione di Pisa, Largo B. Pontecorvo, 3 I-56127 Pisa, Italy}

\author{Albino \surname{Perego}}
\affiliation{Dipartimento di Fisica, Università di Trento, Via Sommarive 14, 38123 Trento, Italy}
\affiliation{INFN-TIFPA, Trento Institute for Fundamental Physics and Applications, Via Sommarive 14, I-38123 Trento, Italy}

\author{Rahul \surname{Kashyap}}
\psuigc
\psuphys

\begin{abstract}
A direct detection of black hole formation in neutron star mergers would provide invaluable information about matter in neutron star cores and finite temperature effects on the nuclear equation of state. We study black hole formation in neutron star mergers using a set of 190 numerical relativity simulations consisting of long-lived and black-hole-forming remnants. 
The postmerger gravitational-wave spectrum of a long-lived remnant has greatly reduced power at a frequency $f$ greater than $f_{\rm peak}$, for $f \gtrsim 4\,\rm kHz$, with $f_{\rm peak} \in [2.5, 4]\,\rm kHz$. On the other hand, black-hole-forming remnants exhibit excess power in the same large $f$ region and manifest exponential damping in the time domain characteristic of a quasinormal mode. We demonstrate that the gravitational-wave signal from a collapsed remnant is indeed a quasinormal ringing. We report on the opportunity for direct detections of black hole formation with next-generation gravitational-wave detectors such as Cosmic Explorer and Einstein Telescope and set forth the tantalizing prospect of such observations up to a distance of 100 Mpc for an optimally oriented and located source with an SNR of 4. 
\end{abstract}

\keywords{Gravitational waves, binary neutron star, binary black hole, ringdown, quasi-normal modes}
\maketitle

\section{Introduction.} 
In \ac{NS} mergers, 
neutron-rich matter is compressed to densities in excess of several times nuclear saturation and heated to temperatures of tens of MeV \cite{Perego:2019adq}. The physics of matter in these conditions is very poorly understood, due to the nonperturbative nature of the strong interaction at the densities relevant for NSs \cite{Drischler:2021kxf, Lattimer:2021emm, MUSES:2023hyz}. Experimental results from heavy-ion collisions (HICs) constrain the properties of symmetric matter up to ${\sim}3$ times nuclear saturation density~\cite{Danielewicz:2002pu}, with constraints on the symmetry energy~\cite{Horowitz:2014bja}  
coming from HICs~\cite{LeFevre:2015paj,Russotto:2016ucm}, for 
densities ranging from well below
up to 2 times saturation density, 
and from parity-violating electron studies on 
nuclei~\cite{Reed:2021nqk}. Radio and x-ray observations of \acp{NS} \cite{Steiner:2012xt, Ozel:2016oaf, Miller:2019cac, Pang:2021jta, Annala:2021gom}, \ac{GW} and multimessenger measurements of the tidal deformability of \acp{NS} in GW170817 \cite{LIGOScientific:2017fdd, Margalit:2017dij, Bauswein:2017vtn, Radice:2017lry, De:2018uhw, LIGOScientific:2018hze, LIGOScientific:2018cki, Radice:2018ozg, Capano:2019eae, Dietrich:2020efo, Breschi:2021tbm} provide strong bounds on the pressure of cold, beta-equilibrated \ac{NS} matter 
at several times saturation density. Additional constraints have been derived from the observation of cooling curves for newborn \acp{NS} \cite{Page:2004fy, Yakovlev:2004iq, Page:2010aw, Beloin:2018fyp}. However, none of these methods can access the conditions realized in the aftermath of \ac{NS} mergers.

\Ac{NR} simulations reveal that the postmerger dynamics of binary \acp{NS} is very sensitive to the \ac{EOS} of dense matter \cite{Radice:2020ddv}. 
The degree of incompressibility of nuclear matter \cite{Perego:2021mkd}, finite-temperature corrections to the pressure \cite{Raithel:2021hye, Hammond:2021vtv, Most:2022yhe, Loffredo:2022prq, Fields:2023bhs, Blacker:2023onp}, the possible appearance of new particles, such as hyperons \cite{Sekiguchi:2011mc, Radice:2016rys} or nonstandard-model ones~\cite{Berryman:2021jjt}, or of pion/kaon condensates \cite{Vijayan:2023qrt}, or QCD phase transitions \cite{Most:2018eaw, Most:2019onn, Bauswein:2018bma, Weih:2019xvw, Liebling:2020dhf, Prakash:2021wpz, Fujimoto:2022xhv, Kedia:2022nns, Espino:2023llj}, all have a direct impact on the postmerger. In particular, the physics of dense matter determines whether and when the remnant collapses to a \ac{BH}. 
Hence the information about BH formation can potentially be used to constrain the EOS of strongly interacting matter. Typically the postmerger signal lays at frequencies beyond the maximum sensitivity region of current \ac{GW} detectors 
LIGO, Virgo, KAGRA, prompting further improvements~\cite{Miller:2014kma}. Its detection is a key science goal \cite{Evans:2021gyd, Lovato:2022vgq} for next-generation observatories, such as the Einstein Telescope (ET) \cite{Punturo:2010zz}, the Cosmic Explorer (CE) \cite{Reitze:2019iox}, and NEMO \cite{Ackley:2020atn}. 

Most work thus far has focused on the physics that can be extracted from the \ac{GW} signal of a postmerger remnant prior to its collapse to a \ac{BH}. Such signals are not well understood \cite{Radice:2020ddv, Bernuzzi:2020tgt}.  However, there is consensus in the literature that the postmerger \ac{GW} spectrum has broad features, or peaks. The dominant feature is called $f_2$, or $f_{\rm peak}$, and is directly related to the angular velocity of the remnant \cite{Shibata:2005xz, Hotokezaka:2011dh, Bauswein:2013jpa, Takami:2014zpa, Bernuzzi:2015rla}. The location of these peaks, as well as their amplitudes, can constrain the \ac{EOS} of \acp{NS} and possibly reveal the presence of QCD phase transitions \cite{Bose:2017jvk, Breschi:2021xrx, Wijngaarden:2022sah}. A parallel line of investigation has focused on the immediate outcome of the merger, studying the implications of ``prompt'' \ac{BH} formation (or lack thereof) for dense matter \cite{Shibata:2005ss, Hotokezaka:2011dh, Bauswein:2013jpa, Zappa:2017xba, Agathos:2019sah, Koppel:2019pys, Bauswein:2020aag, Bauswein:2020xlt, Perego:2021mkd, Kashyap:2021wzs, Kolsch:2021lub, Cokluk:2023xio}. Zhang et al.~\cite{Zhang:2020qlh} considered the detectability of the ringdown signal produced by the remnant \ac{BH}, as it relaxes to a Kerr \ac{BH}, but employed an analytical ansatz for the signal, so its applicability to \ac{NS} mergers is unclear. Finally, some authors have also attempted to develop methods to measure the lifetime of the remnant by identifying the termination of the postmerger \ac{GW} signal \cite{Easter:2021wlb, Breschi:2022xnc}. 
Since the GW signal decays on 
a timescale of ${\sim}10{-}20\ {\rm ms}$ \cite{Bernuzzi:2015opx}, it
may be that this method 
gives a lower bound on the lifetime of the remnant 
rather than yield  
a detection of BH formation. 
It is the latter, though,
that would provide a 
particularly discriminating
probe of the EoS, as in, 
e.g., Ref.~\cite{Sekiguchi:2011mc, Radice:2016rys, Most:2018eaw}. 

In this \emph{Paper}, we use data from 190 \ac{NR} simulations to show that \ac{BH} formation can be revealed by \emph{direct} observation of the \acp{GW} generated by the perturbed \ac{BH} at birth and by oscillations of the remnant shortly before collapse. Such signal extends to frequencies up to $10\,$kHz, well above the 
 sensitivity window of current generation observatories. However, it would be detectable with an SNR $>$ 3 at distances ranging from 15 Mpc to 250 Mpc with a median distance of ${\sim}70$~Mpc for the binaries considered here with the proposed CE observatory. Even a single measurement of the delay time to \ac{BH} formation for a binary \ac{NS} system would provide strong constraints on the properties of extreme matter. Our work motivates the development of new technology for improved \ac{GW} sensitivity in the multi-kHz band, as discussed in this context by \cite{Zhang:2020qlh} and more broadly~\cite{Punturo:2010zz, Reitze:2019iox, Ackley:2020atn, Aggarwal:2020olq}, that we consider further below.

\begin{figure}
    \centering
    \includegraphics[width=\columnwidth]{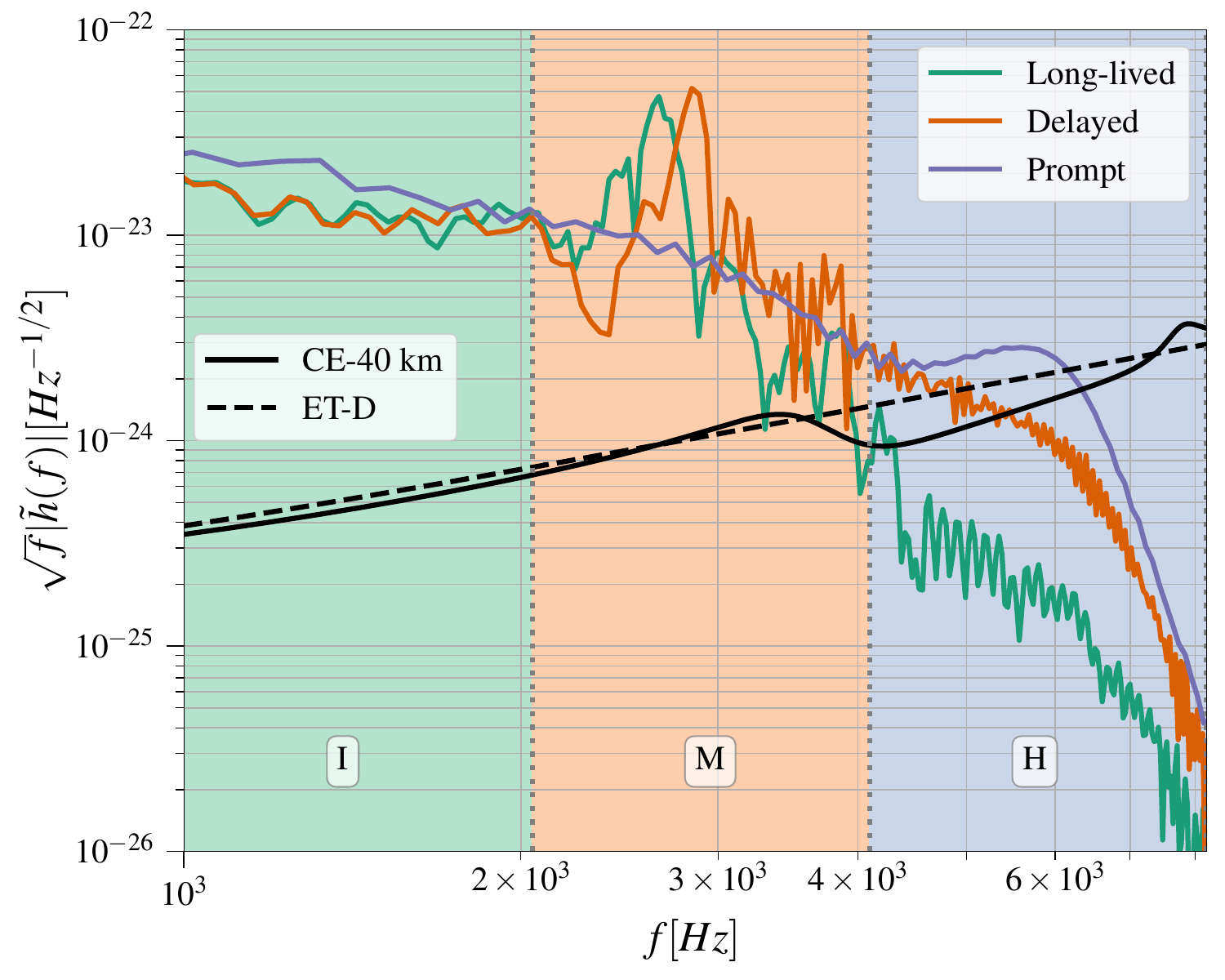}
    \caption{Typical strain profiles for three binary neutron star mergers that forms distinctive remnants. The systems are placed at a distance of 100 Mpc. The sensitivity curves of two proposed next-generation observatories \cite{Reitze:2019iox,Punturo:2010zza} are shown with solid and dashed black curves. The tinted backgrounds, orange and violet, denoted by ``M'' and ``H'', indicate the frequency bands used to define $\rho_M$ and $\rho_H$, respectively, as per the text, 
    while the green background denoted by ``I'' represents the inspiral. The long-lived and delayed collapse remnants shows the unmistakable f-mode peak. This feature is absent in the prompt collapse merger which instead has a prominent peak near the fundamental quasi-normal mode of the remnant BH. The delayed collapse case also has excess power at these frequencies, though not as elevated as the prompt collapse case.
    }
    \label{fig:strain}
\end{figure}

\section{Simulations.} 
We consider a total of 190 binary \ac{NS} merger simulations corresponding to 150 unique binary configurations performed with the \texttt{WhiskyTHC} code \cite{Radice:2012cu, Radice:2013hxh, Radice:2013xpa}. Of the unique configurations, 48 produced long-lived remnants that did not collapse to \ac{BH} within the simulation time, 26 produced remnants that collapsed to \ac{BH} within the simulation time, and 76 underwent prompt \ac{BH} formation. Additionally, we use 10 long-lived, 8 delayed collapse, and 22 prompt collapse remnant simulations at lower resolutions to estimate the errors in our results. The remnants are classified using the minimum lapse function, which is a proxy for the gravitational potential in these simulations.
In the case of BH formation, the minimum lapse function suddenly becomes zero, whereas 
a rapidly oscillating remnant 
has a similarly oscillating minimum lapse. We use these features to characterize the remnant. That is, 
if the minimum lapse is always at a finite value, the remnant is long-lived; if it drops to zero after a few oscillations following the merger, the remnant is classified as a delayed collapse; and, finally, if it immediately goes to zero without any oscillation, the merger is said to have resulted in a prompt collapse (see Fig. 3 of \cite{Kashyap:2021wzs} for representative plots). Furthermore, all the simulations are
carried out to at least 20 milliseconds postmerger (or up to black hole
formation), and they capture most of the signal (except for a potential
collapse signature at later times). Therefore, all long-lived simulations survive for at least 20
milliseconds after merger, but due 
to 
the time required for
gravitational waves to propagate to the wave extraction region, some long-lived
simulations contain only 18 milliseconds of gravitational-wave data after
merger.
The simulations considered here include publicly available data \cite{Gonzalez:2022mgo}, as well as simulations presented in \cite{Perego:2021mkd, Kashyap:2021wzs}.

Our data spans 17 \acp{EOS}, including piecewise polytropic models, zero-temperature beta-equilibrated, and finite-temperature composition-dependent microphysical \acp{EOS}. Some of the simulations also included neutrinos \cite{Radice:2016dwd, Radice:2018pdn} and turbulent viscosity \cite{Radice:2017zta, Radice:2020ids}. The range of total gravitational mass for the long-lived remnants, delayed collapses, and prompt collapses are $[2.4,3.0] \,\rm M_{\odot}$, $[2.7,3.3] \,\rm M_{\odot}$, and $[2.8,3.8] \,\rm M_{\odot}$, respectively. The corresponding mass ratio ranges are $[0.6,1.0]$, $[0.7,1.0]$, and $[0.6,1.0]$, respectively. A figure illustrating all the considered binary configurations is included in Appendix A.

\section{Results.} 
Figure~\ref{fig:strain} shows the frequency domain strain of three typical BNS systems with total gravitational mass of approximately $3 \,\rm M_{\odot}$ depicting the signal from a long-lived, delayed collapse, and prompt collapse remnant. Both the long-lived remnant and the delayed collapse system have a characteristic $f_{\rm peak}$ at $\approx2.5$ kHz. The similarity of their post-merger power spectrum in this band ($f_M\in[2048,4096]$~Hz makes it difficult to determine if a system collapses to form a BH. However, the prompt collapse merger 
does not contain this spectral feature, and it can easily be distinguished in this canonical example. On the contrary, the delayed and prompt collapse remnants have excess signal power at higher frequencies of $f_H\in[4096,8192]$~Hz, while the long-lived remnant has none in this band, making it possible to decisively identify BH formation in this example (see Appendix B, 
where an outlier case of a long-lived remnant with significant power in $f_H$ band is discussed.). In fact, the fundamental quasinormal mode (QNM) frequency for the prompt collapse merger calculated using the remnant quantities at $\approx6.9$ kHz correlates well with the high-frequency spectral feature seen in this case. We remark that such signal is expected to be present even if the remnant collapses with significant delay from merger \cite{Baiotti:2005vi, Baiotti:2007np}.

\begin{figure}
    \centering
    \includegraphics[width=\columnwidth]{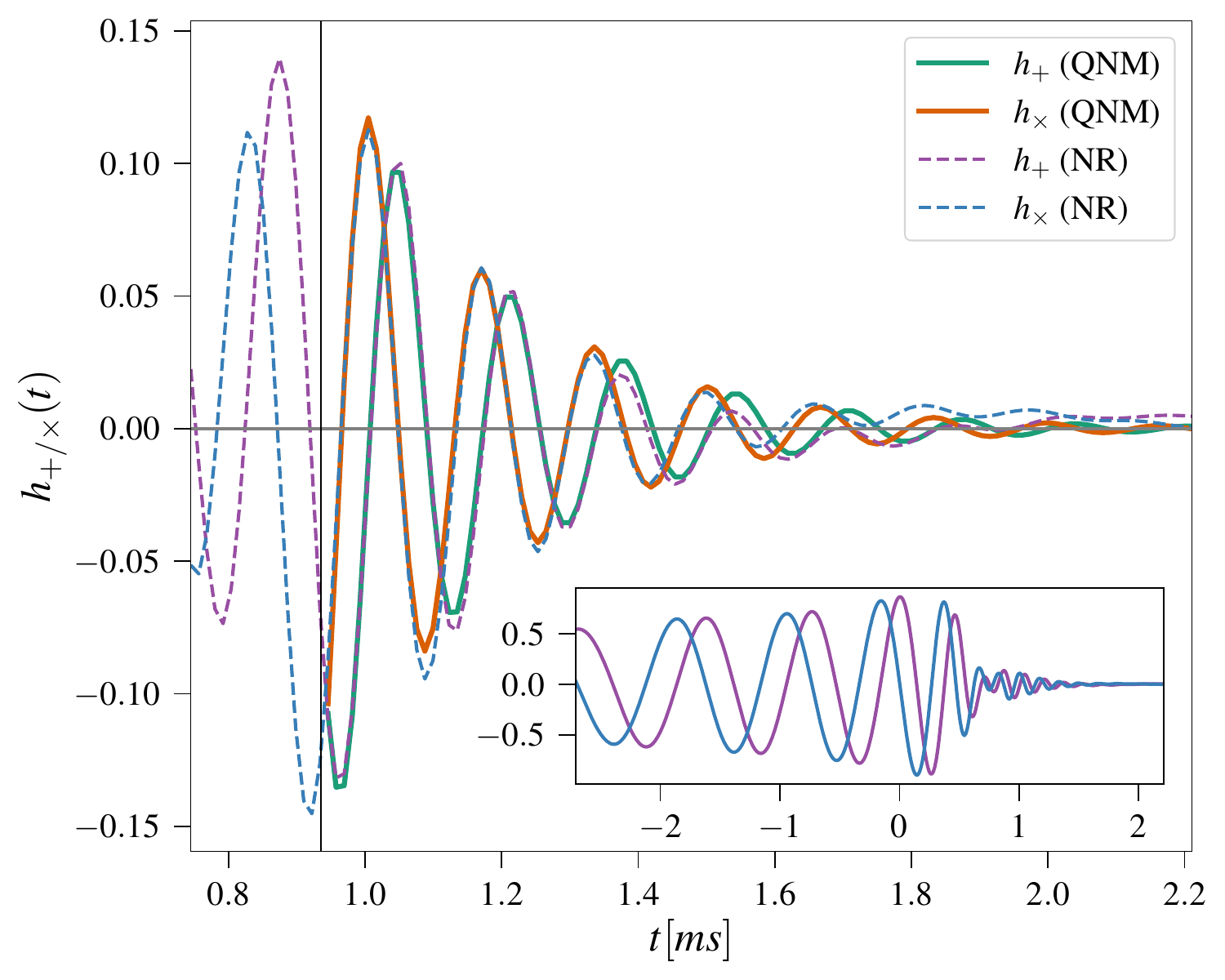} 
    \caption{The time domain gravitational polarizations for the $(l,m)=(2,2)$ mode, $h_{22}$, of the prompt collapse BNS merger shown in Fig.~\ref{fig:strain} with the peak of the $h_{22}$ taken as the reference for the time coordinate. The inset shows the complete waveform including the inspiral and the ringdown while the main part of the figure illustrates the ringdown signal. The dashed lines in this part is the NR data while the solid lines portray the the best-fit QNM model. The vertical line shows the start time for the QNM model which is defined as the time at which the mismatch between the NR data and the QNM model is minimum. It can be seen that the QNM fit worsens at late times. This is due to numerical noise as can be seen 
by noting that the mean of the NR data at late times is no longer zero.}
    \label{fig:qnm_fit}
\end{figure}

To demonstrate that the signal from a collapsed remnant is indeed consistent with a ringing BH, the post-merger signal for the prompt collapse case is modeled using a QNM and illustrated in Fig.~\ref{fig:qnm_fit}. The inset shows the $h_{22}$ mode from the \ac{NR} simulation with its peak taken as the reference for the time coordinate. The post-merger signal is then fit using nonlinear least squares to the model $h_{\rm QNM}=\mathcal{C}\exp({i\omega (t-t_{\rm start}))}$ where $\mathcal{C}$ is the complex amplitude, $\omega$ is the complex QNM frequency, and $t_{\rm start}$ is the start time for the model. For varying start times, the complex amplitude and frequency are fit to NR data,  and the fidelity of the fit is quantified using the inner-product $\mathcal{M}=1-\int_{t_{\rm start}}^{t_{\rm end}} \hat{h}_{\rm NR}^*\hat{h}_{\rm QNM}\rm dt$, where $t_{\rm end} = 2.2$ ms from the peak of the strain, 
determined empirically to ensure the signal has decayed sufficiently but with controlled
numerical noise, 
and $\hat{h}_X$is a normalised waveform given by $\hat{h}_X=h_X/\sqrt{\int_{t_{\rm start}}^{t_{\rm end}} h_X^*h_X\rm dt}$. The best-fit model that minimizes $\mathcal{M}(t_{\rm start})$ is shown with solid lines in Fig.~\ref{fig:qnm_fit}. The percentage difference in the best-fit oscillation frequency and damping frequency compared to the fundamental QNM frequency calculated from the remnant mass and spin is 13.5\% and 11\%, respectively. These values are typical among the set of 20 equal mass binaries for which this analysis is possible: the minimum, maximum, and median errors are 7.3\%, 17.4\%, and 14.4\%, respectively, for the oscillation frequency and 1.3\%, 23.5\%, and 7.9\% for the damping frequency. Our analysis confirms that the signal from the collapsed remnant is truly a BH ringdown.

\begin{figure*}
    \centering
    \includegraphics[width=2\columnwidth]{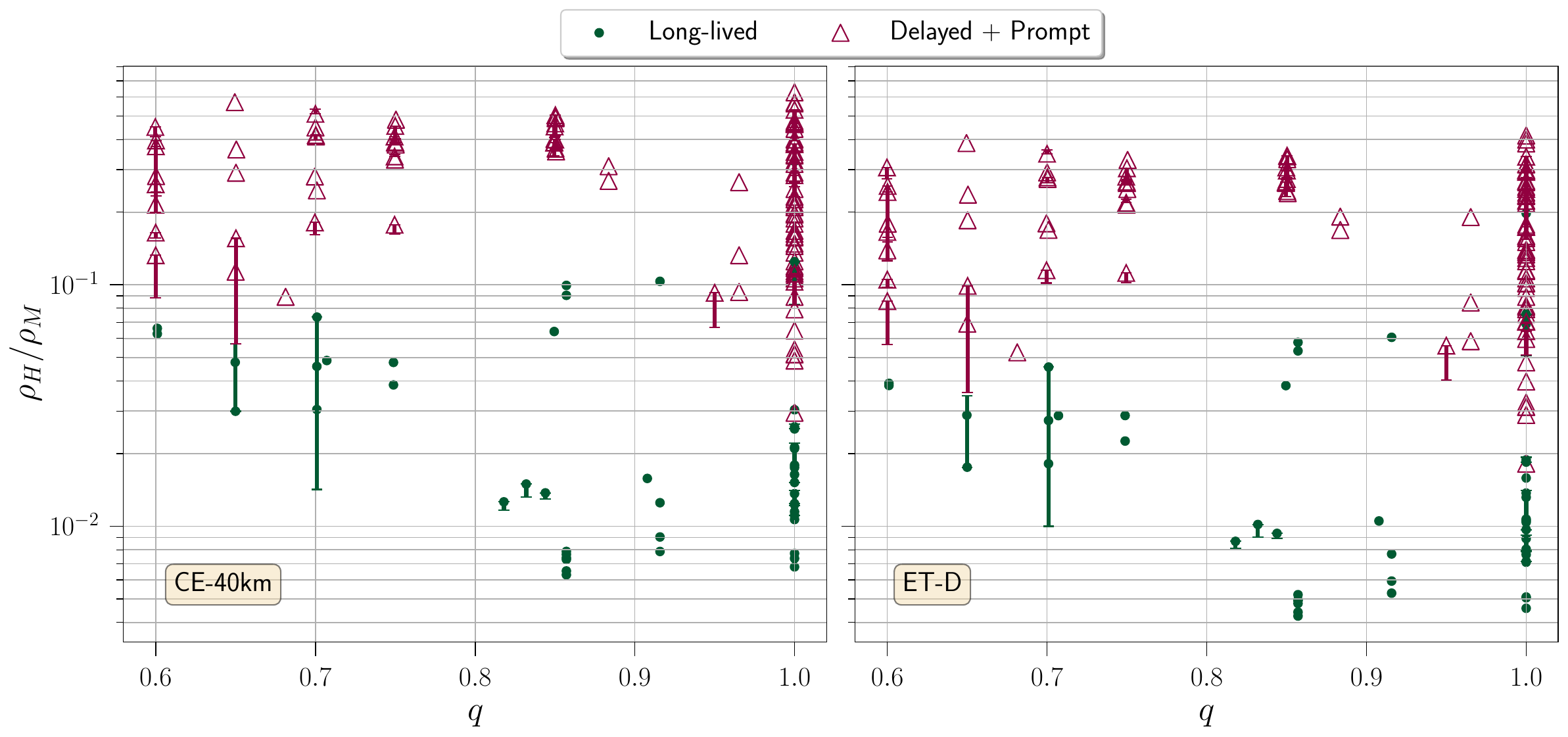}
    \caption{The ratios of the SNRs in the $f_H$ and $f_M$ frequency bands, $\rho_H / \rho_M$, as a function of the mass ratios of the binaries. The long-lived remnants are shown in green ($\bullet$) while the BH forming cases are plotted in red ($\triangle$). The results for the 40 km Cosmic Explorer detector are shown in the left panel, while the right panel shows that for the Einstein Telescope.}
    \label{fig:snr}
\end{figure*}

In order to gauge how useful the difference in the SNR in the  intermediate
($f_M$) and upper ($f_H$) frequency bands is to distinguish long-lived systems from those that collapse to a black hole (delayed and prompt collapse), we compute the SNR in the two bands for all the systems in our catalog. The systems are optimally located and oriented with respect to the detector and placed at a distance of 40 Mpc. The square of the SNR is given by
\begin{equation}\tag{1}
    \rho^2 = 4 \int_{f_{\rm low}}^{f_{high}} \frac{|\Tilde{h}(f)|^2}{S_n(f)} {\rm d}f
    \label{eq:rhosq_def}
\end{equation}
where $S_n(f)$ is the noise power spectral density and $\Tilde{h}(f)$ is the frequency domain strain. Note that we use the optimal SNR here, since we want to quantify the intrinsic properties of the signal. In an observational context, one would estimate the SNR by comparing the observed signal to a postmerger waveform model. Inevitably, this will lead to a loss in SNR due to inaccuracies in the model. Nevertheless, the current state-of-the art waveform models of BNS postmerger~\cite{Soultanis:2021oia,Breschi:2022xnc} have typical matches of $\sim95\%$ with NR simulations, which implies a $~5\%$ reduction in the matched-filter SNR. By the time the detectors considered here become operational, these differences are expected to reduce considerably; and we do not expect any material difference between the matched-filter SNR and the optimal SNR.

We remark here that for the delayed collapse remnants, the SNR in the $H$-band need not be purely due to the ringdown of the collapsing remnant (see in Appendix C 
). It has been shown that the remnants of binaries that are close to the threshold for collapse undergo strong radial oscillations~\cite{Bauswein:2015yca,DePietri:2019mti}. This results in the GW emission from the remnant having prominent power at the beating frequencies $f_{2-0}$ and $f_{2+0}$. We observe that the $f_{2+0}$ mode has a greater chance of lying in the $H$-band for delayed collapses. This is because the $f_2$ frequency and, consequently, $f_{2+0}$ frequency, increases with the total mass and binaries that result in delayed collapses that are heavier than those that lead to long-lived remnants. When this mode falls in the $M$-band, it is dwarfed by the amplitude of the $f_2$ mode and does not contribute materially to the SNR in this band. However, when lying in the $H$-band, it is the dominant contributor to this band's SNR and appears to drive the few outliers that we see.

The ratio of the SNRs in the $f_H$ and $f_M$ bands ($\rho_H$ and $\rho_M$, respectively) are shown in Fig.~\ref{fig:snr} as a function of the mass ratio of the system for two proposed next-generation detectors---Cosmic Explorer \cite{Reitze:2019iox} and Einstein Telescope \cite{Punturo:2010zz} with triangular configuration. For systems where a lower resolution is available, the figure depicts the highest resolution with a marker and the lower resolution as a one-sided errorbar with the endpoint representing the lower resolution. It is observed that $\sim80\%$ ($\sim90\%$) of long-lived remnants have a value of $\rho_H/\rho_M$ smaller than 0.05 while it is greater than that for $\sim99\%$ ($\sim95\%$) of BH forming remnants with a CE (ET) detector. The robustness of the distinguishability characteristic introduced here, $\rho_H/\rho_M$, is also tested by varying the separating frequency of 4096 Hz between 3800 Hz and 4200 Hz. This does not change the qualitative features seen in Fig.~\ref{fig:snr}. We emphasize here that our data set spans a very wide range of the physically allowed parameter space and also includes targeted simulations to GW170817 and GW190425 \cite{Nedora:2020hxc, Camilletti:2022jms} and to the known galactic binary systems \cite{Bernuzzi:2015opx}. Thus, it can be considered to be a representative sample of the population of binaries in nature as we know it today~\cite{Bauswein:2020aag, Breschi:2021xrx, Perego:2021mkd}.

We envisage the following detection scenarios for a GW170817-like event in a future detector. In \textit{Scenario 1}, there would be a direct detection (defined by some detection characteristic; typically $\rho>\rho_{\rm thr}$, for some threshold SNR, $\rho_{\rm thr}$) of a signal in the $f_H$ band which would conclusively reveal BH formation. In this case, one would calculate the fidelity of the signal by calculating the Bayes factor for the signal modeled using a waveform model that includes BH formation against (Gaussian) noise.
In \textit{Scenario 2}, the SNR in the $f_H$ band would not cross the detection threshold $\rho_{\rm thr}$ but rather one would be able to place an upper limit on $\rho_H$ due to its non-detection. In this case, one could set an upper limit on $\rho_H/\rho_M$ and assign a probability for BH formation \cite{Agathos:2019sah}. If accurate postmerger waveform models are available, one could also calculate the Bayes factor between a long-lived remnant model that models the spectral features in the $f_M$ band to a BH formation model that would include a ringdown. We report the absolute SNRs in the two bands $f_M$ and $f_H$ in the supplement.

\section{Detection prospects.}
BNS mergers are expected to be scarce in the local Universe with a median rate of $2-3 \ \rm yr^{-1}$ within a radius of 100 Mpc, and a GW170817-like event happening once every ten years~\cite{Borhanian:2022czq}. However, the uncertainty in the rates are large, and they can be as much as 5 times more (or 10 times less) frequent \cite{KAGRA:2021duu}. As such, maximizing the science potential of every nearby BNS merger is imperative,  making the development of enabling technology vital. 
Indeed a robust R\&D program for such work 
is already in place, with the 
planned evolution of aLIGO to A+~\cite{Miller:2014kma} and then on 
to the proposed LIGO-Voyager project~\cite{LIGO:2020xsf}. 
Such developments are typically cast in terms of 
the detection rate of localized GWs and 
their associated source depth, but particular improvements can grossly enable detection in 
the $1$-$10\,$kHz 
band, 
driving
new scientific opportunities~\cite{Torres-Rivas:2018svp,Martynov:2019gvu,maggiore2020science,Evans:2021gyd}. In this $f$ region
reducing the effects of quantum noise is 
key, and given implemented and 
anticipated technical improvements~\cite{LIGOScientific:2013pcc,Miller:2014kma,LIGO:2020xsf}, increasing the optical 
power in the interferometer 
arms, which is limited by 
the test mass material and its 
coatings at a given laser wavelength~\cite{Zhou:2022xav},  
could yield improved 
sensitivities~\cite{LIGO:2020xsf}. 
We note that laser interferometers
could also be particularly designed to achieve maximal 
sensitivity in the crucial $f > f_{\rm peak}$ range, 
as in \cite{Zhang:2020qlh,srivastava2022science}. 
These plans should ultimately be able to supply 
the experimental sensitivity to discriminate between
the possible scenarios for $f > f_{\rm peak}$ shown in 
Fig.~\ref{fig:strain}. 

There are also 
non-interferometer experiments under development
with a sensitivity window in 
the $1-10\,$kHz range~\cite{Aggarwal:2020olq}, 
offering the prospect of GW detection in this range 
on a more rapid time scale and at
substantially lower cost. We now turn to a brief
overview of such 
possibilities. 
Generally these employ 
the conversion of gravitational to a
mechanical or electromagnetic signal. 
Possibilities include 
(i) superconducting 
cavities~\cite{Braginskii:1973vm,Pegoraro:1978gv,Pegoraro_1978,Caves:1979kq,Reece:1982sc,Reece:1984gv, 
Bernard:2001kp,Bernard:2002ci,Ballantini:2003nt,Ballantini:2005am,Berlin:2023grv}, (ii) resonant mass
detectors~\cite{PhysRev.117.306,PhysRevLett.17.1228,Forward:1971mel,Liccardo:2023nzv,Gottardi:2007zn,Cerdonio:1997hz,Vinante:2006uk,Harry:1996gh,Aguiar:2009zzb}, (iii) optically
levitated sensors~\cite{Arvanitaki:2012cn,Aggarwal:2020umq}, (iv) atomic 
interferometers, either ground-based~\cite{Graham:2017pmn} or in space~\cite{AEDGE:2019nxb,Badurina:2019hst}, 
 or (v)
atomic clocks in space~\cite{Loeb:2015ffa,Vutha:2015aza,Kolkowitz:2016wyg}. In the atomic experiments the sensitivity window is about 1 mHz to 10 Hz, but extracting 
information in the few kHz regime should be 
possible~\cite{Bringmann:2023gba}. 

In pumped superconducting cavities,
an incoming GW can deform the cavity walls and induce a transition from the pump mode to a signal mode. This idea was developed in \cite{Braginskii:1973vm,Pegoraro:1978gv,Pegoraro_1978,Caves:1979kq,Reece:1982sc,Reece:1984gv} 
and a prototype, known as MAGO, was 
built~\cite{Bernard:2001kp,Bernard:2002ci,Ballantini:2003nt,Ballantini:2005am}, though it
was never put into operation. 
Recently MAGO-like detectors have regained interest~\cite{Berlin:2023grv}. 
Since the expected signal from the post merger phase 
is relatively broad, i.e., 
on the order of a kHz, the detector has to be used in its broadband mode, with an expected thermal noise dominated strain sensitivity 
on the order of $10^{-18}$-$10^{-19}\,\sqrt{\rm Hz}^{-1}$ in the $6$-$8$ kHz regime~\cite{Berlin:2023grv}. 
In the case of 
resonant mass detectors~\cite{PhysRev.117.306,PhysRevLett.17.1228} 
an incoming GW deforms the walls of the detector, and the deformation is then converted to an electromagnetic signal by several transducers~\cite{Forward:1971mel}. 
Examples such as the Schenberg spherical resonant mass antenna~\cite{Liccardo:2023nzv}, the MiniGrail experiment~\cite{Gottardi:2007zn} or the AURIGA experiment~\cite{Cerdonio:1997hz,Vinante:2006uk} are 
focused on the 
frequency regime below 4 kHz,
though higher  frequency studies  yield an estimated sensitivity of about $4\times 10^{-23}\,\sqrt{\rm Hz}^{-1}$~\cite{Harry:1996gh} and $10^{-22}\,\sqrt{\rm Hz}^{-1}$~\cite{Aguiar:2009zzb} at $5\,$kHz and $6\,$kHz respectively. 
It has been shown theoretically \cite{Harry:1996gh}
that a bandwidth of the order of a $1\,$kHz can be achieved with a single sphere, with 
the use of multiple spheres 
enabling still broader sensitivity~\cite{Aguiar:2009zzb}, or the ability to probe
either the $f_M$ {\it or} $f_H$
regions shown in Fig.~\ref{fig:strain}, or both. 
 In the case 
 of 
optically-levitated sensors 
the estimated strain sensitivity is $3\times 10^{-22}\,\sqrt{\rm Hz}^{-1}$ at $10$ kHz~\cite{Arvanitaki:2012cn,Aggarwal:2020umq}. By using a larger levitated mass the sensitivity could potentially be increased even further~\cite{Aggarwal:2020umq}.
The sensitivity of atomic clocks at $10\,$kHz was estimated for a specific experimental scenario in \cite{Bringmann:2023gba} as $10^{-17}\,\sqrt{\rm Hz}^{-1}$.
Future technological developments could well increase the sensitivity of 
non-interferometer methods, 
but for now interferometers such as CE and ET seem to have at least one order of magnitude better sensitivity in the $1-10\,$kHz regime. 

\section{Discussion.}
We have shown that it might be possible to use the presence/absence of a high-frequency signal in the $[4096, 8192]~{\rm Hz}$-band to confidently confirm/exclude \ac{BH} formation for binary mergers with sufficiently high $\rho_M \gtrsim 10$, yielding $\rho_H \gtrsim 1$ for \ac{BH} forming binaries. This corresponds to distances ranging from 40~Mpc to 100~Mpc for 90\% of the systems considered. Combined with the measurement of the binary parameters from the inspiral signal, the knowledge of the fate of the binary would strongly constrain the physics of matter at extreme densities. We remark that binaries that collapse within ~15
milliseconds do so due to the loss of angular momentum to gravitational
waves. For longer-lived remnants, the gravitational-wave losses are
negligible, so the mechanism for collapse is qualitatively different. As
such, it is an important result of our work that short- and long-lived
remnants can be distinguished by means of gravitational-wave
observations. Knowing if and when a \ac{BH} is formed in a \ac{NS} merger would also impact our understanding of the central engine of short gamma-ray bursts.

Considerable theoretical and practical challenges need to be addressed to enable the type of measurements and inference we are proposing. On the experimental side, future development of new techniques to improve the sensitivity of \ac{GW} observatories at high frequency, or the development of new high-frequency optimized detectors might be required to extend the detection horizon of the \ac{BH} ringdown in \ac{NS} mergers. On the theory side, substantial advances in 
\ac{NR}
is required to produce reliable and accurate \ac{BH} formation time predictions for different binary parameters and \acp{EOS}. We believe that this work strongly motivates research in these directions. 

We briefly discuss the extensions and caveats of this work that we plan to take up in the future. In Fig.~\ref{fig:delayed_excised}, we show that by windowing the signal to exclude the collapse, the high-frequency contribution can be separated out. A dedicated data analysis study to quantify the feasibility of such a procedure for a wide range of simulations is left for the future. In this study we have not quantified the relative probability of occurrence of systems with different masses as well as the probabilities for the \ac{EOS}. This is due to wide uncertainties in the \ac{NS} mass distribution as well as for the \ac{EOS}. In the future, when there are greater number of detections, one can reweight the correlations that we report on with specific mass distributions and \ac{EOS} probabilities.

\section*{Acknowledgements}
We thank Kevin Kuns for the Cosmic Explorer sensitivity curve. We thank Torsten Bringmann, Valerie Domcke, Sebastian Ellis, Elina Fuchs and Joachim Kopp for helpful conversations on the detection prospects section. A.~D.~was partially supported by the National Science Foundation (NSF) grant No.~PHY-2012083. D.~R.~acknowledges funding from the U.S. Department of Energy, Office of Science, Division of Nuclear Physics under Award Number(s) DE-SC0021177 and from the NSF under Grants No.~PHY-2011725, No.~PHY-2020275, No.~PHY-2116686, and No.~AST-2108467.  The work of J.~S.~E.~is supported in part by DOE grant No. DE-SC0015655. S.~G. acknowledges partial support from the  U.S. Department of Energy, Office of Science, Office of Nuclear Physics  under contract DE-FG02-96ER40989 and from the  Universities Research Association in support  of a sabbatical visit to Fermilab and thanks   the Theoretical Division  for kind hospitality.  B.~S.~S.~was supported in part by NSF Grants No. PHY-2308886, No. PHY-2207638, PHY-2012083 and No. AST-2006384.  A.~P. and D.~L. acknowledges PRACE for awarding them access to Joliot-Curie at GENCI@CEA. A.~P. also  acknowledges the usage of computer resources under a CINECA-INFN agreement (allocation INF21\_teongrav and INF22\_teongrav).

\appendix
\label{sec:supp}

\section{Parameter space}
\begin{figure}
    \centering
    \includegraphics[width=\columnwidth]{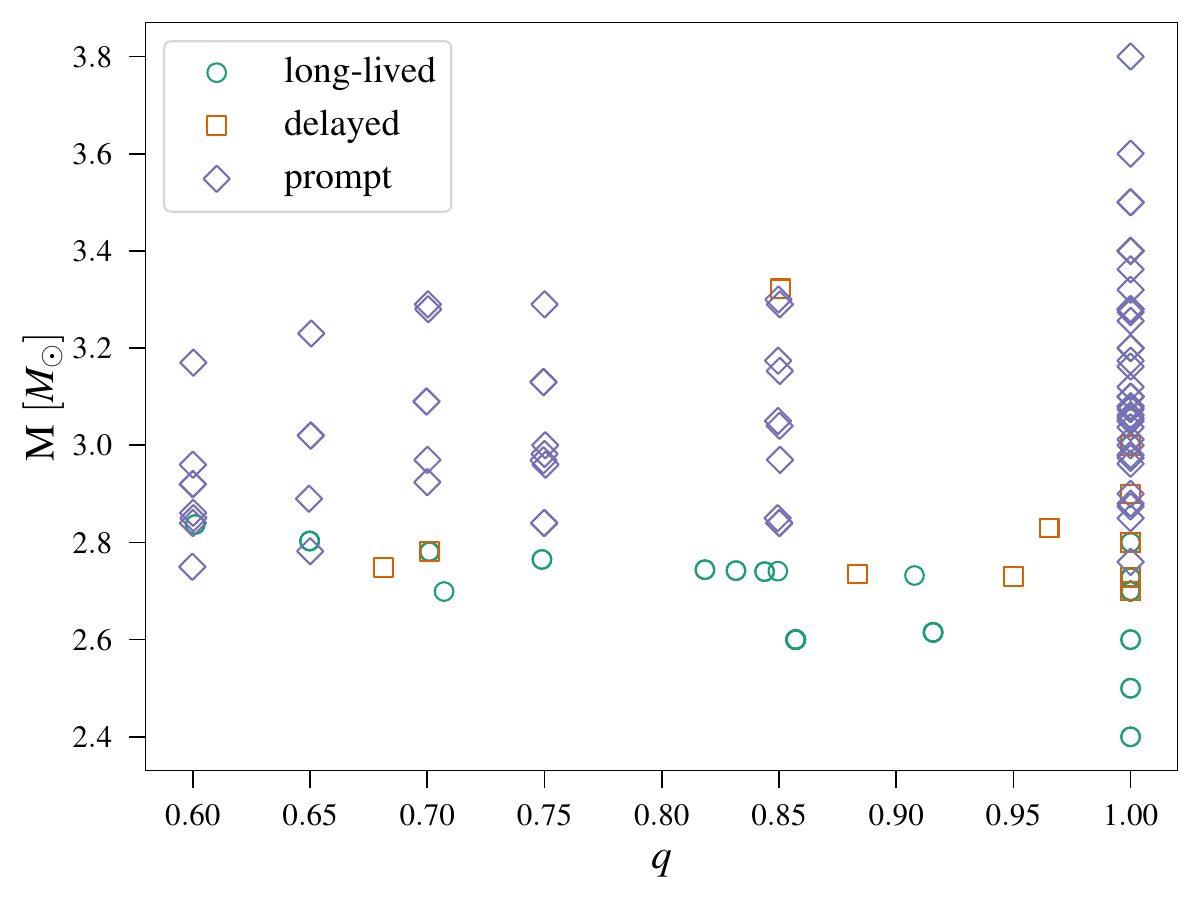}
    \caption{Total mass and mass ratio for the simulations considered in this study. Each simulation is classified according to the merger outcome: prompt or delayed BH formation, or no BH formation within simulation time (long-lived remnants). Note that many simulation share the same total mass and mass ratio, but differ in the choice of EOS, so that 
    there are fewer than 152 points.}
    \label{fig:sims}
\end{figure}

Figure \ref{fig:sims} shows the parameter space of the simulations used in this work. They include simulations with masses taken from the known galactic double \ac{NS} systems \cite{Bernuzzi:2015opx}, as well as simulations targeted to GW170817 \cite{Nedora:2020hxc} and GW190425 \cite{Camilletti:2022jms}. Moreover, our data set covers a wide range of mass ratios and total masses. To avoid biases, we have included all 
the simulations performed by our groups to date.

\section{Long-lived outlier}
\begin{figure*}
    \centering
    \includegraphics[width=2\columnwidth]{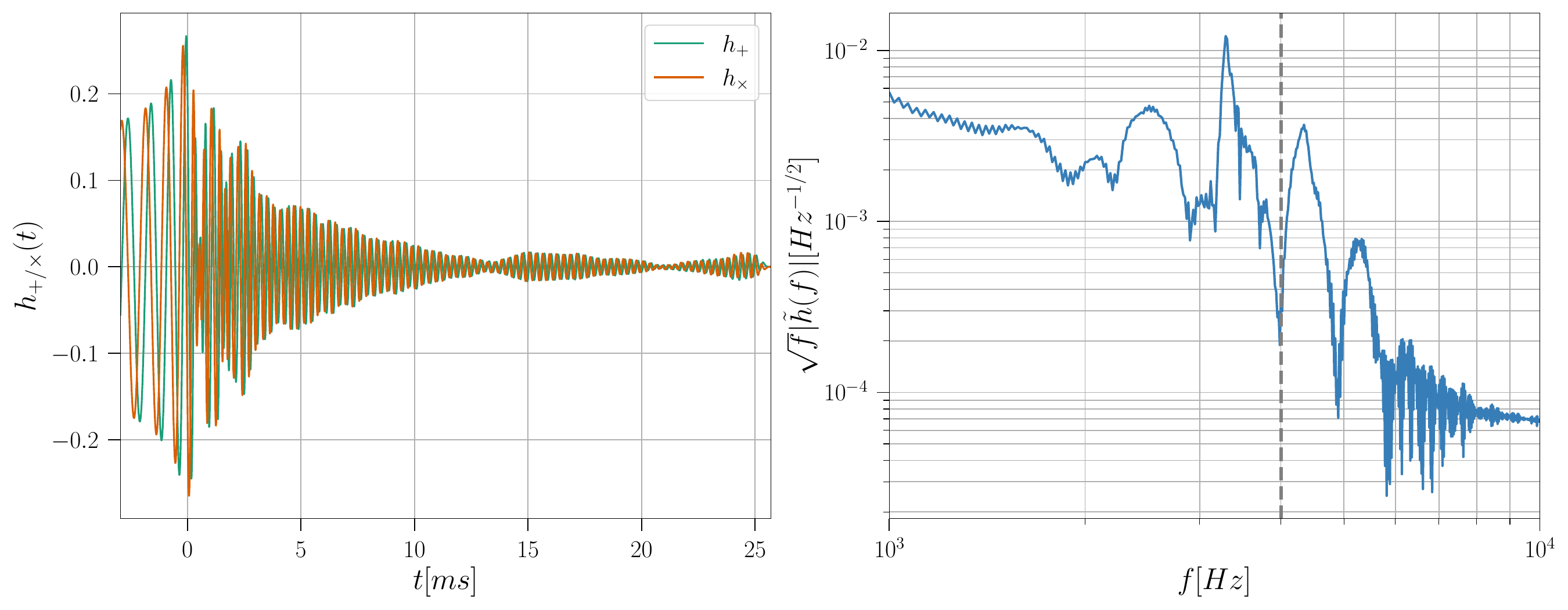}
    \caption{An example of a long-lived remnant with significant power in the $f_H$ band.
    \textit{Left:} 
    The time domain strain of the post-merger waveform showing the two GW polarizations.  \textit{Right:} The spectrum corresponding to the strain on the left panel.}
    \label{fig:long_lived_outlier}
\end{figure*}

Figure \ref{fig:long_lived_outlier} shows a long-lived remnant with a prominent spectral feature at $f>4$ kHz leading to 
significant power in the $f_H$ band. It is an equal mass simulation with component masses 1.35 $M_{\odot}$ 
using the 
SLy EoS. This is one of the outliers in the Fig.~\ref{fig:snr} with a large $\rho_H/\rho_M$ compared to other long-lived remnants. This binary shows strong modulations of the \ac{GW} signal due to radial oscillations. As a result of this, there is a high-frequency peak in the spectrum at the frequency $f_{2+0}$, which is due to the beating between the radial pulsation mode of the star ($f_0$) and the main postmerger peak frequency ($f_2$). All the long-lived outliers present these strong modulations. In some cases, the $f_{2+0}$ mode is in the $M$-band, while in other cases it is in the $H$-band, as in the example presented here. We remark that such amplitude modulations have been reported also by other groups, e.g.~\cite{DePietri:2019mti}, particularly for remnants close to the threshold for collapse.

\section{Delayed collapse with/without ringdown}
\begin{figure*}
    \centering
    \includegraphics[width=2\columnwidth]{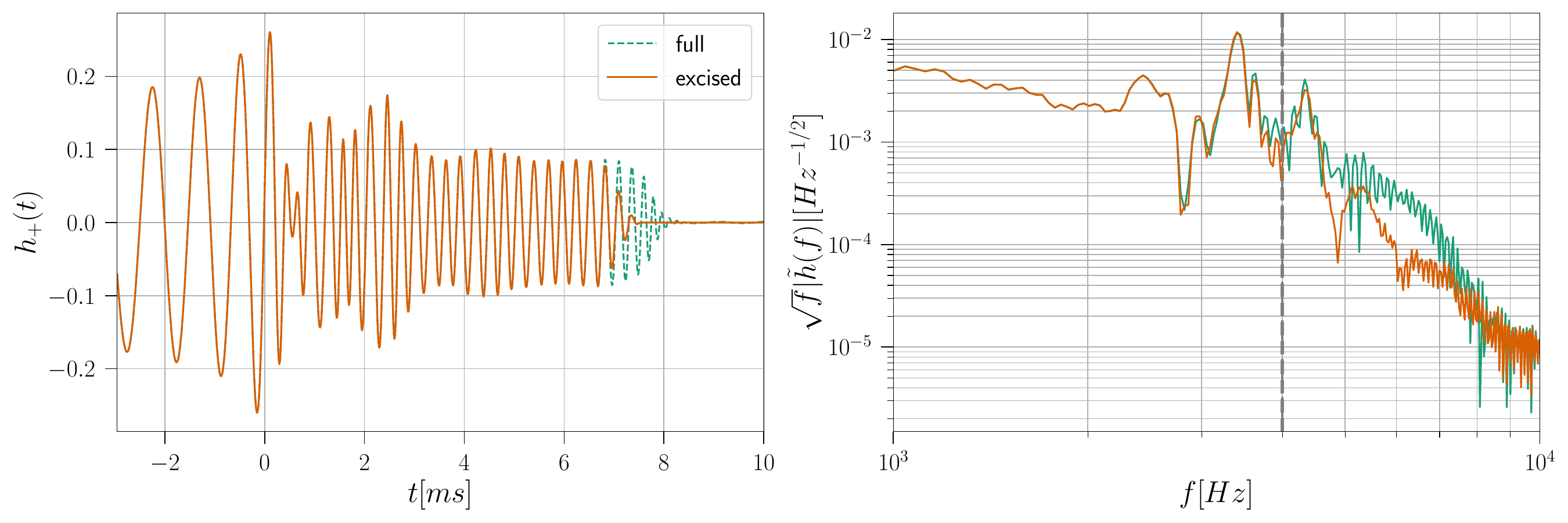}
    \caption{An example of a delayed collapse remnant with equal component masses of 1.35 $M_{\odot}$ and employing SFHo EoS. \textit{Left:} The time domain strain of the full postmerger waveform (green) and a section of the postmerger with the ringdown excised out. \textit{Right:} The spectrum corresponding to the strain in the left panel.}
    \label{fig:delayed_excised}
\end{figure*}

Figure \ref{fig:delayed_excised} shows the spectrum of a delayed collapse remnant when its ringdown is excised out. It can be observed that the power in the $H$-band is strongly suppressed, particularly at the frequencies corresponding to the quasi-normal ringdown of the final \ac{BH}. However, there is still power in the $f_{2+0}$ mode at a frequency of ${\sim}4.5\ {\rm kHz}$. This shows that the $H$-band 
probes not only the ringdown signal, but also the presence of strong radial pulsations in the remnant, which are associated with the \ac{BH} formation.

\section{Absolute SNRs}
\label{subsec:absolute_snrs}
\begin{figure*}
    \centering
    \includegraphics[width=2\columnwidth]{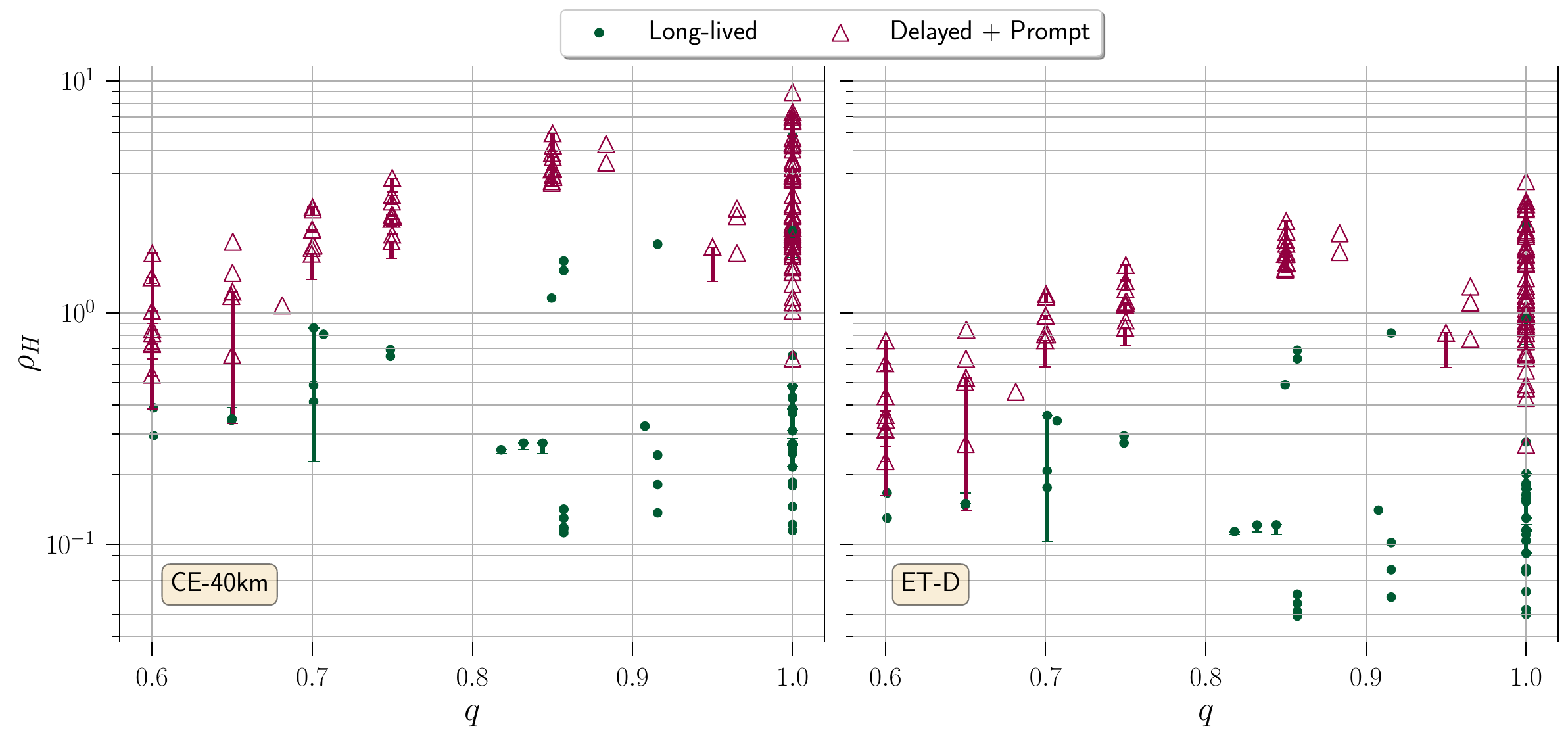}
    \includegraphics[width=2\columnwidth]{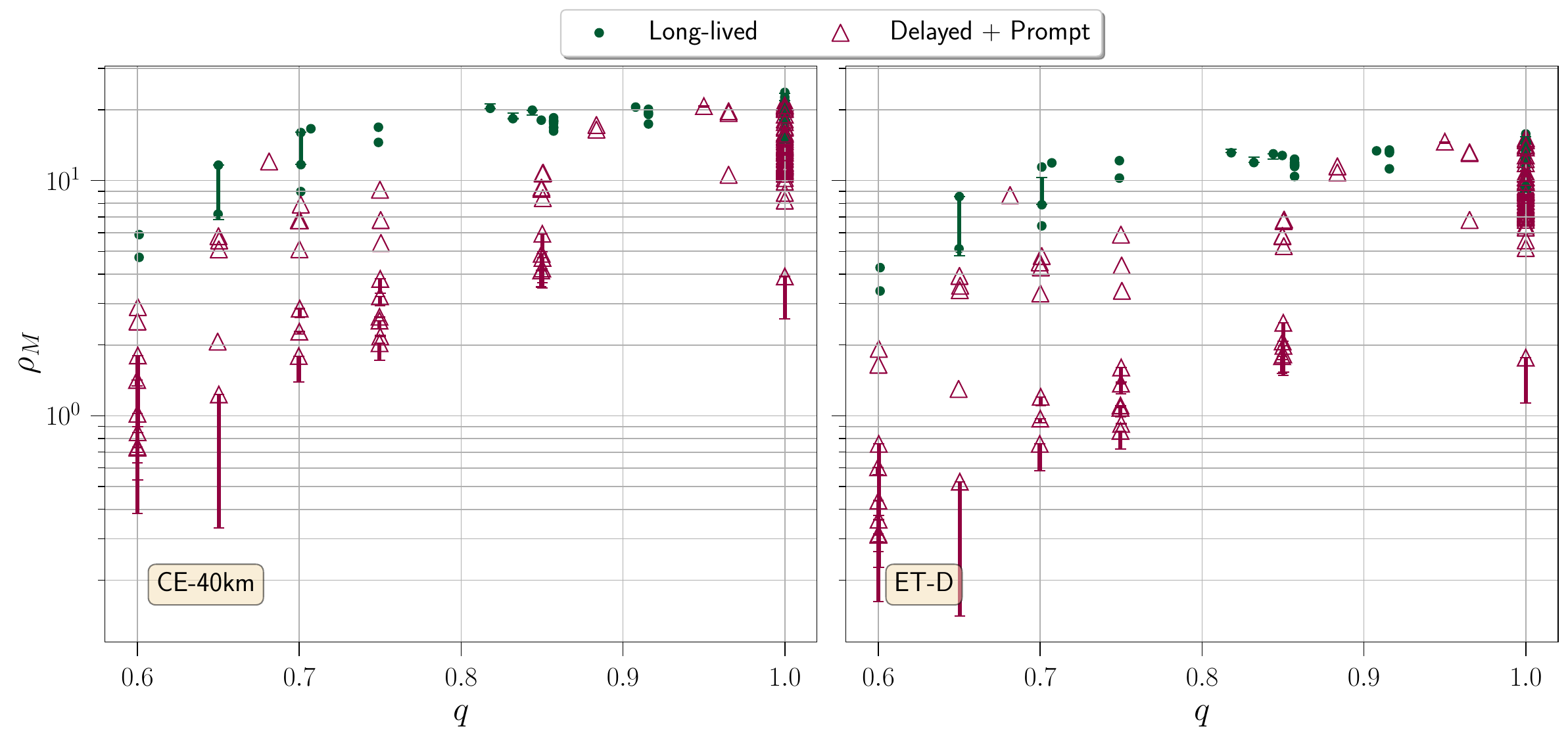}
    \caption{SNRs of long-lived and BH forming binaries in the frequency ranges 
    $f_H\in[4096, 8192]$ Hz (top panel) and 
    $f_M\in[2048, 4096]$ Hz (bottom panel) assuming optimal orientation and sky position at a distance of 40 Mpc.}
    \label{fig:snr_rd_pm}
\end{figure*}

The absolute SNRs for the long-lived remnants and BH-forming remnants are shown in Fig.~\ref{fig:snr_rd_pm}. It can be seen that a majority of the long-lived remnants have an SNR greater than 10 in the $f_M$ band at 40 Mpc and will have an SNR $\gtrsim 1$ for distances ranging from 400 - 1000 Mpc. It is also clear that BH-forming remnants have smaller SNRs in general. The $q=1$ cases where the two groups have similar SNRs are mostly composed 
of delayed collapse remnants that have the $f_{\rm peak}$ peak as discussed earlier in the section. On the other hand, the SNRs for the BH-forming remnants are greater in the $f_H$ band, albeit the values are lower than the $f_M$ band as expected. In this band, most of the BH-forming remnants have $\rho_H \gtrsim 1$ at 40 Mpc in the CE detector with the loudest having $\rho_H \gtrsim 1$ up to a distance of 350 Mpc.

We elucidate our rationale for using $\rho=1$ in the above arguments. An SNR = 1 assigns a 1-$\sigma$ ($\sim 68\%$) confidence that the signal in a given segment of data is not a (Gaussian) noise artifact. In standard GW data analysis, one does not know a priori which data segment contains a signal and, therefore, the chance of noise imitating a signal (false alarm rate) for $\rho=1$ is very large when one considers the full data. However, the signals of interest here will be preceded by their inspiral-merger signals that would have SNRs of $\mathcal{O}(10^2-10^3)$. Hence, one knows a priori where the signal lies.

\bibliographystyle{apsrev4-1}
\bibliography{bns_ringdown}{}

\begin{thebibliography}{133}%
\makeatletter
\providecommand \@ifxundefined [1]{%
 \@ifx{#1\undefined}
}%
\providecommand \@ifnum [1]{%
 \ifnum #1\expandafter \@firstoftwo
 \else \expandafter \@secondoftwo
 \fi
}%
\providecommand \@ifx [1]{%
 \ifx #1\expandafter \@firstoftwo
 \else \expandafter \@secondoftwo
 \fi
}%
\providecommand \natexlab [1]{#1}%
\providecommand \enquote  [1]{``#1''}%
\providecommand \bibnamefont  [1]{#1}%
\providecommand \bibfnamefont [1]{#1}%
\providecommand \citenamefont [1]{#1}%
\providecommand \href@noop [0]{\@secondoftwo}%
\providecommand \href [0]{\begingroup \@sanitize@url \@href}%
\providecommand \@href[1]{\@@startlink{#1}\@@href}%
\providecommand \@@href[1]{\endgroup#1\@@endlink}%
\providecommand \@sanitize@url [0]{\catcode `\\12\catcode `\$12\catcode
  `\&12\catcode `\#12\catcode `\^12\catcode `\_12\catcode `\%12\relax}%
\providecommand \@@startlink[1]{}%
\providecommand \@@endlink[0]{}%
\providecommand \url  [0]{\begingroup\@sanitize@url \@url }%
\providecommand \@url [1]{\endgroup\@href {#1}{\urlprefix }}%
\providecommand \urlprefix  [0]{URL }%
\providecommand \Eprint [0]{\href }%
\providecommand \doibase [0]{http://dx.doi.org/}%
\providecommand \selectlanguage [0]{\@gobble}%
\providecommand \bibinfo  [0]{\@secondoftwo}%
\providecommand \bibfield  [0]{\@secondoftwo}%
\providecommand \translation [1]{[#1]}%
\providecommand \BibitemOpen [0]{}%
\providecommand \bibitemStop [0]{}%
\providecommand \bibitemNoStop [0]{.\EOS\space}%
\providecommand \EOS [0]{\spacefactor3000\relax}%
\providecommand \BibitemShut  [1]{\csname bibitem#1\endcsname}%
\let\auto@bib@innerbib\@empty
\bibitem [{\citenamefont {Perego}\ \emph {et~al.}(2019)\citenamefont {Perego},
  \citenamefont {Bernuzzi},\ and\ \citenamefont {Radice}}]{Perego:2019adq}%
  \BibitemOpen
  \bibfield  {author} {\bibinfo {author} {\bibfnamefont {A.}~\bibnamefont
  {Perego}}, \bibinfo {author} {\bibfnamefont {S.}~\bibnamefont {Bernuzzi}}, \
  and\ \bibinfo {author} {\bibfnamefont {D.}~\bibnamefont {Radice}},\ }\href
  {\doibase 10.1140/epja/i2019-12810-7} {\bibfield  {journal} {\bibinfo
  {journal} {Eur. Phys. J. A}\ }\textbf {\bibinfo {volume} {55}},\ \bibinfo
  {pages} {124} (\bibinfo {year} {2019})},\ \Eprint
  {http://arxiv.org/abs/1903.07898} {arXiv:1903.07898 [gr-qc]} \BibitemShut
  {NoStop}%
\bibitem [{\citenamefont {Drischler}\ \emph {et~al.}(2021)\citenamefont
  {Drischler}, \citenamefont {Holt},\ and\ \citenamefont
  {Wellenhofer}}]{Drischler:2021kxf}%
  \BibitemOpen
  \bibfield  {author} {\bibinfo {author} {\bibfnamefont {C.}~\bibnamefont
  {Drischler}}, \bibinfo {author} {\bibfnamefont {J.~W.}\ \bibnamefont {Holt}},
  \ and\ \bibinfo {author} {\bibfnamefont {C.}~\bibnamefont {Wellenhofer}},\
  }\href {\doibase 10.1146/annurev-nucl-102419-041903} {\bibfield  {journal}
  {\bibinfo  {journal} {Ann. Rev. Nucl. Part. Sci.}\ }\textbf {\bibinfo
  {volume} {71}},\ \bibinfo {pages} {403} (\bibinfo {year} {2021})},\ \Eprint
  {http://arxiv.org/abs/2101.01709} {arXiv:2101.01709 [nucl-th]} \BibitemShut
  {NoStop}%
\bibitem [{\citenamefont {Lattimer}(2021)}]{Lattimer:2021emm}%
  \BibitemOpen
  \bibfield  {author} {\bibinfo {author} {\bibfnamefont {J.~M.}\ \bibnamefont
  {Lattimer}},\ }\href {\doibase 10.1146/annurev-nucl-102419-124827} {\bibfield
   {journal} {\bibinfo  {journal} {Ann. Rev. Nucl. Part. Sci.}\ }\textbf
  {\bibinfo {volume} {71}},\ \bibinfo {pages} {433} (\bibinfo {year}
  {2021})}\BibitemShut {NoStop}%
\bibitem [{\citenamefont {Kumar}\ \emph {et~al.}(2023)\citenamefont {Kumar}
  \emph {et~al.}}]{MUSES:2023hyz}%
  \BibitemOpen
  \bibfield  {author} {\bibinfo {author} {\bibfnamefont {R.}~\bibnamefont
  {Kumar}} \emph {et~al.} (\bibinfo {collaboration} {MUSES}),\ }\href@noop {}
  {\  (\bibinfo {year} {2023})},\ \Eprint {http://arxiv.org/abs/2303.17021}
  {arXiv:2303.17021 [nucl-th]} \BibitemShut {NoStop}%
\bibitem [{\citenamefont {Danielewicz}\ \emph {et~al.}(2002)\citenamefont
  {Danielewicz}, \citenamefont {Lacey},\ and\ \citenamefont
  {Lynch}}]{Danielewicz:2002pu}%
  \BibitemOpen
  \bibfield  {author} {\bibinfo {author} {\bibfnamefont {P.}~\bibnamefont
  {Danielewicz}}, \bibinfo {author} {\bibfnamefont {R.}~\bibnamefont {Lacey}},
  \ and\ \bibinfo {author} {\bibfnamefont {W.~G.}\ \bibnamefont {Lynch}},\
  }\href {\doibase 10.1126/science.1078070} {\bibfield  {journal} {\bibinfo
  {journal} {Science}\ }\textbf {\bibinfo {volume} {298}},\ \bibinfo {pages}
  {1592} (\bibinfo {year} {2002})},\ \Eprint
  {http://arxiv.org/abs/nucl-th/0208016} {arXiv:nucl-th/0208016} \BibitemShut
  {NoStop}%
\bibitem [{\citenamefont {Horowitz}\ \emph {et~al.}(2014)\citenamefont
  {Horowitz}, \citenamefont {Brown}, \citenamefont {Kim}, \citenamefont
  {Lynch}, \citenamefont {Michaels}, \citenamefont {Ono}, \citenamefont
  {Piekarewicz}, \citenamefont {Tsang},\ and\ \citenamefont
  {Wolter}}]{Horowitz:2014bja}%
  \BibitemOpen
  \bibfield  {author} {\bibinfo {author} {\bibfnamefont {C.~J.}\ \bibnamefont
  {Horowitz}}, \bibinfo {author} {\bibfnamefont {E.~F.}\ \bibnamefont {Brown}},
  \bibinfo {author} {\bibfnamefont {Y.}~\bibnamefont {Kim}}, \bibinfo {author}
  {\bibfnamefont {W.~G.}\ \bibnamefont {Lynch}}, \bibinfo {author}
  {\bibfnamefont {R.}~\bibnamefont {Michaels}}, \bibinfo {author}
  {\bibfnamefont {A.}~\bibnamefont {Ono}}, \bibinfo {author} {\bibfnamefont
  {J.}~\bibnamefont {Piekarewicz}}, \bibinfo {author} {\bibfnamefont {M.~B.}\
  \bibnamefont {Tsang}}, \ and\ \bibinfo {author} {\bibfnamefont {H.~H.}\
  \bibnamefont {Wolter}},\ }\href {\doibase 10.1088/0954-3899/41/9/093001}
  {\bibfield  {journal} {\bibinfo  {journal} {J. Phys. G}\ }\textbf {\bibinfo
  {volume} {41}},\ \bibinfo {pages} {093001} (\bibinfo {year} {2014})},\
  \Eprint {http://arxiv.org/abs/1401.5839} {arXiv:1401.5839 [nucl-th]}
  \BibitemShut {NoStop}%
\bibitem [{\citenamefont {Le~F\`evre}\ \emph {et~al.}(2016)\citenamefont
  {Le~F\`evre}, \citenamefont {Leifels}, \citenamefont {Reisdorf},
  \citenamefont {Aichelin},\ and\ \citenamefont {Hartnack}}]{LeFevre:2015paj}%
  \BibitemOpen
  \bibfield  {author} {\bibinfo {author} {\bibfnamefont {A.}~\bibnamefont
  {Le~F\`evre}}, \bibinfo {author} {\bibfnamefont {Y.}~\bibnamefont {Leifels}},
  \bibinfo {author} {\bibfnamefont {W.}~\bibnamefont {Reisdorf}}, \bibinfo
  {author} {\bibfnamefont {J.}~\bibnamefont {Aichelin}}, \ and\ \bibinfo
  {author} {\bibfnamefont {C.}~\bibnamefont {Hartnack}},\ }\href {\doibase
  10.1016/j.nuclphysa.2015.09.015} {\bibfield  {journal} {\bibinfo  {journal}
  {Nucl. Phys. A}\ }\textbf {\bibinfo {volume} {945}},\ \bibinfo {pages} {112}
  (\bibinfo {year} {2016})},\ \Eprint {http://arxiv.org/abs/1501.05246}
  {arXiv:1501.05246 [nucl-ex]} \BibitemShut {NoStop}%
\bibitem [{\citenamefont {Russotto}\ \emph {et~al.}(2016)\citenamefont
  {Russotto} \emph {et~al.}}]{Russotto:2016ucm}%
  \BibitemOpen
  \bibfield  {author} {\bibinfo {author} {\bibfnamefont {P.}~\bibnamefont
  {Russotto}} \emph {et~al.},\ }\href {\doibase 10.1103/PhysRevC.94.034608}
  {\bibfield  {journal} {\bibinfo  {journal} {Phys. Rev. C}\ }\textbf {\bibinfo
  {volume} {94}},\ \bibinfo {pages} {034608} (\bibinfo {year} {2016})},\
  \Eprint {http://arxiv.org/abs/1608.04332} {arXiv:1608.04332 [nucl-ex]}
  \BibitemShut {NoStop}%
\bibitem [{\citenamefont {Reed}\ \emph {et~al.}(2021)\citenamefont {Reed},
  \citenamefont {Fattoyev}, \citenamefont {Horowitz},\ and\ \citenamefont
  {Piekarewicz}}]{Reed:2021nqk}%
  \BibitemOpen
  \bibfield  {author} {\bibinfo {author} {\bibfnamefont {B.~T.}\ \bibnamefont
  {Reed}}, \bibinfo {author} {\bibfnamefont {F.~J.}\ \bibnamefont {Fattoyev}},
  \bibinfo {author} {\bibfnamefont {C.~J.}\ \bibnamefont {Horowitz}}, \ and\
  \bibinfo {author} {\bibfnamefont {J.}~\bibnamefont {Piekarewicz}},\ }\href
  {\doibase 10.1103/PhysRevLett.126.172503} {\bibfield  {journal} {\bibinfo
  {journal} {Phys. Rev. Lett.}\ }\textbf {\bibinfo {volume} {126}},\ \bibinfo
  {pages} {172503} (\bibinfo {year} {2021})},\ \Eprint
  {http://arxiv.org/abs/2101.03193} {arXiv:2101.03193 [nucl-th]} \BibitemShut
  {NoStop}%
\bibitem [{\citenamefont {Steiner}\ \emph {et~al.}(2013)\citenamefont
  {Steiner}, \citenamefont {Lattimer},\ and\ \citenamefont
  {Brown}}]{Steiner:2012xt}%
  \BibitemOpen
  \bibfield  {author} {\bibinfo {author} {\bibfnamefont {A.~W.}\ \bibnamefont
  {Steiner}}, \bibinfo {author} {\bibfnamefont {J.~M.}\ \bibnamefont
  {Lattimer}}, \ and\ \bibinfo {author} {\bibfnamefont {E.~F.}\ \bibnamefont
  {Brown}},\ }\href {\doibase 10.1088/2041-8205/765/1/L5} {\bibfield  {journal}
  {\bibinfo  {journal} {Astrophys. J. Lett.}\ }\textbf {\bibinfo {volume}
  {765}},\ \bibinfo {pages} {L5} (\bibinfo {year} {2013})},\ \Eprint
  {http://arxiv.org/abs/1205.6871} {arXiv:1205.6871 [nucl-th]} \BibitemShut
  {NoStop}%
\bibitem [{\citenamefont {\"Ozel}\ and\ \citenamefont
  {Freire}(2016)}]{Ozel:2016oaf}%
  \BibitemOpen
  \bibfield  {author} {\bibinfo {author} {\bibfnamefont {F.}~\bibnamefont
  {\"Ozel}}\ and\ \bibinfo {author} {\bibfnamefont {P.}~\bibnamefont
  {Freire}},\ }\href {\doibase 10.1146/annurev-astro-081915-023322} {\bibfield
  {journal} {\bibinfo  {journal} {Ann. Rev. Astron. Astrophys.}\ }\textbf
  {\bibinfo {volume} {54}},\ \bibinfo {pages} {401} (\bibinfo {year} {2016})},\
  \Eprint {http://arxiv.org/abs/1603.02698} {arXiv:1603.02698 [astro-ph.HE]}
  \BibitemShut {NoStop}%
\bibitem [{\citenamefont {Miller}\ \emph {et~al.}(2019)\citenamefont {Miller}
  \emph {et~al.}}]{Miller:2019cac}%
  \BibitemOpen
  \bibfield  {author} {\bibinfo {author} {\bibfnamefont {M.~C.}\ \bibnamefont
  {Miller}} \emph {et~al.},\ }\href {\doibase 10.3847/2041-8213/ab50c5}
  {\bibfield  {journal} {\bibinfo  {journal} {Astrophys. J. Lett.}\ }\textbf
  {\bibinfo {volume} {887}},\ \bibinfo {pages} {L24} (\bibinfo {year}
  {2019})},\ \Eprint {http://arxiv.org/abs/1912.05705} {arXiv:1912.05705
  [astro-ph.HE]} \BibitemShut {NoStop}%
\bibitem [{\citenamefont {Pang}\ \emph {et~al.}(2021)\citenamefont {Pang},
  \citenamefont {Tews}, \citenamefont {Coughlin}, \citenamefont {Bulla},
  \citenamefont {Van Den~Broeck},\ and\ \citenamefont
  {Dietrich}}]{Pang:2021jta}%
  \BibitemOpen
  \bibfield  {author} {\bibinfo {author} {\bibfnamefont {P.~T.~H.}\
  \bibnamefont {Pang}}, \bibinfo {author} {\bibfnamefont {I.}~\bibnamefont
  {Tews}}, \bibinfo {author} {\bibfnamefont {M.~W.}\ \bibnamefont {Coughlin}},
  \bibinfo {author} {\bibfnamefont {M.}~\bibnamefont {Bulla}}, \bibinfo
  {author} {\bibfnamefont {C.}~\bibnamefont {Van Den~Broeck}}, \ and\ \bibinfo
  {author} {\bibfnamefont {T.}~\bibnamefont {Dietrich}},\ }\href {\doibase
  10.3847/1538-4357/ac19ab} {\bibfield  {journal} {\bibinfo  {journal}
  {Astrophys. J.}\ }\textbf {\bibinfo {volume} {922}},\ \bibinfo {pages} {14}
  (\bibinfo {year} {2021})},\ \Eprint {http://arxiv.org/abs/2105.08688}
  {arXiv:2105.08688 [astro-ph.HE]} \BibitemShut {NoStop}%
\bibitem [{\citenamefont {Annala}\ \emph {et~al.}(2022)\citenamefont {Annala},
  \citenamefont {Gorda}, \citenamefont {Katerini}, \citenamefont {Kurkela},
  \citenamefont {N\"attil\"a}, \citenamefont {Paschalidis},\ and\ \citenamefont
  {Vuorinen}}]{Annala:2021gom}%
  \BibitemOpen
  \bibfield  {author} {\bibinfo {author} {\bibfnamefont {E.}~\bibnamefont
  {Annala}}, \bibinfo {author} {\bibfnamefont {T.}~\bibnamefont {Gorda}},
  \bibinfo {author} {\bibfnamefont {E.}~\bibnamefont {Katerini}}, \bibinfo
  {author} {\bibfnamefont {A.}~\bibnamefont {Kurkela}}, \bibinfo {author}
  {\bibfnamefont {J.}~\bibnamefont {N\"attil\"a}}, \bibinfo {author}
  {\bibfnamefont {V.}~\bibnamefont {Paschalidis}}, \ and\ \bibinfo {author}
  {\bibfnamefont {A.}~\bibnamefont {Vuorinen}},\ }\href {\doibase
  10.1103/PhysRevX.12.011058} {\bibfield  {journal} {\bibinfo  {journal} {Phys.
  Rev. X}\ }\textbf {\bibinfo {volume} {12}},\ \bibinfo {pages} {011058}
  (\bibinfo {year} {2022})},\ \Eprint {http://arxiv.org/abs/2105.05132}
  {arXiv:2105.05132 [astro-ph.HE]} \BibitemShut {NoStop}%
\bibitem [{\citenamefont {Abbott}\ \emph {et~al.}(2017)\citenamefont {Abbott}
  \emph {et~al.}}]{LIGOScientific:2017fdd}%
  \BibitemOpen
  \bibfield  {author} {\bibinfo {author} {\bibfnamefont {B.~P.}\ \bibnamefont
  {Abbott}} \emph {et~al.} (\bibinfo {collaboration} {LIGO Scientific,
  Virgo}),\ }\href {\doibase 10.3847/2041-8213/aa9a35} {\bibfield  {journal}
  {\bibinfo  {journal} {Astrophys. J. Lett.}\ }\textbf {\bibinfo {volume}
  {851}},\ \bibinfo {pages} {L16} (\bibinfo {year} {2017})},\ \Eprint
  {http://arxiv.org/abs/1710.09320} {arXiv:1710.09320 [astro-ph.HE]}
  \BibitemShut {NoStop}%
\bibitem [{\citenamefont {Margalit}\ and\ \citenamefont
  {Metzger}(2017)}]{Margalit:2017dij}%
  \BibitemOpen
  \bibfield  {author} {\bibinfo {author} {\bibfnamefont {B.}~\bibnamefont
  {Margalit}}\ and\ \bibinfo {author} {\bibfnamefont {B.~D.}\ \bibnamefont
  {Metzger}},\ }\href {\doibase 10.3847/2041-8213/aa991c} {\bibfield  {journal}
  {\bibinfo  {journal} {Astrophys. J. Lett.}\ }\textbf {\bibinfo {volume}
  {850}},\ \bibinfo {pages} {L19} (\bibinfo {year} {2017})},\ \Eprint
  {http://arxiv.org/abs/1710.05938} {arXiv:1710.05938 [astro-ph.HE]}
  \BibitemShut {NoStop}%
\bibitem [{\citenamefont {Bauswein}\ \emph {et~al.}(2017)\citenamefont
  {Bauswein}, \citenamefont {Just}, \citenamefont {Janka},\ and\ \citenamefont
  {Stergioulas}}]{Bauswein:2017vtn}%
  \BibitemOpen
  \bibfield  {author} {\bibinfo {author} {\bibfnamefont {A.}~\bibnamefont
  {Bauswein}}, \bibinfo {author} {\bibfnamefont {O.}~\bibnamefont {Just}},
  \bibinfo {author} {\bibfnamefont {H.-T.}\ \bibnamefont {Janka}}, \ and\
  \bibinfo {author} {\bibfnamefont {N.}~\bibnamefont {Stergioulas}},\ }\href
  {\doibase 10.3847/2041-8213/aa9994} {\bibfield  {journal} {\bibinfo
  {journal} {Astrophys. J. Lett.}\ }\textbf {\bibinfo {volume} {850}},\
  \bibinfo {pages} {L34} (\bibinfo {year} {2017})},\ \Eprint
  {http://arxiv.org/abs/1710.06843} {arXiv:1710.06843 [astro-ph.HE]}
  \BibitemShut {NoStop}%
\bibitem [{\citenamefont {Radice}\ \emph
  {et~al.}(2018{\natexlab{a}})\citenamefont {Radice}, \citenamefont {Perego},
  \citenamefont {Zappa},\ and\ \citenamefont {Bernuzzi}}]{Radice:2017lry}%
  \BibitemOpen
  \bibfield  {author} {\bibinfo {author} {\bibfnamefont {D.}~\bibnamefont
  {Radice}}, \bibinfo {author} {\bibfnamefont {A.}~\bibnamefont {Perego}},
  \bibinfo {author} {\bibfnamefont {F.}~\bibnamefont {Zappa}}, \ and\ \bibinfo
  {author} {\bibfnamefont {S.}~\bibnamefont {Bernuzzi}},\ }\href {\doibase
  10.3847/2041-8213/aaa402} {\bibfield  {journal} {\bibinfo  {journal}
  {Astrophys. J. Lett.}\ }\textbf {\bibinfo {volume} {852}},\ \bibinfo {pages}
  {L29} (\bibinfo {year} {2018}{\natexlab{a}})},\ \Eprint
  {http://arxiv.org/abs/1711.03647} {arXiv:1711.03647 [astro-ph.HE]}
  \BibitemShut {NoStop}%
\bibitem [{\citenamefont {De}\ \emph {et~al.}(2018)\citenamefont {De},
  \citenamefont {Finstad}, \citenamefont {Lattimer}, \citenamefont {Brown},
  \citenamefont {Berger},\ and\ \citenamefont {Biwer}}]{De:2018uhw}%
  \BibitemOpen
  \bibfield  {author} {\bibinfo {author} {\bibfnamefont {S.}~\bibnamefont
  {De}}, \bibinfo {author} {\bibfnamefont {D.}~\bibnamefont {Finstad}},
  \bibinfo {author} {\bibfnamefont {J.~M.}\ \bibnamefont {Lattimer}}, \bibinfo
  {author} {\bibfnamefont {D.~A.}\ \bibnamefont {Brown}}, \bibinfo {author}
  {\bibfnamefont {E.}~\bibnamefont {Berger}}, \ and\ \bibinfo {author}
  {\bibfnamefont {C.~M.}\ \bibnamefont {Biwer}},\ }\href {\doibase
  10.1103/PhysRevLett.121.091102} {\bibfield  {journal} {\bibinfo  {journal}
  {Phys. Rev. Lett.}\ }\textbf {\bibinfo {volume} {121}},\ \bibinfo {pages}
  {091102} (\bibinfo {year} {2018})},\ \bibinfo {note} {[Erratum:
  Phys.Rev.Lett. 121, 259902 (2018)]},\ \Eprint
  {http://arxiv.org/abs/1804.08583} {arXiv:1804.08583 [astro-ph.HE]}
  \BibitemShut {NoStop}%
\bibitem [{\citenamefont {Abbott}\ \emph {et~al.}(2019)\citenamefont {Abbott}
  \emph {et~al.}}]{LIGOScientific:2018hze}%
  \BibitemOpen
  \bibfield  {author} {\bibinfo {author} {\bibfnamefont {B.~P.}\ \bibnamefont
  {Abbott}} \emph {et~al.} (\bibinfo {collaboration} {LIGO Scientific,
  Virgo}),\ }\href {\doibase 10.1103/PhysRevX.9.011001} {\bibfield  {journal}
  {\bibinfo  {journal} {Phys. Rev. X}\ }\textbf {\bibinfo {volume} {9}},\
  \bibinfo {pages} {011001} (\bibinfo {year} {2019})},\ \Eprint
  {http://arxiv.org/abs/1805.11579} {arXiv:1805.11579 [gr-qc]} \BibitemShut
  {NoStop}%
\bibitem [{\citenamefont {Abbott}\ \emph {et~al.}(2018)\citenamefont {Abbott}
  \emph {et~al.}}]{LIGOScientific:2018cki}%
  \BibitemOpen
  \bibfield  {author} {\bibinfo {author} {\bibfnamefont {B.~P.}\ \bibnamefont
  {Abbott}} \emph {et~al.} (\bibinfo {collaboration} {LIGO Scientific,
  Virgo}),\ }\href {\doibase 10.1103/PhysRevLett.121.161101} {\bibfield
  {journal} {\bibinfo  {journal} {Phys. Rev. Lett.}\ }\textbf {\bibinfo
  {volume} {121}},\ \bibinfo {pages} {161101} (\bibinfo {year} {2018})},\
  \Eprint {http://arxiv.org/abs/1805.11581} {arXiv:1805.11581 [gr-qc]}
  \BibitemShut {NoStop}%
\bibitem [{\citenamefont {Radice}\ and\ \citenamefont
  {Dai}(2019)}]{Radice:2018ozg}%
  \BibitemOpen
  \bibfield  {author} {\bibinfo {author} {\bibfnamefont {D.}~\bibnamefont
  {Radice}}\ and\ \bibinfo {author} {\bibfnamefont {L.}~\bibnamefont {Dai}},\
  }\href {\doibase 10.1140/epja/i2019-12716-4} {\bibfield  {journal} {\bibinfo
  {journal} {Eur. Phys. J. A}\ }\textbf {\bibinfo {volume} {55}},\ \bibinfo
  {pages} {50} (\bibinfo {year} {2019})},\ \Eprint
  {http://arxiv.org/abs/1810.12917} {arXiv:1810.12917 [astro-ph.HE]}
  \BibitemShut {NoStop}%
\bibitem [{\citenamefont {Capano}\ \emph {et~al.}(2020)\citenamefont {Capano},
  \citenamefont {Tews}, \citenamefont {Brown}, \citenamefont {Margalit},
  \citenamefont {De}, \citenamefont {Kumar}, \citenamefont {Brown},
  \citenamefont {Krishnan},\ and\ \citenamefont {Reddy}}]{Capano:2019eae}%
  \BibitemOpen
  \bibfield  {author} {\bibinfo {author} {\bibfnamefont {C.~D.}\ \bibnamefont
  {Capano}}, \bibinfo {author} {\bibfnamefont {I.}~\bibnamefont {Tews}},
  \bibinfo {author} {\bibfnamefont {S.~M.}\ \bibnamefont {Brown}}, \bibinfo
  {author} {\bibfnamefont {B.}~\bibnamefont {Margalit}}, \bibinfo {author}
  {\bibfnamefont {S.}~\bibnamefont {De}}, \bibinfo {author} {\bibfnamefont
  {S.}~\bibnamefont {Kumar}}, \bibinfo {author} {\bibfnamefont {D.~A.}\
  \bibnamefont {Brown}}, \bibinfo {author} {\bibfnamefont {B.}~\bibnamefont
  {Krishnan}}, \ and\ \bibinfo {author} {\bibfnamefont {S.}~\bibnamefont
  {Reddy}},\ }\href {\doibase 10.1038/s41550-020-1014-6} {\bibfield  {journal}
  {\bibinfo  {journal} {Nature Astron.}\ }\textbf {\bibinfo {volume} {4}},\
  \bibinfo {pages} {625} (\bibinfo {year} {2020})},\ \Eprint
  {http://arxiv.org/abs/1908.10352} {arXiv:1908.10352 [astro-ph.HE]}
  \BibitemShut {NoStop}%
\bibitem [{\citenamefont {Dietrich}\ \emph {et~al.}(2020)\citenamefont
  {Dietrich}, \citenamefont {Coughlin}, \citenamefont {Pang}, \citenamefont
  {Bulla}, \citenamefont {Heinzel}, \citenamefont {Issa}, \citenamefont
  {Tews},\ and\ \citenamefont {Antier}}]{Dietrich:2020efo}%
  \BibitemOpen
  \bibfield  {author} {\bibinfo {author} {\bibfnamefont {T.}~\bibnamefont
  {Dietrich}}, \bibinfo {author} {\bibfnamefont {M.~W.}\ \bibnamefont
  {Coughlin}}, \bibinfo {author} {\bibfnamefont {P.~T.~H.}\ \bibnamefont
  {Pang}}, \bibinfo {author} {\bibfnamefont {M.}~\bibnamefont {Bulla}},
  \bibinfo {author} {\bibfnamefont {J.}~\bibnamefont {Heinzel}}, \bibinfo
  {author} {\bibfnamefont {L.}~\bibnamefont {Issa}}, \bibinfo {author}
  {\bibfnamefont {I.}~\bibnamefont {Tews}}, \ and\ \bibinfo {author}
  {\bibfnamefont {S.}~\bibnamefont {Antier}},\ }\href {\doibase
  10.1126/science.abb4317} {\bibfield  {journal} {\bibinfo  {journal}
  {Science}\ }\textbf {\bibinfo {volume} {370}},\ \bibinfo {pages} {1450}
  (\bibinfo {year} {2020})},\ \Eprint {http://arxiv.org/abs/2002.11355}
  {arXiv:2002.11355 [astro-ph.HE]} \BibitemShut {NoStop}%
\bibitem [{\citenamefont {Breschi}\ \emph {et~al.}(2021)\citenamefont
  {Breschi}, \citenamefont {Perego}, \citenamefont {Bernuzzi}, \citenamefont
  {Del~Pozzo}, \citenamefont {Nedora}, \citenamefont {Radice},\ and\
  \citenamefont {Vescovi}}]{Breschi:2021tbm}%
  \BibitemOpen
  \bibfield  {author} {\bibinfo {author} {\bibfnamefont {M.}~\bibnamefont
  {Breschi}}, \bibinfo {author} {\bibfnamefont {A.}~\bibnamefont {Perego}},
  \bibinfo {author} {\bibfnamefont {S.}~\bibnamefont {Bernuzzi}}, \bibinfo
  {author} {\bibfnamefont {W.}~\bibnamefont {Del~Pozzo}}, \bibinfo {author}
  {\bibfnamefont {V.}~\bibnamefont {Nedora}}, \bibinfo {author} {\bibfnamefont
  {D.}~\bibnamefont {Radice}}, \ and\ \bibinfo {author} {\bibfnamefont
  {D.}~\bibnamefont {Vescovi}},\ }\href {\doibase 10.1093/mnras/stab1287}
  {\bibfield  {journal} {\bibinfo  {journal} {Mon. Not. Roy. Astron. Soc.}\
  }\textbf {\bibinfo {volume} {505}},\ \bibinfo {pages} {1661} (\bibinfo {year}
  {2021})},\ \Eprint {http://arxiv.org/abs/2101.01201} {arXiv:2101.01201
  [astro-ph.HE]} \BibitemShut {NoStop}%
\bibitem [{\citenamefont {Page}\ \emph {et~al.}(2004)\citenamefont {Page},
  \citenamefont {Lattimer}, \citenamefont {Prakash},\ and\ \citenamefont
  {Steiner}}]{Page:2004fy}%
  \BibitemOpen
  \bibfield  {author} {\bibinfo {author} {\bibfnamefont {D.}~\bibnamefont
  {Page}}, \bibinfo {author} {\bibfnamefont {J.~M.}\ \bibnamefont {Lattimer}},
  \bibinfo {author} {\bibfnamefont {M.}~\bibnamefont {Prakash}}, \ and\
  \bibinfo {author} {\bibfnamefont {A.~W.}\ \bibnamefont {Steiner}},\ }\href
  {\doibase 10.1086/424844} {\bibfield  {journal} {\bibinfo  {journal}
  {Astrophys. J. Suppl.}\ }\textbf {\bibinfo {volume} {155}},\ \bibinfo {pages}
  {623} (\bibinfo {year} {2004})},\ \Eprint
  {http://arxiv.org/abs/astro-ph/0403657} {arXiv:astro-ph/0403657} \BibitemShut
  {NoStop}%
\bibitem [{\citenamefont {Yakovlev}\ and\ \citenamefont
  {Pethick}(2004)}]{Yakovlev:2004iq}%
  \BibitemOpen
  \bibfield  {author} {\bibinfo {author} {\bibfnamefont {D.~G.}\ \bibnamefont
  {Yakovlev}}\ and\ \bibinfo {author} {\bibfnamefont {C.~J.}\ \bibnamefont
  {Pethick}},\ }\href {\doibase 10.1146/annurev.astro.42.053102.134013}
  {\bibfield  {journal} {\bibinfo  {journal} {Ann. Rev. Astron. Astrophys.}\
  }\textbf {\bibinfo {volume} {42}},\ \bibinfo {pages} {169} (\bibinfo {year}
  {2004})},\ \Eprint {http://arxiv.org/abs/astro-ph/0402143}
  {arXiv:astro-ph/0402143} \BibitemShut {NoStop}%
\bibitem [{\citenamefont {Page}\ \emph {et~al.}(2011)\citenamefont {Page},
  \citenamefont {Prakash}, \citenamefont {Lattimer},\ and\ \citenamefont
  {Steiner}}]{Page:2010aw}%
  \BibitemOpen
  \bibfield  {author} {\bibinfo {author} {\bibfnamefont {D.}~\bibnamefont
  {Page}}, \bibinfo {author} {\bibfnamefont {M.}~\bibnamefont {Prakash}},
  \bibinfo {author} {\bibfnamefont {J.~M.}\ \bibnamefont {Lattimer}}, \ and\
  \bibinfo {author} {\bibfnamefont {A.~W.}\ \bibnamefont {Steiner}},\ }\href
  {\doibase 10.1103/PhysRevLett.106.081101} {\bibfield  {journal} {\bibinfo
  {journal} {Phys. Rev. Lett.}\ }\textbf {\bibinfo {volume} {106}},\ \bibinfo
  {pages} {081101} (\bibinfo {year} {2011})},\ \Eprint
  {http://arxiv.org/abs/1011.6142} {arXiv:1011.6142 [astro-ph.HE]} \BibitemShut
  {NoStop}%
\bibitem [{\citenamefont {Beloin}\ \emph {et~al.}(2019)\citenamefont {Beloin},
  \citenamefont {Han}, \citenamefont {Steiner},\ and\ \citenamefont
  {Odbadrakh}}]{Beloin:2018fyp}%
  \BibitemOpen
  \bibfield  {author} {\bibinfo {author} {\bibfnamefont {S.}~\bibnamefont
  {Beloin}}, \bibinfo {author} {\bibfnamefont {S.}~\bibnamefont {Han}},
  \bibinfo {author} {\bibfnamefont {A.~W.}\ \bibnamefont {Steiner}}, \ and\
  \bibinfo {author} {\bibfnamefont {K.}~\bibnamefont {Odbadrakh}},\ }\href
  {\doibase 10.1103/PhysRevC.100.055801} {\bibfield  {journal} {\bibinfo
  {journal} {Phys. Rev. C}\ }\textbf {\bibinfo {volume} {100}},\ \bibinfo
  {pages} {055801} (\bibinfo {year} {2019})},\ \Eprint
  {http://arxiv.org/abs/1812.00494} {arXiv:1812.00494 [nucl-th]} \BibitemShut
  {NoStop}%
\bibitem [{\citenamefont {Radice}\ \emph {et~al.}(2020)\citenamefont {Radice},
  \citenamefont {Bernuzzi},\ and\ \citenamefont {Perego}}]{Radice:2020ddv}%
  \BibitemOpen
  \bibfield  {author} {\bibinfo {author} {\bibfnamefont {D.}~\bibnamefont
  {Radice}}, \bibinfo {author} {\bibfnamefont {S.}~\bibnamefont {Bernuzzi}}, \
  and\ \bibinfo {author} {\bibfnamefont {A.}~\bibnamefont {Perego}},\ }\href
  {\doibase 10.1146/annurev-nucl-013120-114541} {\bibfield  {journal} {\bibinfo
   {journal} {Ann. Rev. Nucl. Part. Sci.}\ }\textbf {\bibinfo {volume} {70}},\
  \bibinfo {pages} {95} (\bibinfo {year} {2020})},\ \Eprint
  {http://arxiv.org/abs/2002.03863} {arXiv:2002.03863 [astro-ph.HE]}
  \BibitemShut {NoStop}%
\bibitem [{\citenamefont {Perego}\ \emph {et~al.}(2022)\citenamefont {Perego},
  \citenamefont {Logoteta}, \citenamefont {Radice}, \citenamefont {Bernuzzi},
  \citenamefont {Kashyap}, \citenamefont {Das}, \citenamefont {Padamata},\ and\
  \citenamefont {Prakash}}]{Perego:2021mkd}%
  \BibitemOpen
  \bibfield  {author} {\bibinfo {author} {\bibfnamefont {A.}~\bibnamefont
  {Perego}}, \bibinfo {author} {\bibfnamefont {D.}~\bibnamefont {Logoteta}},
  \bibinfo {author} {\bibfnamefont {D.}~\bibnamefont {Radice}}, \bibinfo
  {author} {\bibfnamefont {S.}~\bibnamefont {Bernuzzi}}, \bibinfo {author}
  {\bibfnamefont {R.}~\bibnamefont {Kashyap}}, \bibinfo {author} {\bibfnamefont
  {A.}~\bibnamefont {Das}}, \bibinfo {author} {\bibfnamefont {S.}~\bibnamefont
  {Padamata}}, \ and\ \bibinfo {author} {\bibfnamefont {A.}~\bibnamefont
  {Prakash}},\ }\href {\doibase 10.1103/PhysRevLett.129.032701} {\bibfield
  {journal} {\bibinfo  {journal} {Phys. Rev. Lett.}\ }\textbf {\bibinfo
  {volume} {129}},\ \bibinfo {pages} {032701} (\bibinfo {year} {2022})},\
  \Eprint {http://arxiv.org/abs/2112.05864} {arXiv:2112.05864 [astro-ph.HE]}
  \BibitemShut {NoStop}%
\bibitem [{\citenamefont {Raithel}\ \emph {et~al.}(2021)\citenamefont
  {Raithel}, \citenamefont {Paschalidis},\ and\ \citenamefont
  {\"Ozel}}]{Raithel:2021hye}%
  \BibitemOpen
  \bibfield  {author} {\bibinfo {author} {\bibfnamefont {C.}~\bibnamefont
  {Raithel}}, \bibinfo {author} {\bibfnamefont {V.}~\bibnamefont
  {Paschalidis}}, \ and\ \bibinfo {author} {\bibfnamefont {F.}~\bibnamefont
  {\"Ozel}},\ }\href {\doibase 10.1103/PhysRevD.104.063016} {\bibfield
  {journal} {\bibinfo  {journal} {Phys. Rev. D}\ }\textbf {\bibinfo {volume}
  {104}},\ \bibinfo {pages} {063016} (\bibinfo {year} {2021})},\ \Eprint
  {http://arxiv.org/abs/2104.07226} {arXiv:2104.07226 [astro-ph.HE]}
  \BibitemShut {NoStop}%
\bibitem [{\citenamefont {Hammond}\ \emph {et~al.}(2021)\citenamefont
  {Hammond}, \citenamefont {Hawke},\ and\ \citenamefont
  {Andersson}}]{Hammond:2021vtv}%
  \BibitemOpen
  \bibfield  {author} {\bibinfo {author} {\bibfnamefont {P.}~\bibnamefont
  {Hammond}}, \bibinfo {author} {\bibfnamefont {I.}~\bibnamefont {Hawke}}, \
  and\ \bibinfo {author} {\bibfnamefont {N.}~\bibnamefont {Andersson}},\ }\href
  {\doibase 10.1103/PhysRevD.104.103006} {\bibfield  {journal} {\bibinfo
  {journal} {Phys. Rev. D}\ }\textbf {\bibinfo {volume} {104}},\ \bibinfo
  {pages} {103006} (\bibinfo {year} {2021})},\ \Eprint
  {http://arxiv.org/abs/2108.08649} {arXiv:2108.08649 [astro-ph.HE]}
  \BibitemShut {NoStop}%
\bibitem [{\citenamefont {Most}\ \emph {et~al.}(2022)\citenamefont {Most},
  \citenamefont {Haber}, \citenamefont {Harris}, \citenamefont {Zhang},
  \citenamefont {Alford},\ and\ \citenamefont {Noronha}}]{Most:2022yhe}%
  \BibitemOpen
  \bibfield  {author} {\bibinfo {author} {\bibfnamefont {E.~R.}\ \bibnamefont
  {Most}}, \bibinfo {author} {\bibfnamefont {A.}~\bibnamefont {Haber}},
  \bibinfo {author} {\bibfnamefont {S.~P.}\ \bibnamefont {Harris}}, \bibinfo
  {author} {\bibfnamefont {Z.}~\bibnamefont {Zhang}}, \bibinfo {author}
  {\bibfnamefont {M.~G.}\ \bibnamefont {Alford}}, \ and\ \bibinfo {author}
  {\bibfnamefont {J.}~\bibnamefont {Noronha}},\ }\href@noop {} {\  (\bibinfo
  {year} {2022})},\ \Eprint {http://arxiv.org/abs/2207.00442} {arXiv:2207.00442
  [astro-ph.HE]} \BibitemShut {NoStop}%
\bibitem [{\citenamefont {Loffredo}\ \emph {et~al.}(2023)\citenamefont
  {Loffredo}, \citenamefont {Perego}, \citenamefont {Logoteta},\ and\
  \citenamefont {Branchesi}}]{Loffredo:2022prq}%
  \BibitemOpen
  \bibfield  {author} {\bibinfo {author} {\bibfnamefont {E.}~\bibnamefont
  {Loffredo}}, \bibinfo {author} {\bibfnamefont {A.}~\bibnamefont {Perego}},
  \bibinfo {author} {\bibfnamefont {D.}~\bibnamefont {Logoteta}}, \ and\
  \bibinfo {author} {\bibfnamefont {M.}~\bibnamefont {Branchesi}},\ }\href
  {\doibase 10.1051/0004-6361/202244927} {\bibfield  {journal} {\bibinfo
  {journal} {Astron. Astrophys.}\ }\textbf {\bibinfo {volume} {672}},\ \bibinfo
  {pages} {A124} (\bibinfo {year} {2023})},\ \Eprint
  {http://arxiv.org/abs/2209.04458} {arXiv:2209.04458 [astro-ph.HE]}
  \BibitemShut {NoStop}%
\bibitem [{\citenamefont {Fields}\ \emph {et~al.}(2023)\citenamefont {Fields},
  \citenamefont {Prakash}, \citenamefont {Breschi}, \citenamefont {Radice},
  \citenamefont {Bernuzzi},\ and\ \citenamefont {Schneider}}]{Fields:2023bhs}%
  \BibitemOpen
  \bibfield  {author} {\bibinfo {author} {\bibfnamefont {J.}~\bibnamefont
  {Fields}}, \bibinfo {author} {\bibfnamefont {A.}~\bibnamefont {Prakash}},
  \bibinfo {author} {\bibfnamefont {M.}~\bibnamefont {Breschi}}, \bibinfo
  {author} {\bibfnamefont {D.}~\bibnamefont {Radice}}, \bibinfo {author}
  {\bibfnamefont {S.}~\bibnamefont {Bernuzzi}}, \ and\ \bibinfo {author}
  {\bibfnamefont {A.~d.~S.}\ \bibnamefont {Schneider}},\ }\href@noop {} {\
  (\bibinfo {year} {2023})},\ \Eprint {http://arxiv.org/abs/2302.11359}
  {arXiv:2302.11359 [astro-ph.HE]} \BibitemShut {NoStop}%
\bibitem [{\citenamefont {Blacker}\ \emph {et~al.}(2023)\citenamefont
  {Blacker}, \citenamefont {Bauswein},\ and\ \citenamefont
  {Typel}}]{Blacker:2023onp}%
  \BibitemOpen
  \bibfield  {author} {\bibinfo {author} {\bibfnamefont {S.}~\bibnamefont
  {Blacker}}, \bibinfo {author} {\bibfnamefont {A.}~\bibnamefont {Bauswein}}, \
  and\ \bibinfo {author} {\bibfnamefont {S.}~\bibnamefont {Typel}},\
  }\href@noop {} {\  (\bibinfo {year} {2023})},\ \Eprint
  {http://arxiv.org/abs/2304.01971} {arXiv:2304.01971 [astro-ph.HE]}
  \BibitemShut {NoStop}%
\bibitem [{\citenamefont {Sekiguchi}\ \emph {et~al.}(2011)\citenamefont
  {Sekiguchi}, \citenamefont {Kiuchi}, \citenamefont {Kyutoku},\ and\
  \citenamefont {Shibata}}]{Sekiguchi:2011mc}%
  \BibitemOpen
  \bibfield  {author} {\bibinfo {author} {\bibfnamefont {Y.}~\bibnamefont
  {Sekiguchi}}, \bibinfo {author} {\bibfnamefont {K.}~\bibnamefont {Kiuchi}},
  \bibinfo {author} {\bibfnamefont {K.}~\bibnamefont {Kyutoku}}, \ and\
  \bibinfo {author} {\bibfnamefont {M.}~\bibnamefont {Shibata}},\ }\href
  {\doibase 10.1103/PhysRevLett.107.211101} {\bibfield  {journal} {\bibinfo
  {journal} {Phys. Rev. Lett.}\ }\textbf {\bibinfo {volume} {107}},\ \bibinfo
  {pages} {211101} (\bibinfo {year} {2011})},\ \Eprint
  {http://arxiv.org/abs/1110.4442} {arXiv:1110.4442 [astro-ph.HE]} \BibitemShut
  {NoStop}%
\bibitem [{\citenamefont {Radice}\ \emph {et~al.}(2017)\citenamefont {Radice},
  \citenamefont {Bernuzzi}, \citenamefont {Del~Pozzo}, \citenamefont
  {Roberts},\ and\ \citenamefont {Ott}}]{Radice:2016rys}%
  \BibitemOpen
  \bibfield  {author} {\bibinfo {author} {\bibfnamefont {D.}~\bibnamefont
  {Radice}}, \bibinfo {author} {\bibfnamefont {S.}~\bibnamefont {Bernuzzi}},
  \bibinfo {author} {\bibfnamefont {W.}~\bibnamefont {Del~Pozzo}}, \bibinfo
  {author} {\bibfnamefont {L.~F.}\ \bibnamefont {Roberts}}, \ and\ \bibinfo
  {author} {\bibfnamefont {C.~D.}\ \bibnamefont {Ott}},\ }\href {\doibase
  10.3847/2041-8213/aa775f} {\bibfield  {journal} {\bibinfo  {journal}
  {Astrophys. J. Lett.}\ }\textbf {\bibinfo {volume} {842}},\ \bibinfo {pages}
  {L10} (\bibinfo {year} {2017})},\ \Eprint {http://arxiv.org/abs/1612.06429}
  {arXiv:1612.06429 [astro-ph.HE]} \BibitemShut {NoStop}%
\bibitem [{\citenamefont {Berryman}\ and\ \citenamefont
  {Gardner}(2021)}]{Berryman:2021jjt}%
  \BibitemOpen
  \bibfield  {author} {\bibinfo {author} {\bibfnamefont {J.~M.}\ \bibnamefont
  {Berryman}}\ and\ \bibinfo {author} {\bibfnamefont {S.}~\bibnamefont
  {Gardner}},\ }\href {\doibase 10.1103/PhysRevC.104.045802} {\bibfield
  {journal} {\bibinfo  {journal} {Phys. Rev. C}\ }\textbf {\bibinfo {volume}
  {104}},\ \bibinfo {pages} {045802} (\bibinfo {year} {2021})},\ \Eprint
  {http://arxiv.org/abs/2105.05254} {arXiv:2105.05254 [hep-ph]} \BibitemShut
  {NoStop}%
\bibitem [{\citenamefont {Vijayan}\ \emph {et~al.}(2023)\citenamefont
  {Vijayan}, \citenamefont {Rahman}, \citenamefont {Bauswein}, \citenamefont
  {Mart\'\i{}nez-Pinedo},\ and\ \citenamefont {Arbina}}]{Vijayan:2023qrt}%
  \BibitemOpen
  \bibfield  {author} {\bibinfo {author} {\bibfnamefont {V.}~\bibnamefont
  {Vijayan}}, \bibinfo {author} {\bibfnamefont {N.}~\bibnamefont {Rahman}},
  \bibinfo {author} {\bibfnamefont {A.}~\bibnamefont {Bauswein}}, \bibinfo
  {author} {\bibfnamefont {G.}~\bibnamefont {Mart\'\i{}nez-Pinedo}}, \ and\
  \bibinfo {author} {\bibfnamefont {I.~L.}\ \bibnamefont {Arbina}},\
  }\href@noop {} {\  (\bibinfo {year} {2023})},\ \Eprint
  {http://arxiv.org/abs/2302.12055} {arXiv:2302.12055 [astro-ph.HE]}
  \BibitemShut {NoStop}%
\bibitem [{\citenamefont {Most}\ \emph {et~al.}(2019)\citenamefont {Most},
  \citenamefont {Papenfort}, \citenamefont {Dexheimer}, \citenamefont
  {Hanauske}, \citenamefont {Schramm}, \citenamefont {St\"ocker},\ and\
  \citenamefont {Rezzolla}}]{Most:2018eaw}%
  \BibitemOpen
  \bibfield  {author} {\bibinfo {author} {\bibfnamefont {E.~R.}\ \bibnamefont
  {Most}}, \bibinfo {author} {\bibfnamefont {L.~J.}\ \bibnamefont {Papenfort}},
  \bibinfo {author} {\bibfnamefont {V.}~\bibnamefont {Dexheimer}}, \bibinfo
  {author} {\bibfnamefont {M.}~\bibnamefont {Hanauske}}, \bibinfo {author}
  {\bibfnamefont {S.}~\bibnamefont {Schramm}}, \bibinfo {author} {\bibfnamefont
  {H.}~\bibnamefont {St\"ocker}}, \ and\ \bibinfo {author} {\bibfnamefont
  {L.}~\bibnamefont {Rezzolla}},\ }\href {\doibase
  10.1103/PhysRevLett.122.061101} {\bibfield  {journal} {\bibinfo  {journal}
  {Phys. Rev. Lett.}\ }\textbf {\bibinfo {volume} {122}},\ \bibinfo {pages}
  {061101} (\bibinfo {year} {2019})},\ \Eprint
  {http://arxiv.org/abs/1807.03684} {arXiv:1807.03684 [astro-ph.HE]}
  \BibitemShut {NoStop}%
\bibitem [{\citenamefont {Most}\ \emph {et~al.}(2020)\citenamefont {Most},
  \citenamefont {Jens~Papenfort}, \citenamefont {Dexheimer}, \citenamefont
  {Hanauske}, \citenamefont {Stoecker},\ and\ \citenamefont
  {Rezzolla}}]{Most:2019onn}%
  \BibitemOpen
  \bibfield  {author} {\bibinfo {author} {\bibfnamefont {E.~R.}\ \bibnamefont
  {Most}}, \bibinfo {author} {\bibfnamefont {L.}~\bibnamefont
  {Jens~Papenfort}}, \bibinfo {author} {\bibfnamefont {V.}~\bibnamefont
  {Dexheimer}}, \bibinfo {author} {\bibfnamefont {M.}~\bibnamefont {Hanauske}},
  \bibinfo {author} {\bibfnamefont {H.}~\bibnamefont {Stoecker}}, \ and\
  \bibinfo {author} {\bibfnamefont {L.}~\bibnamefont {Rezzolla}},\ }\href
  {\doibase 10.1140/epja/s10050-020-00073-4} {\bibfield  {journal} {\bibinfo
  {journal} {Eur. Phys. J. A}\ }\textbf {\bibinfo {volume} {56}},\ \bibinfo
  {pages} {59} (\bibinfo {year} {2020})},\ \Eprint
  {http://arxiv.org/abs/1910.13893} {arXiv:1910.13893 [astro-ph.HE]}
  \BibitemShut {NoStop}%
\bibitem [{\citenamefont {Bauswein}\ \emph {et~al.}(2019)\citenamefont
  {Bauswein}, \citenamefont {Bastian}, \citenamefont {Blaschke}, \citenamefont
  {Chatziioannou}, \citenamefont {Clark}, \citenamefont {Fischer},\ and\
  \citenamefont {Oertel}}]{Bauswein:2018bma}%
  \BibitemOpen
  \bibfield  {author} {\bibinfo {author} {\bibfnamefont {A.}~\bibnamefont
  {Bauswein}}, \bibinfo {author} {\bibfnamefont {N.-U.~F.}\ \bibnamefont
  {Bastian}}, \bibinfo {author} {\bibfnamefont {D.~B.}\ \bibnamefont
  {Blaschke}}, \bibinfo {author} {\bibfnamefont {K.}~\bibnamefont
  {Chatziioannou}}, \bibinfo {author} {\bibfnamefont {J.~A.}\ \bibnamefont
  {Clark}}, \bibinfo {author} {\bibfnamefont {T.}~\bibnamefont {Fischer}}, \
  and\ \bibinfo {author} {\bibfnamefont {M.}~\bibnamefont {Oertel}},\ }\href
  {\doibase 10.1103/PhysRevLett.122.061102} {\bibfield  {journal} {\bibinfo
  {journal} {Phys. Rev. Lett.}\ }\textbf {\bibinfo {volume} {122}},\ \bibinfo
  {pages} {061102} (\bibinfo {year} {2019})},\ \Eprint
  {http://arxiv.org/abs/1809.01116} {arXiv:1809.01116 [astro-ph.HE]}
  \BibitemShut {NoStop}%
\bibitem [{\citenamefont {Weih}\ \emph {et~al.}(2020)\citenamefont {Weih},
  \citenamefont {Hanauske},\ and\ \citenamefont {Rezzolla}}]{Weih:2019xvw}%
  \BibitemOpen
  \bibfield  {author} {\bibinfo {author} {\bibfnamefont {L.~R.}\ \bibnamefont
  {Weih}}, \bibinfo {author} {\bibfnamefont {M.}~\bibnamefont {Hanauske}}, \
  and\ \bibinfo {author} {\bibfnamefont {L.}~\bibnamefont {Rezzolla}},\ }\href
  {\doibase 10.1103/PhysRevLett.124.171103} {\bibfield  {journal} {\bibinfo
  {journal} {Phys. Rev. Lett.}\ }\textbf {\bibinfo {volume} {124}},\ \bibinfo
  {pages} {171103} (\bibinfo {year} {2020})},\ \Eprint
  {http://arxiv.org/abs/1912.09340} {arXiv:1912.09340 [gr-qc]} \BibitemShut
  {NoStop}%
\bibitem [{\citenamefont {Liebling}\ \emph {et~al.}(2021)\citenamefont
  {Liebling}, \citenamefont {Palenzuela},\ and\ \citenamefont
  {Lehner}}]{Liebling:2020dhf}%
  \BibitemOpen
  \bibfield  {author} {\bibinfo {author} {\bibfnamefont {S.~L.}\ \bibnamefont
  {Liebling}}, \bibinfo {author} {\bibfnamefont {C.}~\bibnamefont
  {Palenzuela}}, \ and\ \bibinfo {author} {\bibfnamefont {L.}~\bibnamefont
  {Lehner}},\ }\href {\doibase 10.1088/1361-6382/abf898} {\bibfield  {journal}
  {\bibinfo  {journal} {Class. Quant. Grav.}\ }\textbf {\bibinfo {volume}
  {38}},\ \bibinfo {pages} {115007} (\bibinfo {year} {2021})},\ \Eprint
  {http://arxiv.org/abs/2010.12567} {arXiv:2010.12567 [gr-qc]} \BibitemShut
  {NoStop}%
\bibitem [{\citenamefont {Prakash}\ \emph {et~al.}(2021)\citenamefont
  {Prakash}, \citenamefont {Radice}, \citenamefont {Logoteta}, \citenamefont
  {Perego}, \citenamefont {Nedora}, \citenamefont {Bombaci}, \citenamefont
  {Kashyap}, \citenamefont {Bernuzzi},\ and\ \citenamefont
  {Endrizzi}}]{Prakash:2021wpz}%
  \BibitemOpen
  \bibfield  {author} {\bibinfo {author} {\bibfnamefont {A.}~\bibnamefont
  {Prakash}}, \bibinfo {author} {\bibfnamefont {D.}~\bibnamefont {Radice}},
  \bibinfo {author} {\bibfnamefont {D.}~\bibnamefont {Logoteta}}, \bibinfo
  {author} {\bibfnamefont {A.}~\bibnamefont {Perego}}, \bibinfo {author}
  {\bibfnamefont {V.}~\bibnamefont {Nedora}}, \bibinfo {author} {\bibfnamefont
  {I.}~\bibnamefont {Bombaci}}, \bibinfo {author} {\bibfnamefont
  {R.}~\bibnamefont {Kashyap}}, \bibinfo {author} {\bibfnamefont
  {S.}~\bibnamefont {Bernuzzi}}, \ and\ \bibinfo {author} {\bibfnamefont
  {A.}~\bibnamefont {Endrizzi}},\ }\href {\doibase 10.1103/PhysRevD.104.083029}
  {\bibfield  {journal} {\bibinfo  {journal} {Phys. Rev. D}\ }\textbf {\bibinfo
  {volume} {104}},\ \bibinfo {pages} {083029} (\bibinfo {year} {2021})},\
  \Eprint {http://arxiv.org/abs/2106.07885} {arXiv:2106.07885 [astro-ph.HE]}
  \BibitemShut {NoStop}%
\bibitem [{\citenamefont {Fujimoto}\ \emph {et~al.}(2023)\citenamefont
  {Fujimoto}, \citenamefont {Fukushima}, \citenamefont {Hotokezaka},\ and\
  \citenamefont {Kyutoku}}]{Fujimoto:2022xhv}%
  \BibitemOpen
  \bibfield  {author} {\bibinfo {author} {\bibfnamefont {Y.}~\bibnamefont
  {Fujimoto}}, \bibinfo {author} {\bibfnamefont {K.}~\bibnamefont {Fukushima}},
  \bibinfo {author} {\bibfnamefont {K.}~\bibnamefont {Hotokezaka}}, \ and\
  \bibinfo {author} {\bibfnamefont {K.}~\bibnamefont {Kyutoku}},\ }\href
  {\doibase 10.1103/PhysRevLett.130.091404} {\bibfield  {journal} {\bibinfo
  {journal} {Phys. Rev. Lett.}\ }\textbf {\bibinfo {volume} {130}},\ \bibinfo
  {pages} {091404} (\bibinfo {year} {2023})},\ \Eprint
  {http://arxiv.org/abs/2205.03882} {arXiv:2205.03882 [astro-ph.HE]}
  \BibitemShut {NoStop}%
\bibitem [{\citenamefont {Kedia}\ \emph {et~al.}(2022)\citenamefont {Kedia},
  \citenamefont {Kim}, \citenamefont {Suh},\ and\ \citenamefont
  {Mathews}}]{Kedia:2022nns}%
  \BibitemOpen
  \bibfield  {author} {\bibinfo {author} {\bibfnamefont {A.}~\bibnamefont
  {Kedia}}, \bibinfo {author} {\bibfnamefont {H.~I.}\ \bibnamefont {Kim}},
  \bibinfo {author} {\bibfnamefont {I.-S.}\ \bibnamefont {Suh}}, \ and\
  \bibinfo {author} {\bibfnamefont {G.~J.}\ \bibnamefont {Mathews}},\ }\href
  {\doibase 10.1103/PhysRevD.106.103027} {\bibfield  {journal} {\bibinfo
  {journal} {Phys. Rev. D}\ }\textbf {\bibinfo {volume} {106}},\ \bibinfo
  {pages} {103027} (\bibinfo {year} {2022})},\ \Eprint
  {http://arxiv.org/abs/2203.05461} {arXiv:2203.05461 [gr-qc]} \BibitemShut
  {NoStop}%
\bibitem [{\citenamefont {Espino}\ \emph {et~al.}(2023)\citenamefont {Espino},
  \citenamefont {Prakash}, \citenamefont {Radice},\ and\ \citenamefont
  {Logoteta}}]{Espino:2023llj}%
  \BibitemOpen
  \bibfield  {author} {\bibinfo {author} {\bibfnamefont {P.~L.}\ \bibnamefont
  {Espino}}, \bibinfo {author} {\bibfnamefont {A.}~\bibnamefont {Prakash}},
  \bibinfo {author} {\bibfnamefont {D.}~\bibnamefont {Radice}}, \ and\ \bibinfo
  {author} {\bibfnamefont {D.}~\bibnamefont {Logoteta}},\ }\href@noop {} {\
  (\bibinfo {year} {2023})},\ \Eprint {http://arxiv.org/abs/2301.03619}
  {arXiv:2301.03619 [astro-ph.HE]} \BibitemShut {NoStop}%
\bibitem [{\citenamefont {Miller}\ \emph {et~al.}(2015)\citenamefont {Miller},
  \citenamefont {Barsotti}, \citenamefont {Vitale}, \citenamefont {Fritschel},
  \citenamefont {Evans},\ and\ \citenamefont {Sigg}}]{Miller:2014kma}%
  \BibitemOpen
  \bibfield  {author} {\bibinfo {author} {\bibfnamefont {J.}~\bibnamefont
  {Miller}}, \bibinfo {author} {\bibfnamefont {L.}~\bibnamefont {Barsotti}},
  \bibinfo {author} {\bibfnamefont {S.}~\bibnamefont {Vitale}}, \bibinfo
  {author} {\bibfnamefont {P.}~\bibnamefont {Fritschel}}, \bibinfo {author}
  {\bibfnamefont {M.}~\bibnamefont {Evans}}, \ and\ \bibinfo {author}
  {\bibfnamefont {D.}~\bibnamefont {Sigg}},\ }\href {\doibase
  10.1103/PhysRevD.91.062005} {\bibfield  {journal} {\bibinfo  {journal} {Phys.
  Rev. D}\ }\textbf {\bibinfo {volume} {91}},\ \bibinfo {pages} {062005}
  (\bibinfo {year} {2015})},\ \Eprint {http://arxiv.org/abs/1410.5882}
  {arXiv:1410.5882 [gr-qc]} \BibitemShut {NoStop}%
\bibitem [{\citenamefont {Evans}\ \emph {et~al.}(2021)\citenamefont {Evans}
  \emph {et~al.}}]{Evans:2021gyd}%
  \BibitemOpen
  \bibfield  {author} {\bibinfo {author} {\bibfnamefont {M.}~\bibnamefont
  {Evans}} \emph {et~al.},\ }\href@noop {} {\  (\bibinfo {year} {2021})},\
  \Eprint {http://arxiv.org/abs/2109.09882} {arXiv:2109.09882 [astro-ph.IM]}
  \BibitemShut {NoStop}%
\bibitem [{\citenamefont {Lovato}\ \emph {et~al.}(2022)\citenamefont {Lovato}
  \emph {et~al.}}]{Lovato:2022vgq}%
  \BibitemOpen
  \bibfield  {author} {\bibinfo {author} {\bibfnamefont {A.}~\bibnamefont
  {Lovato}} \emph {et~al.},\ }\href@noop {} {\  (\bibinfo {year} {2022})},\
  \Eprint {http://arxiv.org/abs/2211.02224} {arXiv:2211.02224 [nucl-th]}
  \BibitemShut {NoStop}%
\bibitem [{\citenamefont {Punturo}\ \emph
  {et~al.}(2010{\natexlab{a}})\citenamefont {Punturo} \emph
  {et~al.}}]{Punturo:2010zz}%
  \BibitemOpen
  \bibfield  {author} {\bibinfo {author} {\bibfnamefont {M.}~\bibnamefont
  {Punturo}} \emph {et~al.},\ }\href {\doibase 10.1088/0264-9381/27/19/194002}
  {\bibfield  {journal} {\bibinfo  {journal} {Class. Quant. Grav.}\ }\textbf
  {\bibinfo {volume} {27}},\ \bibinfo {pages} {194002} (\bibinfo {year}
  {2010}{\natexlab{a}})}\BibitemShut {NoStop}%
\bibitem [{\citenamefont {Reitze}\ \emph {et~al.}(2019)\citenamefont {Reitze}
  \emph {et~al.}}]{Reitze:2019iox}%
  \BibitemOpen
  \bibfield  {author} {\bibinfo {author} {\bibfnamefont {D.}~\bibnamefont
  {Reitze}} \emph {et~al.},\ }\href@noop {} {\bibfield  {journal} {\bibinfo
  {journal} {Bull. Am. Astron. Soc.}\ }\textbf {\bibinfo {volume} {51}},\
  \bibinfo {pages} {035} (\bibinfo {year} {2019})},\ \Eprint
  {http://arxiv.org/abs/1907.04833} {arXiv:1907.04833 [astro-ph.IM]}
  \BibitemShut {NoStop}%
\bibitem [{\citenamefont {Ackley}\ \emph {et~al.}(2020)\citenamefont {Ackley}
  \emph {et~al.}}]{Ackley:2020atn}%
  \BibitemOpen
  \bibfield  {author} {\bibinfo {author} {\bibfnamefont {K.}~\bibnamefont
  {Ackley}} \emph {et~al.},\ }\href {\doibase 10.1017/pasa.2020.39} {\bibfield
  {journal} {\bibinfo  {journal} {Publ. Astron. Soc. Austral.}\ }\textbf
  {\bibinfo {volume} {37}},\ \bibinfo {pages} {e047} (\bibinfo {year}
  {2020})},\ \Eprint {http://arxiv.org/abs/2007.03128} {arXiv:2007.03128
  [astro-ph.HE]} \BibitemShut {NoStop}%
\bibitem [{\citenamefont {Bernuzzi}(2020)}]{Bernuzzi:2020tgt}%
  \BibitemOpen
  \bibfield  {author} {\bibinfo {author} {\bibfnamefont {S.}~\bibnamefont
  {Bernuzzi}},\ }\href {\doibase 10.1007/s10714-020-02752-5} {\bibfield
  {journal} {\bibinfo  {journal} {Gen. Rel. Grav.}\ }\textbf {\bibinfo {volume}
  {52}},\ \bibinfo {pages} {108} (\bibinfo {year} {2020})},\ \Eprint
  {http://arxiv.org/abs/2004.06419} {arXiv:2004.06419 [astro-ph.HE]}
  \BibitemShut {NoStop}%
\bibitem [{\citenamefont {Shibata}(2005)}]{Shibata:2005xz}%
  \BibitemOpen
  \bibfield  {author} {\bibinfo {author} {\bibfnamefont {M.}~\bibnamefont
  {Shibata}},\ }\href {\doibase 10.1103/PhysRevLett.94.201101} {\bibfield
  {journal} {\bibinfo  {journal} {Phys. Rev. Lett.}\ }\textbf {\bibinfo
  {volume} {94}},\ \bibinfo {pages} {201101} (\bibinfo {year} {2005})},\
  \Eprint {http://arxiv.org/abs/gr-qc/0504082} {arXiv:gr-qc/0504082}
  \BibitemShut {NoStop}%
\bibitem [{\citenamefont {Hotokezaka}\ \emph {et~al.}(2011)\citenamefont
  {Hotokezaka}, \citenamefont {Kyutoku}, \citenamefont {Okawa}, \citenamefont
  {Shibata},\ and\ \citenamefont {Kiuchi}}]{Hotokezaka:2011dh}%
  \BibitemOpen
  \bibfield  {author} {\bibinfo {author} {\bibfnamefont {K.}~\bibnamefont
  {Hotokezaka}}, \bibinfo {author} {\bibfnamefont {K.}~\bibnamefont {Kyutoku}},
  \bibinfo {author} {\bibfnamefont {H.}~\bibnamefont {Okawa}}, \bibinfo
  {author} {\bibfnamefont {M.}~\bibnamefont {Shibata}}, \ and\ \bibinfo
  {author} {\bibfnamefont {K.}~\bibnamefont {Kiuchi}},\ }\href {\doibase
  10.1103/PhysRevD.83.124008} {\bibfield  {journal} {\bibinfo  {journal} {Phys.
  Rev. D}\ }\textbf {\bibinfo {volume} {83}},\ \bibinfo {pages} {124008}
  (\bibinfo {year} {2011})},\ \Eprint {http://arxiv.org/abs/1105.4370}
  {arXiv:1105.4370 [astro-ph.HE]} \BibitemShut {NoStop}%
\bibitem [{\citenamefont {Bauswein}\ \emph {et~al.}(2013)\citenamefont
  {Bauswein}, \citenamefont {Baumgarte},\ and\ \citenamefont
  {Janka}}]{Bauswein:2013jpa}%
  \BibitemOpen
  \bibfield  {author} {\bibinfo {author} {\bibfnamefont {A.}~\bibnamefont
  {Bauswein}}, \bibinfo {author} {\bibfnamefont {T.~W.}\ \bibnamefont
  {Baumgarte}}, \ and\ \bibinfo {author} {\bibfnamefont {H.~T.}\ \bibnamefont
  {Janka}},\ }\href {\doibase 10.1103/PhysRevLett.111.131101} {\bibfield
  {journal} {\bibinfo  {journal} {Phys. Rev. Lett.}\ }\textbf {\bibinfo
  {volume} {111}},\ \bibinfo {pages} {131101} (\bibinfo {year} {2013})},\
  \Eprint {http://arxiv.org/abs/1307.5191} {arXiv:1307.5191 [astro-ph.SR]}
  \BibitemShut {NoStop}%
\bibitem [{\citenamefont {Takami}\ \emph {et~al.}(2014)\citenamefont {Takami},
  \citenamefont {Rezzolla},\ and\ \citenamefont {Baiotti}}]{Takami:2014zpa}%
  \BibitemOpen
  \bibfield  {author} {\bibinfo {author} {\bibfnamefont {K.}~\bibnamefont
  {Takami}}, \bibinfo {author} {\bibfnamefont {L.}~\bibnamefont {Rezzolla}}, \
  and\ \bibinfo {author} {\bibfnamefont {L.}~\bibnamefont {Baiotti}},\ }\href
  {\doibase 10.1103/PhysRevLett.113.091104} {\bibfield  {journal} {\bibinfo
  {journal} {Phys. Rev. Lett.}\ }\textbf {\bibinfo {volume} {113}},\ \bibinfo
  {pages} {091104} (\bibinfo {year} {2014})},\ \Eprint
  {http://arxiv.org/abs/1403.5672} {arXiv:1403.5672 [gr-qc]} \BibitemShut
  {NoStop}%
\bibitem [{\citenamefont {Bernuzzi}\ \emph {et~al.}(2015)\citenamefont
  {Bernuzzi}, \citenamefont {Dietrich},\ and\ \citenamefont
  {Nagar}}]{Bernuzzi:2015rla}%
  \BibitemOpen
  \bibfield  {author} {\bibinfo {author} {\bibfnamefont {S.}~\bibnamefont
  {Bernuzzi}}, \bibinfo {author} {\bibfnamefont {T.}~\bibnamefont {Dietrich}},
  \ and\ \bibinfo {author} {\bibfnamefont {A.}~\bibnamefont {Nagar}},\ }\href
  {\doibase 10.1103/PhysRevLett.115.091101} {\bibfield  {journal} {\bibinfo
  {journal} {Phys. Rev. Lett.}\ }\textbf {\bibinfo {volume} {115}},\ \bibinfo
  {pages} {091101} (\bibinfo {year} {2015})},\ \Eprint
  {http://arxiv.org/abs/1504.01764} {arXiv:1504.01764 [gr-qc]} \BibitemShut
  {NoStop}%
\bibitem [{\citenamefont {Bose}\ \emph {et~al.}(2018)\citenamefont {Bose},
  \citenamefont {Chakravarti}, \citenamefont {Rezzolla}, \citenamefont
  {Sathyaprakash},\ and\ \citenamefont {Takami}}]{Bose:2017jvk}%
  \BibitemOpen
  \bibfield  {author} {\bibinfo {author} {\bibfnamefont {S.}~\bibnamefont
  {Bose}}, \bibinfo {author} {\bibfnamefont {K.}~\bibnamefont {Chakravarti}},
  \bibinfo {author} {\bibfnamefont {L.}~\bibnamefont {Rezzolla}}, \bibinfo
  {author} {\bibfnamefont {B.~S.}\ \bibnamefont {Sathyaprakash}}, \ and\
  \bibinfo {author} {\bibfnamefont {K.}~\bibnamefont {Takami}},\ }\href
  {\doibase 10.1103/PhysRevLett.120.031102} {\bibfield  {journal} {\bibinfo
  {journal} {Phys. Rev. Lett.}\ }\textbf {\bibinfo {volume} {120}},\ \bibinfo
  {pages} {031102} (\bibinfo {year} {2018})},\ \Eprint
  {http://arxiv.org/abs/1705.10850} {arXiv:1705.10850 [gr-qc]} \BibitemShut
  {NoStop}%
\bibitem [{\citenamefont {Breschi}\ \emph
  {et~al.}(2022{\natexlab{a}})\citenamefont {Breschi}, \citenamefont
  {Bernuzzi}, \citenamefont {Godzieba}, \citenamefont {Perego},\ and\
  \citenamefont {Radice}}]{Breschi:2021xrx}%
  \BibitemOpen
  \bibfield  {author} {\bibinfo {author} {\bibfnamefont {M.}~\bibnamefont
  {Breschi}}, \bibinfo {author} {\bibfnamefont {S.}~\bibnamefont {Bernuzzi}},
  \bibinfo {author} {\bibfnamefont {D.}~\bibnamefont {Godzieba}}, \bibinfo
  {author} {\bibfnamefont {A.}~\bibnamefont {Perego}}, \ and\ \bibinfo {author}
  {\bibfnamefont {D.}~\bibnamefont {Radice}},\ }\href {\doibase
  10.1103/PhysRevLett.128.161102} {\bibfield  {journal} {\bibinfo  {journal}
  {Phys. Rev. Lett.}\ }\textbf {\bibinfo {volume} {128}},\ \bibinfo {pages}
  {161102} (\bibinfo {year} {2022}{\natexlab{a}})},\ \Eprint
  {http://arxiv.org/abs/2110.06957} {arXiv:2110.06957 [gr-qc]} \BibitemShut
  {NoStop}%
\bibitem [{\citenamefont {Wijngaarden}\ \emph {et~al.}(2022)\citenamefont
  {Wijngaarden}, \citenamefont {Chatziioannou}, \citenamefont {Bauswein},
  \citenamefont {Clark},\ and\ \citenamefont {Cornish}}]{Wijngaarden:2022sah}%
  \BibitemOpen
  \bibfield  {author} {\bibinfo {author} {\bibfnamefont {M.}~\bibnamefont
  {Wijngaarden}}, \bibinfo {author} {\bibfnamefont {K.}~\bibnamefont
  {Chatziioannou}}, \bibinfo {author} {\bibfnamefont {A.}~\bibnamefont
  {Bauswein}}, \bibinfo {author} {\bibfnamefont {J.~A.}\ \bibnamefont {Clark}},
  \ and\ \bibinfo {author} {\bibfnamefont {N.~J.}\ \bibnamefont {Cornish}},\
  }\href {\doibase 10.1103/PhysRevD.105.104019} {\bibfield  {journal} {\bibinfo
   {journal} {Phys. Rev. D}\ }\textbf {\bibinfo {volume} {105}},\ \bibinfo
  {pages} {104019} (\bibinfo {year} {2022})},\ \Eprint
  {http://arxiv.org/abs/2202.09382} {arXiv:2202.09382 [gr-qc]} \BibitemShut
  {NoStop}%
\bibitem [{\citenamefont {Shibata}\ \emph {et~al.}(2005)\citenamefont
  {Shibata}, \citenamefont {Taniguchi},\ and\ \citenamefont
  {Uryu}}]{Shibata:2005ss}%
  \BibitemOpen
  \bibfield  {author} {\bibinfo {author} {\bibfnamefont {M.}~\bibnamefont
  {Shibata}}, \bibinfo {author} {\bibfnamefont {K.}~\bibnamefont {Taniguchi}},
  \ and\ \bibinfo {author} {\bibfnamefont {K.}~\bibnamefont {Uryu}},\ }\href
  {\doibase 10.1103/PhysRevD.71.084021} {\bibfield  {journal} {\bibinfo
  {journal} {Phys. Rev. D}\ }\textbf {\bibinfo {volume} {71}},\ \bibinfo
  {pages} {084021} (\bibinfo {year} {2005})},\ \Eprint
  {http://arxiv.org/abs/gr-qc/0503119} {arXiv:gr-qc/0503119} \BibitemShut
  {NoStop}%
\bibitem [{\citenamefont {Zappa}\ \emph {et~al.}(2018)\citenamefont {Zappa},
  \citenamefont {Bernuzzi}, \citenamefont {Radice}, \citenamefont {Perego},\
  and\ \citenamefont {Dietrich}}]{Zappa:2017xba}%
  \BibitemOpen
  \bibfield  {author} {\bibinfo {author} {\bibfnamefont {F.}~\bibnamefont
  {Zappa}}, \bibinfo {author} {\bibfnamefont {S.}~\bibnamefont {Bernuzzi}},
  \bibinfo {author} {\bibfnamefont {D.}~\bibnamefont {Radice}}, \bibinfo
  {author} {\bibfnamefont {A.}~\bibnamefont {Perego}}, \ and\ \bibinfo {author}
  {\bibfnamefont {T.}~\bibnamefont {Dietrich}},\ }\href {\doibase
  10.1103/PhysRevLett.120.111101} {\bibfield  {journal} {\bibinfo  {journal}
  {Phys. Rev. Lett.}\ }\textbf {\bibinfo {volume} {120}},\ \bibinfo {pages}
  {111101} (\bibinfo {year} {2018})},\ \Eprint
  {http://arxiv.org/abs/1712.04267} {arXiv:1712.04267 [gr-qc]} \BibitemShut
  {NoStop}%
\bibitem [{\citenamefont {Agathos}\ \emph {et~al.}(2020)\citenamefont
  {Agathos}, \citenamefont {Zappa}, \citenamefont {Bernuzzi}, \citenamefont
  {Perego}, \citenamefont {Breschi},\ and\ \citenamefont
  {Radice}}]{Agathos:2019sah}%
  \BibitemOpen
  \bibfield  {author} {\bibinfo {author} {\bibfnamefont {M.}~\bibnamefont
  {Agathos}}, \bibinfo {author} {\bibfnamefont {F.}~\bibnamefont {Zappa}},
  \bibinfo {author} {\bibfnamefont {S.}~\bibnamefont {Bernuzzi}}, \bibinfo
  {author} {\bibfnamefont {A.}~\bibnamefont {Perego}}, \bibinfo {author}
  {\bibfnamefont {M.}~\bibnamefont {Breschi}}, \ and\ \bibinfo {author}
  {\bibfnamefont {D.}~\bibnamefont {Radice}},\ }\href {\doibase
  10.1103/PhysRevD.101.044006} {\bibfield  {journal} {\bibinfo  {journal}
  {Phys. Rev. D}\ }\textbf {\bibinfo {volume} {101}},\ \bibinfo {pages}
  {044006} (\bibinfo {year} {2020})},\ \Eprint
  {http://arxiv.org/abs/1908.05442} {arXiv:1908.05442 [gr-qc]} \BibitemShut
  {NoStop}%
\bibitem [{\citenamefont {K\"oppel}\ \emph {et~al.}(2019)\citenamefont
  {K\"oppel}, \citenamefont {Bovard},\ and\ \citenamefont
  {Rezzolla}}]{Koppel:2019pys}%
  \BibitemOpen
  \bibfield  {author} {\bibinfo {author} {\bibfnamefont {S.}~\bibnamefont
  {K\"oppel}}, \bibinfo {author} {\bibfnamefont {L.}~\bibnamefont {Bovard}}, \
  and\ \bibinfo {author} {\bibfnamefont {L.}~\bibnamefont {Rezzolla}},\ }\href
  {\doibase 10.3847/2041-8213/ab0210} {\bibfield  {journal} {\bibinfo
  {journal} {Astrophys. J. Lett.}\ }\textbf {\bibinfo {volume} {872}},\
  \bibinfo {pages} {L16} (\bibinfo {year} {2019})},\ \Eprint
  {http://arxiv.org/abs/1901.09977} {arXiv:1901.09977 [gr-qc]} \BibitemShut
  {NoStop}%
\bibitem [{\citenamefont {Bauswein}\ \emph {et~al.}(2020)\citenamefont
  {Bauswein}, \citenamefont {Blacker}, \citenamefont {Vijayan}, \citenamefont
  {Stergioulas}, \citenamefont {Chatziioannou}, \citenamefont {Clark},
  \citenamefont {Bastian}, \citenamefont {Blaschke}, \citenamefont {Cierniak},\
  and\ \citenamefont {Fischer}}]{Bauswein:2020aag}%
  \BibitemOpen
  \bibfield  {author} {\bibinfo {author} {\bibfnamefont {A.}~\bibnamefont
  {Bauswein}}, \bibinfo {author} {\bibfnamefont {S.}~\bibnamefont {Blacker}},
  \bibinfo {author} {\bibfnamefont {V.}~\bibnamefont {Vijayan}}, \bibinfo
  {author} {\bibfnamefont {N.}~\bibnamefont {Stergioulas}}, \bibinfo {author}
  {\bibfnamefont {K.}~\bibnamefont {Chatziioannou}}, \bibinfo {author}
  {\bibfnamefont {J.~A.}\ \bibnamefont {Clark}}, \bibinfo {author}
  {\bibfnamefont {N.-U.~F.}\ \bibnamefont {Bastian}}, \bibinfo {author}
  {\bibfnamefont {D.~B.}\ \bibnamefont {Blaschke}}, \bibinfo {author}
  {\bibfnamefont {M.}~\bibnamefont {Cierniak}}, \ and\ \bibinfo {author}
  {\bibfnamefont {T.}~\bibnamefont {Fischer}},\ }\href {\doibase
  10.1103/PhysRevLett.125.141103} {\bibfield  {journal} {\bibinfo  {journal}
  {Phys. Rev. Lett.}\ }\textbf {\bibinfo {volume} {125}},\ \bibinfo {pages}
  {141103} (\bibinfo {year} {2020})},\ \Eprint
  {http://arxiv.org/abs/2004.00846} {arXiv:2004.00846 [astro-ph.HE]}
  \BibitemShut {NoStop}%
\bibitem [{\citenamefont {Bauswein}\ \emph {et~al.}(2021)\citenamefont
  {Bauswein}, \citenamefont {Blacker}, \citenamefont {Lioutas}, \citenamefont
  {Soultanis}, \citenamefont {Vijayan},\ and\ \citenamefont
  {Stergioulas}}]{Bauswein:2020xlt}%
  \BibitemOpen
  \bibfield  {author} {\bibinfo {author} {\bibfnamefont {A.}~\bibnamefont
  {Bauswein}}, \bibinfo {author} {\bibfnamefont {S.}~\bibnamefont {Blacker}},
  \bibinfo {author} {\bibfnamefont {G.}~\bibnamefont {Lioutas}}, \bibinfo
  {author} {\bibfnamefont {T.}~\bibnamefont {Soultanis}}, \bibinfo {author}
  {\bibfnamefont {V.}~\bibnamefont {Vijayan}}, \ and\ \bibinfo {author}
  {\bibfnamefont {N.}~\bibnamefont {Stergioulas}},\ }\href {\doibase
  10.1103/PhysRevD.103.123004} {\bibfield  {journal} {\bibinfo  {journal}
  {Phys. Rev. D}\ }\textbf {\bibinfo {volume} {103}},\ \bibinfo {pages}
  {123004} (\bibinfo {year} {2021})},\ \Eprint
  {http://arxiv.org/abs/2010.04461} {arXiv:2010.04461 [astro-ph.HE]}
  \BibitemShut {NoStop}%
\bibitem [{\citenamefont {Kashyap}\ \emph {et~al.}(2022)\citenamefont {Kashyap}
  \emph {et~al.}}]{Kashyap:2021wzs}%
  \BibitemOpen
  \bibfield  {author} {\bibinfo {author} {\bibfnamefont {R.}~\bibnamefont
  {Kashyap}} \emph {et~al.},\ }\href {\doibase 10.1103/PhysRevD.105.103022}
  {\bibfield  {journal} {\bibinfo  {journal} {Phys. Rev. D}\ }\textbf {\bibinfo
  {volume} {105}},\ \bibinfo {pages} {103022} (\bibinfo {year} {2022})},\
  \Eprint {http://arxiv.org/abs/2111.05183} {arXiv:2111.05183 [astro-ph.HE]}
  \BibitemShut {NoStop}%
\bibitem [{\citenamefont {K\"olsch}\ \emph {et~al.}(2022)\citenamefont
  {K\"olsch}, \citenamefont {Dietrich}, \citenamefont {Ujevic},\ and\
  \citenamefont {Bruegmann}}]{Kolsch:2021lub}%
  \BibitemOpen
  \bibfield  {author} {\bibinfo {author} {\bibfnamefont {M.}~\bibnamefont
  {K\"olsch}}, \bibinfo {author} {\bibfnamefont {T.}~\bibnamefont {Dietrich}},
  \bibinfo {author} {\bibfnamefont {M.}~\bibnamefont {Ujevic}}, \ and\ \bibinfo
  {author} {\bibfnamefont {B.}~\bibnamefont {Bruegmann}},\ }\href {\doibase
  10.1103/PhysRevD.106.044026} {\bibfield  {journal} {\bibinfo  {journal}
  {Phys. Rev. D}\ }\textbf {\bibinfo {volume} {106}},\ \bibinfo {pages}
  {044026} (\bibinfo {year} {2022})},\ \Eprint
  {http://arxiv.org/abs/2112.11851} {arXiv:2112.11851 [gr-qc]} \BibitemShut
  {NoStop}%
\bibitem [{\citenamefont {\c{C}okluk}\ \emph {et~al.}(2023)\citenamefont
  {\c{C}okluk}, \citenamefont {Yakut},\ and\ \citenamefont
  {Giacomazzo}}]{Cokluk:2023xio}%
  \BibitemOpen
  \bibfield  {author} {\bibinfo {author} {\bibfnamefont {K.~A.}\ \bibnamefont
  {\c{C}okluk}}, \bibinfo {author} {\bibfnamefont {K.}~\bibnamefont {Yakut}}, \
  and\ \bibinfo {author} {\bibfnamefont {B.}~\bibnamefont {Giacomazzo}},\
  }\href@noop {} {\  (\bibinfo {year} {2023})},\ \Eprint
  {http://arxiv.org/abs/2301.09635} {arXiv:2301.09635 [astro-ph.HE]}
  \BibitemShut {NoStop}%
\bibitem [{\citenamefont {Zhang}\ \emph {et~al.}(2021)\citenamefont {Zhang},
  \citenamefont {Smetana}, \citenamefont {Chen}, \citenamefont {Bentley},
  \citenamefont {Martynov}, \citenamefont {Miao}, \citenamefont {East},\ and\
  \citenamefont {Yang}}]{Zhang:2020qlh}%
  \BibitemOpen
  \bibfield  {author} {\bibinfo {author} {\bibfnamefont {T.}~\bibnamefont
  {Zhang}}, \bibinfo {author} {\bibfnamefont {J.}~\bibnamefont {Smetana}},
  \bibinfo {author} {\bibfnamefont {Y.}~\bibnamefont {Chen}}, \bibinfo {author}
  {\bibfnamefont {J.}~\bibnamefont {Bentley}}, \bibinfo {author} {\bibfnamefont
  {D.}~\bibnamefont {Martynov}}, \bibinfo {author} {\bibfnamefont
  {H.}~\bibnamefont {Miao}}, \bibinfo {author} {\bibfnamefont {W.~E.}\
  \bibnamefont {East}}, \ and\ \bibinfo {author} {\bibfnamefont
  {H.}~\bibnamefont {Yang}},\ }\href {\doibase 10.1103/PhysRevD.103.044063}
  {\bibfield  {journal} {\bibinfo  {journal} {Phys. Rev. D}\ }\textbf {\bibinfo
  {volume} {103}},\ \bibinfo {pages} {044063} (\bibinfo {year} {2021})},\
  \Eprint {http://arxiv.org/abs/2011.06705} {arXiv:2011.06705 [gr-qc]}
  \BibitemShut {NoStop}%
\bibitem [{\citenamefont {Easter}\ \emph {et~al.}(2021)\citenamefont {Easter},
  \citenamefont {Lasky},\ and\ \citenamefont {Casey}}]{Easter:2021wlb}%
  \BibitemOpen
  \bibfield  {author} {\bibinfo {author} {\bibfnamefont {P.~J.}\ \bibnamefont
  {Easter}}, \bibinfo {author} {\bibfnamefont {P.~D.}\ \bibnamefont {Lasky}}, \
  and\ \bibinfo {author} {\bibfnamefont {A.~R.}\ \bibnamefont {Casey}},\
  }\href@noop {} {\  (\bibinfo {year} {2021})},\ \Eprint
  {http://arxiv.org/abs/2106.04064} {arXiv:2106.04064 [astro-ph.HE]}
  \BibitemShut {NoStop}%
\bibitem [{\citenamefont {Breschi}\ \emph
  {et~al.}(2022{\natexlab{b}})\citenamefont {Breschi}, \citenamefont
  {Bernuzzi}, \citenamefont {Chakravarti}, \citenamefont {Camilletti},
  \citenamefont {Prakash},\ and\ \citenamefont {Perego}}]{Breschi:2022xnc}%
  \BibitemOpen
  \bibfield  {author} {\bibinfo {author} {\bibfnamefont {M.}~\bibnamefont
  {Breschi}}, \bibinfo {author} {\bibfnamefont {S.}~\bibnamefont {Bernuzzi}},
  \bibinfo {author} {\bibfnamefont {K.}~\bibnamefont {Chakravarti}}, \bibinfo
  {author} {\bibfnamefont {A.}~\bibnamefont {Camilletti}}, \bibinfo {author}
  {\bibfnamefont {A.}~\bibnamefont {Prakash}}, \ and\ \bibinfo {author}
  {\bibfnamefont {A.}~\bibnamefont {Perego}},\ }\href@noop {} {\  (\bibinfo
  {year} {2022}{\natexlab{b}})},\ \Eprint {http://arxiv.org/abs/2205.09112}
  {arXiv:2205.09112 [gr-qc]} \BibitemShut {NoStop}%
\bibitem [{\citenamefont {Bernuzzi}\ \emph {et~al.}(2016)\citenamefont
  {Bernuzzi}, \citenamefont {Radice}, \citenamefont {Ott}, \citenamefont
  {Roberts}, \citenamefont {Moesta},\ and\ \citenamefont
  {Galeazzi}}]{Bernuzzi:2015opx}%
  \BibitemOpen
  \bibfield  {author} {\bibinfo {author} {\bibfnamefont {S.}~\bibnamefont
  {Bernuzzi}}, \bibinfo {author} {\bibfnamefont {D.}~\bibnamefont {Radice}},
  \bibinfo {author} {\bibfnamefont {C.~D.}\ \bibnamefont {Ott}}, \bibinfo
  {author} {\bibfnamefont {L.~F.}\ \bibnamefont {Roberts}}, \bibinfo {author}
  {\bibfnamefont {P.}~\bibnamefont {Moesta}}, \ and\ \bibinfo {author}
  {\bibfnamefont {F.}~\bibnamefont {Galeazzi}},\ }\href {\doibase
  10.1103/PhysRevD.94.024023} {\bibfield  {journal} {\bibinfo  {journal} {Phys.
  Rev. D}\ }\textbf {\bibinfo {volume} {94}},\ \bibinfo {pages} {024023}
  (\bibinfo {year} {2016})},\ \Eprint {http://arxiv.org/abs/1512.06397}
  {arXiv:1512.06397 [gr-qc]} \BibitemShut {NoStop}%
\bibitem [{\citenamefont {Aggarwal}\ \emph {et~al.}(2021)\citenamefont
  {Aggarwal} \emph {et~al.}}]{Aggarwal:2020olq}%
  \BibitemOpen
  \bibfield  {author} {\bibinfo {author} {\bibfnamefont {N.}~\bibnamefont
  {Aggarwal}} \emph {et~al.},\ }\href {\doibase 10.1007/s41114-021-00032-5}
  {\bibfield  {journal} {\bibinfo  {journal} {Living Rev. Rel.}\ }\textbf
  {\bibinfo {volume} {24}},\ \bibinfo {pages} {4} (\bibinfo {year} {2021})},\
  \Eprint {http://arxiv.org/abs/2011.12414} {arXiv:2011.12414 [gr-qc]}
  \BibitemShut {NoStop}%
\bibitem [{\citenamefont {Punturo}\ \emph
  {et~al.}(2010{\natexlab{b}})\citenamefont {Punturo} \emph
  {et~al.}}]{Punturo:2010zza}%
  \BibitemOpen
  \bibfield  {author} {\bibinfo {author} {\bibfnamefont {M.}~\bibnamefont
  {Punturo}} \emph {et~al.},\ }\href {\doibase 10.1088/0264-9381/27/8/084007}
  {\bibfield  {journal} {\bibinfo  {journal} {Class. Quant. Grav.}\ }\textbf
  {\bibinfo {volume} {27}},\ \bibinfo {pages} {084007} (\bibinfo {year}
  {2010}{\natexlab{b}})}\BibitemShut {NoStop}%
\bibitem [{\citenamefont {Radice}\ and\ \citenamefont
  {Rezzolla}(2012)}]{Radice:2012cu}%
  \BibitemOpen
  \bibfield  {author} {\bibinfo {author} {\bibfnamefont {D.}~\bibnamefont
  {Radice}}\ and\ \bibinfo {author} {\bibfnamefont {L.}~\bibnamefont
  {Rezzolla}},\ }\href {\doibase 10.1051/0004-6361/201219735} {\bibfield
  {journal} {\bibinfo  {journal} {Astron. Astrophys.}\ }\textbf {\bibinfo
  {volume} {547}},\ \bibinfo {pages} {A26} (\bibinfo {year} {2012})},\ \Eprint
  {http://arxiv.org/abs/1206.6502} {arXiv:1206.6502 [astro-ph.IM]} \BibitemShut
  {NoStop}%
\bibitem [{\citenamefont {Radice}\ \emph
  {et~al.}(2014{\natexlab{a}})\citenamefont {Radice}, \citenamefont
  {Rezzolla},\ and\ \citenamefont {Galeazzi}}]{Radice:2013hxh}%
  \BibitemOpen
  \bibfield  {author} {\bibinfo {author} {\bibfnamefont {D.}~\bibnamefont
  {Radice}}, \bibinfo {author} {\bibfnamefont {L.}~\bibnamefont {Rezzolla}}, \
  and\ \bibinfo {author} {\bibfnamefont {F.}~\bibnamefont {Galeazzi}},\ }\href
  {\doibase 10.1093/mnrasl/slt137} {\bibfield  {journal} {\bibinfo  {journal}
  {Mon. Not. Roy. Astron. Soc.}\ }\textbf {\bibinfo {volume} {437}},\ \bibinfo
  {pages} {L46} (\bibinfo {year} {2014}{\natexlab{a}})},\ \Eprint
  {http://arxiv.org/abs/1306.6052} {arXiv:1306.6052 [gr-qc]} \BibitemShut
  {NoStop}%
\bibitem [{\citenamefont {Radice}\ \emph
  {et~al.}(2014{\natexlab{b}})\citenamefont {Radice}, \citenamefont
  {Rezzolla},\ and\ \citenamefont {Galeazzi}}]{Radice:2013xpa}%
  \BibitemOpen
  \bibfield  {author} {\bibinfo {author} {\bibfnamefont {D.}~\bibnamefont
  {Radice}}, \bibinfo {author} {\bibfnamefont {L.}~\bibnamefont {Rezzolla}}, \
  and\ \bibinfo {author} {\bibfnamefont {F.}~\bibnamefont {Galeazzi}},\ }\href
  {\doibase 10.1088/0264-9381/31/7/075012} {\bibfield  {journal} {\bibinfo
  {journal} {Class. Quant. Grav.}\ }\textbf {\bibinfo {volume} {31}},\ \bibinfo
  {pages} {075012} (\bibinfo {year} {2014}{\natexlab{b}})},\ \Eprint
  {http://arxiv.org/abs/1312.5004} {arXiv:1312.5004 [gr-qc]} \BibitemShut
  {NoStop}%
\bibitem [{\citenamefont {Gonzalez}\ \emph {et~al.}(2023)\citenamefont
  {Gonzalez} \emph {et~al.}}]{Gonzalez:2022mgo}%
  \BibitemOpen
  \bibfield  {author} {\bibinfo {author} {\bibfnamefont {A.}~\bibnamefont
  {Gonzalez}} \emph {et~al.},\ }\href {\doibase 10.1088/1361-6382/acc231}
  {\bibfield  {journal} {\bibinfo  {journal} {Class. Quant. Grav.}\ }\textbf
  {\bibinfo {volume} {40}},\ \bibinfo {pages} {085011} (\bibinfo {year}
  {2023})},\ \Eprint {http://arxiv.org/abs/2210.16366} {arXiv:2210.16366
  [gr-qc]} \BibitemShut {NoStop}%
\bibitem [{\citenamefont {Radice}\ \emph {et~al.}(2016)\citenamefont {Radice},
  \citenamefont {Galeazzi}, \citenamefont {Lippuner}, \citenamefont {Roberts},
  \citenamefont {Ott},\ and\ \citenamefont {Rezzolla}}]{Radice:2016dwd}%
  \BibitemOpen
  \bibfield  {author} {\bibinfo {author} {\bibfnamefont {D.}~\bibnamefont
  {Radice}}, \bibinfo {author} {\bibfnamefont {F.}~\bibnamefont {Galeazzi}},
  \bibinfo {author} {\bibfnamefont {J.}~\bibnamefont {Lippuner}}, \bibinfo
  {author} {\bibfnamefont {L.~F.}\ \bibnamefont {Roberts}}, \bibinfo {author}
  {\bibfnamefont {C.~D.}\ \bibnamefont {Ott}}, \ and\ \bibinfo {author}
  {\bibfnamefont {L.}~\bibnamefont {Rezzolla}},\ }\href {\doibase
  10.1093/mnras/stw1227} {\bibfield  {journal} {\bibinfo  {journal} {Mon. Not.
  Roy. Astron. Soc.}\ }\textbf {\bibinfo {volume} {460}},\ \bibinfo {pages}
  {3255} (\bibinfo {year} {2016})},\ \Eprint {http://arxiv.org/abs/1601.02426}
  {arXiv:1601.02426 [astro-ph.HE]} \BibitemShut {NoStop}%
\bibitem [{\citenamefont {Radice}\ \emph
  {et~al.}(2018{\natexlab{b}})\citenamefont {Radice}, \citenamefont {Perego},
  \citenamefont {Hotokezaka}, \citenamefont {Fromm}, \citenamefont {Bernuzzi},\
  and\ \citenamefont {Roberts}}]{Radice:2018pdn}%
  \BibitemOpen
  \bibfield  {author} {\bibinfo {author} {\bibfnamefont {D.}~\bibnamefont
  {Radice}}, \bibinfo {author} {\bibfnamefont {A.}~\bibnamefont {Perego}},
  \bibinfo {author} {\bibfnamefont {K.}~\bibnamefont {Hotokezaka}}, \bibinfo
  {author} {\bibfnamefont {S.~A.}\ \bibnamefont {Fromm}}, \bibinfo {author}
  {\bibfnamefont {S.}~\bibnamefont {Bernuzzi}}, \ and\ \bibinfo {author}
  {\bibfnamefont {L.~F.}\ \bibnamefont {Roberts}},\ }\href {\doibase
  10.3847/1538-4357/aaf054} {\bibfield  {journal} {\bibinfo  {journal}
  {Astrophys. J.}\ }\textbf {\bibinfo {volume} {869}},\ \bibinfo {pages} {130}
  (\bibinfo {year} {2018}{\natexlab{b}})},\ \Eprint
  {http://arxiv.org/abs/1809.11161} {arXiv:1809.11161 [astro-ph.HE]}
  \BibitemShut {NoStop}%
\bibitem [{\citenamefont {Radice}(2017)}]{Radice:2017zta}%
  \BibitemOpen
  \bibfield  {author} {\bibinfo {author} {\bibfnamefont {D.}~\bibnamefont
  {Radice}},\ }\href {\doibase 10.3847/2041-8213/aa6483} {\bibfield  {journal}
  {\bibinfo  {journal} {Astrophys. J. Lett.}\ }\textbf {\bibinfo {volume}
  {838}},\ \bibinfo {pages} {L2} (\bibinfo {year} {2017})},\ \Eprint
  {http://arxiv.org/abs/1703.02046} {arXiv:1703.02046 [astro-ph.HE]}
  \BibitemShut {NoStop}%
\bibitem [{\citenamefont {Radice}(2020)}]{Radice:2020ids}%
  \BibitemOpen
  \bibfield  {author} {\bibinfo {author} {\bibfnamefont {D.}~\bibnamefont
  {Radice}},\ }\href {\doibase 10.3390/sym12081249} {\bibfield  {journal}
  {\bibinfo  {journal} {Symmetry}\ }\textbf {\bibinfo {volume} {12}},\ \bibinfo
  {pages} {1249} (\bibinfo {year} {2020})},\ \Eprint
  {http://arxiv.org/abs/2005.09002} {arXiv:2005.09002 [astro-ph.HE]}
  \BibitemShut {NoStop}%
\bibitem [{\citenamefont {Baiotti}\ \emph {et~al.}(2005)\citenamefont
  {Baiotti}, \citenamefont {Hawke}, \citenamefont {Rezzolla},\ and\
  \citenamefont {Schnetter}}]{Baiotti:2005vi}%
  \BibitemOpen
  \bibfield  {author} {\bibinfo {author} {\bibfnamefont {L.}~\bibnamefont
  {Baiotti}}, \bibinfo {author} {\bibfnamefont {I.}~\bibnamefont {Hawke}},
  \bibinfo {author} {\bibfnamefont {L.}~\bibnamefont {Rezzolla}}, \ and\
  \bibinfo {author} {\bibfnamefont {E.}~\bibnamefont {Schnetter}},\ }\href
  {\doibase 10.1103/PhysRevLett.94.131101} {\bibfield  {journal} {\bibinfo
  {journal} {Phys. Rev. Lett.}\ }\textbf {\bibinfo {volume} {94}},\ \bibinfo
  {pages} {131101} (\bibinfo {year} {2005})},\ \Eprint
  {http://arxiv.org/abs/gr-qc/0503016} {arXiv:gr-qc/0503016} \BibitemShut
  {NoStop}%
\bibitem [{\citenamefont {Baiotti}\ \emph {et~al.}(2007)\citenamefont
  {Baiotti}, \citenamefont {Hawke},\ and\ \citenamefont
  {Rezzolla}}]{Baiotti:2007np}%
  \BibitemOpen
  \bibfield  {author} {\bibinfo {author} {\bibfnamefont {L.}~\bibnamefont
  {Baiotti}}, \bibinfo {author} {\bibfnamefont {I.}~\bibnamefont {Hawke}}, \
  and\ \bibinfo {author} {\bibfnamefont {L.}~\bibnamefont {Rezzolla}},\ }\href
  {\doibase 10.1088/0264-9381/24/12/S13} {\bibfield  {journal} {\bibinfo
  {journal} {Class. Quant. Grav.}\ }\textbf {\bibinfo {volume} {24}},\ \bibinfo
  {pages} {S187} (\bibinfo {year} {2007})},\ \Eprint
  {http://arxiv.org/abs/gr-qc/0701043} {arXiv:gr-qc/0701043} \BibitemShut
  {NoStop}%
\bibitem [{\citenamefont {Soultanis}\ \emph {et~al.}(2022)\citenamefont
  {Soultanis}, \citenamefont {Bauswein},\ and\ \citenamefont
  {Stergioulas}}]{Soultanis:2021oia}%
  \BibitemOpen
  \bibfield  {author} {\bibinfo {author} {\bibfnamefont {T.}~\bibnamefont
  {Soultanis}}, \bibinfo {author} {\bibfnamefont {A.}~\bibnamefont {Bauswein}},
  \ and\ \bibinfo {author} {\bibfnamefont {N.}~\bibnamefont {Stergioulas}},\
  }\href {\doibase 10.1103/PhysRevD.105.043020} {\bibfield  {journal} {\bibinfo
   {journal} {Phys. Rev. D}\ }\textbf {\bibinfo {volume} {105}},\ \bibinfo
  {pages} {043020} (\bibinfo {year} {2022})},\ \Eprint
  {http://arxiv.org/abs/2111.08353} {arXiv:2111.08353 [astro-ph.HE]}
  \BibitemShut {NoStop}%
\bibitem [{\citenamefont {Bauswein}\ and\ \citenamefont
  {Stergioulas}(2015)}]{Bauswein:2015yca}%
  \BibitemOpen
  \bibfield  {author} {\bibinfo {author} {\bibfnamefont {A.}~\bibnamefont
  {Bauswein}}\ and\ \bibinfo {author} {\bibfnamefont {N.}~\bibnamefont
  {Stergioulas}},\ }\href {\doibase 10.1103/PhysRevD.91.124056} {\bibfield
  {journal} {\bibinfo  {journal} {Phys. Rev. D}\ }\textbf {\bibinfo {volume}
  {91}},\ \bibinfo {pages} {124056} (\bibinfo {year} {2015})},\ \Eprint
  {http://arxiv.org/abs/1502.03176} {arXiv:1502.03176 [astro-ph.SR]}
  \BibitemShut {NoStop}%
\bibitem [{\citenamefont {De~Pietri}\ \emph {et~al.}(2020)\citenamefont
  {De~Pietri}, \citenamefont {Feo}, \citenamefont {Font}, \citenamefont
  {L\"offler}, \citenamefont {Pasquali},\ and\ \citenamefont
  {Stergioulas}}]{DePietri:2019mti}%
  \BibitemOpen
  \bibfield  {author} {\bibinfo {author} {\bibfnamefont {R.}~\bibnamefont
  {De~Pietri}}, \bibinfo {author} {\bibfnamefont {A.}~\bibnamefont {Feo}},
  \bibinfo {author} {\bibfnamefont {J.~A.}\ \bibnamefont {Font}}, \bibinfo
  {author} {\bibfnamefont {F.}~\bibnamefont {L\"offler}}, \bibinfo {author}
  {\bibfnamefont {M.}~\bibnamefont {Pasquali}}, \ and\ \bibinfo {author}
  {\bibfnamefont {N.}~\bibnamefont {Stergioulas}},\ }\href {\doibase
  10.1103/PhysRevD.101.064052} {\bibfield  {journal} {\bibinfo  {journal}
  {Phys. Rev. D}\ }\textbf {\bibinfo {volume} {101}},\ \bibinfo {pages}
  {064052} (\bibinfo {year} {2020})},\ \Eprint
  {http://arxiv.org/abs/1910.04036} {arXiv:1910.04036 [gr-qc]} \BibitemShut
  {NoStop}%
\bibitem [{\citenamefont {Nedora}\ \emph {et~al.}(2021)\citenamefont {Nedora},
  \citenamefont {Bernuzzi}, \citenamefont {Radice}, \citenamefont {Daszuta},
  \citenamefont {Endrizzi}, \citenamefont {Perego}, \citenamefont {Prakash},
  \citenamefont {Safarzadeh}, \citenamefont {Schianchi},\ and\ \citenamefont
  {Logoteta}}]{Nedora:2020hxc}%
  \BibitemOpen
  \bibfield  {author} {\bibinfo {author} {\bibfnamefont {V.}~\bibnamefont
  {Nedora}}, \bibinfo {author} {\bibfnamefont {S.}~\bibnamefont {Bernuzzi}},
  \bibinfo {author} {\bibfnamefont {D.}~\bibnamefont {Radice}}, \bibinfo
  {author} {\bibfnamefont {B.}~\bibnamefont {Daszuta}}, \bibinfo {author}
  {\bibfnamefont {A.}~\bibnamefont {Endrizzi}}, \bibinfo {author}
  {\bibfnamefont {A.}~\bibnamefont {Perego}}, \bibinfo {author} {\bibfnamefont
  {A.}~\bibnamefont {Prakash}}, \bibinfo {author} {\bibfnamefont
  {M.}~\bibnamefont {Safarzadeh}}, \bibinfo {author} {\bibfnamefont
  {F.}~\bibnamefont {Schianchi}}, \ and\ \bibinfo {author} {\bibfnamefont
  {D.}~\bibnamefont {Logoteta}},\ }\href {\doibase 10.3847/1538-4357/abc9be}
  {\bibfield  {journal} {\bibinfo  {journal} {Astrophys. J.}\ }\textbf
  {\bibinfo {volume} {906}},\ \bibinfo {pages} {98} (\bibinfo {year} {2021})},\
  \Eprint {http://arxiv.org/abs/2008.04333} {arXiv:2008.04333 [astro-ph.HE]}
  \BibitemShut {NoStop}%
\bibitem [{\citenamefont {Camilletti}\ \emph {et~al.}(2022)\citenamefont
  {Camilletti}, \citenamefont {Chiesa}, \citenamefont {Ricigliano},
  \citenamefont {Perego}, \citenamefont {Lippold}, \citenamefont {Padamata},
  \citenamefont {Bernuzzi}, \citenamefont {Radice}, \citenamefont {Logoteta},\
  and\ \citenamefont {Guercilena}}]{Camilletti:2022jms}%
  \BibitemOpen
  \bibfield  {author} {\bibinfo {author} {\bibfnamefont {A.}~\bibnamefont
  {Camilletti}}, \bibinfo {author} {\bibfnamefont {L.}~\bibnamefont {Chiesa}},
  \bibinfo {author} {\bibfnamefont {G.}~\bibnamefont {Ricigliano}}, \bibinfo
  {author} {\bibfnamefont {A.}~\bibnamefont {Perego}}, \bibinfo {author}
  {\bibfnamefont {L.~C.}\ \bibnamefont {Lippold}}, \bibinfo {author}
  {\bibfnamefont {S.}~\bibnamefont {Padamata}}, \bibinfo {author}
  {\bibfnamefont {S.}~\bibnamefont {Bernuzzi}}, \bibinfo {author}
  {\bibfnamefont {D.}~\bibnamefont {Radice}}, \bibinfo {author} {\bibfnamefont
  {D.}~\bibnamefont {Logoteta}}, \ and\ \bibinfo {author} {\bibfnamefont
  {F.~M.}\ \bibnamefont {Guercilena}},\ }\href {\doibase
  10.1093/mnras/stac2333} {\  (\bibinfo {year} {2022}),\
  10.1093/mnras/stac2333},\ \Eprint {http://arxiv.org/abs/2204.05336}
  {arXiv:2204.05336 [astro-ph.HE]} \BibitemShut {NoStop}%
\bibitem [{\citenamefont {Borhanian}\ and\ \citenamefont
  {Sathyaprakash}(2022)}]{Borhanian:2022czq}%
  \BibitemOpen
  \bibfield  {author} {\bibinfo {author} {\bibfnamefont {S.}~\bibnamefont
  {Borhanian}}\ and\ \bibinfo {author} {\bibfnamefont {B.~S.}\ \bibnamefont
  {Sathyaprakash}},\ }\href@noop {} {\  (\bibinfo {year} {2022})},\ \Eprint
  {http://arxiv.org/abs/2202.11048} {arXiv:2202.11048 [gr-qc]} \BibitemShut
  {NoStop}%
\bibitem [{\citenamefont {Abbott}\ \emph {et~al.}(2023)\citenamefont {Abbott}
  \emph {et~al.}}]{KAGRA:2021duu}%
  \BibitemOpen
  \bibfield  {author} {\bibinfo {author} {\bibfnamefont {R.}~\bibnamefont
  {Abbott}} \emph {et~al.} (\bibinfo {collaboration} {KAGRA, VIRGO, LIGO
  Scientific}),\ }\href {\doibase 10.1103/PhysRevX.13.011048} {\bibfield
  {journal} {\bibinfo  {journal} {Phys. Rev. X}\ }\textbf {\bibinfo {volume}
  {13}},\ \bibinfo {pages} {011048} (\bibinfo {year} {2023})},\ \Eprint
  {http://arxiv.org/abs/2111.03634} {arXiv:2111.03634 [astro-ph.HE]}
  \BibitemShut {NoStop}%
\bibitem [{\citenamefont {Adhikari}\ \emph {et~al.}(2020)\citenamefont
  {Adhikari} \emph {et~al.}}]{LIGO:2020xsf}%
  \BibitemOpen
  \bibfield  {author} {\bibinfo {author} {\bibfnamefont {R.~X.}\ \bibnamefont
  {Adhikari}} \emph {et~al.} (\bibinfo {collaboration} {LIGO}),\ }\href
  {\doibase 10.1088/1361-6382/ab9143} {\bibfield  {journal} {\bibinfo
  {journal} {Class. Quant. Grav.}\ }\textbf {\bibinfo {volume} {37}},\ \bibinfo
  {pages} {165003} (\bibinfo {year} {2020})},\ \Eprint
  {http://arxiv.org/abs/2001.11173} {arXiv:2001.11173 [astro-ph.IM]}
  \BibitemShut {NoStop}%
\bibitem [{\citenamefont {Torres-Rivas}\ \emph {et~al.}(2019)\citenamefont
  {Torres-Rivas}, \citenamefont {Chatziioannou}, \citenamefont {Bauswein},\
  and\ \citenamefont {Clark}}]{Torres-Rivas:2018svp}%
  \BibitemOpen
  \bibfield  {author} {\bibinfo {author} {\bibfnamefont {A.}~\bibnamefont
  {Torres-Rivas}}, \bibinfo {author} {\bibfnamefont {K.}~\bibnamefont
  {Chatziioannou}}, \bibinfo {author} {\bibfnamefont {A.}~\bibnamefont
  {Bauswein}}, \ and\ \bibinfo {author} {\bibfnamefont {J.~A.}\ \bibnamefont
  {Clark}},\ }\href {\doibase 10.1103/PhysRevD.99.044014} {\bibfield  {journal}
  {\bibinfo  {journal} {Phys. Rev. D}\ }\textbf {\bibinfo {volume} {99}},\
  \bibinfo {pages} {044014} (\bibinfo {year} {2019})},\ \Eprint
  {http://arxiv.org/abs/1811.08931} {arXiv:1811.08931 [gr-qc]} \BibitemShut
  {NoStop}%
\bibitem [{\citenamefont {Martynov}\ \emph {et~al.}(2019)\citenamefont
  {Martynov} \emph {et~al.}}]{Martynov:2019gvu}%
  \BibitemOpen
  \bibfield  {author} {\bibinfo {author} {\bibfnamefont {D.}~\bibnamefont
  {Martynov}} \emph {et~al.},\ }\href {\doibase 10.1103/PhysRevD.99.102004}
  {\bibfield  {journal} {\bibinfo  {journal} {Phys. Rev. D}\ }\textbf {\bibinfo
  {volume} {99}},\ \bibinfo {pages} {102004} (\bibinfo {year} {2019})},\
  \Eprint {http://arxiv.org/abs/1901.03885} {arXiv:1901.03885 [astro-ph.IM]}
  \BibitemShut {NoStop}%
\bibitem [{\citenamefont {Maggiore}\ \emph {et~al.}(2020)\citenamefont
  {Maggiore}, \citenamefont {Van Den~Broeck}, \citenamefont {Bartolo},
  \citenamefont {Belgacem}, \citenamefont {Bertacca}, \citenamefont {Bizouard},
  \citenamefont {Branchesi}, \citenamefont {Clesse}, \citenamefont {Foffa},
  \citenamefont {Garc{\'\i}a-Bellido} \emph {et~al.}}]{maggiore2020science}%
  \BibitemOpen
  \bibfield  {author} {\bibinfo {author} {\bibfnamefont {M.}~\bibnamefont
  {Maggiore}}, \bibinfo {author} {\bibfnamefont {C.}~\bibnamefont {Van
  Den~Broeck}}, \bibinfo {author} {\bibfnamefont {N.}~\bibnamefont {Bartolo}},
  \bibinfo {author} {\bibfnamefont {E.}~\bibnamefont {Belgacem}}, \bibinfo
  {author} {\bibfnamefont {D.}~\bibnamefont {Bertacca}}, \bibinfo {author}
  {\bibfnamefont {M.~A.}\ \bibnamefont {Bizouard}}, \bibinfo {author}
  {\bibfnamefont {M.}~\bibnamefont {Branchesi}}, \bibinfo {author}
  {\bibfnamefont {S.}~\bibnamefont {Clesse}}, \bibinfo {author} {\bibfnamefont
  {S.}~\bibnamefont {Foffa}}, \bibinfo {author} {\bibfnamefont
  {J.}~\bibnamefont {Garc{\'\i}a-Bellido}},  \emph {et~al.},\ }\href@noop {}
  {\bibfield  {journal} {\bibinfo  {journal} {Journal of Cosmology and
  Astroparticle Physics}\ }\textbf {\bibinfo {volume} {2020}},\ \bibinfo
  {pages} {050} (\bibinfo {year} {2020})}\BibitemShut {NoStop}%
\bibitem [{\citenamefont {Aasi}\ \emph {et~al.}(2013)\citenamefont {Aasi} \emph
  {et~al.}}]{LIGOScientific:2013pcc}%
  \BibitemOpen
  \bibfield  {author} {\bibinfo {author} {\bibfnamefont {J.}~\bibnamefont
  {Aasi}} \emph {et~al.} (\bibinfo {collaboration} {LIGO Scientific}),\ }\href
  {\doibase 10.1038/nphoton.2013.177} {\bibfield  {journal} {\bibinfo
  {journal} {Nature Photon.}\ }\textbf {\bibinfo {volume} {7}},\ \bibinfo
  {pages} {613} (\bibinfo {year} {2013})},\ \Eprint
  {http://arxiv.org/abs/1310.0383} {arXiv:1310.0383 [quant-ph]} \BibitemShut
  {NoStop}%
\bibitem [{\citenamefont {Zhou}\ \emph {et~al.}(2022)\citenamefont {Zhou},
  \citenamefont {Molina-Ruiz},\ and\ \citenamefont {Hellman}}]{Zhou:2022xav}%
  \BibitemOpen
  \bibfield  {author} {\bibinfo {author} {\bibfnamefont {R.}~\bibnamefont
  {Zhou}}, \bibinfo {author} {\bibfnamefont {M.}~\bibnamefont {Molina-Ruiz}}, \
  and\ \bibinfo {author} {\bibfnamefont {F.}~\bibnamefont {Hellman}},\
  }\href@noop {} {\enquote {\bibinfo {title} {{Strategies to reduce the
  thermoelastic loss of multimaterial coated finite substrates}},}\ } (\bibinfo
  {year} {2022}),\ \Eprint {http://arxiv.org/abs/2204.09808} {arXiv:2204.09808
  [cond-mat.mtrl-sci]} \BibitemShut {NoStop}%
\bibitem [{\citenamefont {Srivastava}\ \emph {et~al.}(2022)\citenamefont
  {Srivastava}, \citenamefont {Davis}, \citenamefont {Kuns}, \citenamefont
  {Landry}, \citenamefont {Ballmer}, \citenamefont {Evans}, \citenamefont
  {Hall}, \citenamefont {Read},\ and\ \citenamefont
  {Sathyaprakash}}]{srivastava2022science}%
  \BibitemOpen
  \bibfield  {author} {\bibinfo {author} {\bibfnamefont {V.}~\bibnamefont
  {Srivastava}}, \bibinfo {author} {\bibfnamefont {D.}~\bibnamefont {Davis}},
  \bibinfo {author} {\bibfnamefont {K.}~\bibnamefont {Kuns}}, \bibinfo {author}
  {\bibfnamefont {P.}~\bibnamefont {Landry}}, \bibinfo {author} {\bibfnamefont
  {S.}~\bibnamefont {Ballmer}}, \bibinfo {author} {\bibfnamefont
  {M.}~\bibnamefont {Evans}}, \bibinfo {author} {\bibfnamefont {E.~D.}\
  \bibnamefont {Hall}}, \bibinfo {author} {\bibfnamefont {J.}~\bibnamefont
  {Read}}, \ and\ \bibinfo {author} {\bibfnamefont {B.}~\bibnamefont
  {Sathyaprakash}},\ }\href@noop {} {\bibfield  {journal} {\bibinfo  {journal}
  {The Astrophysical Journal}\ }\textbf {\bibinfo {volume} {931}},\ \bibinfo
  {pages} {22} (\bibinfo {year} {2022})}\BibitemShut {NoStop}%
\bibitem [{\citenamefont {Braginskii}\ \emph {et~al.}(1973)\citenamefont
  {Braginskii}, \citenamefont {Grishchuk}, \citenamefont {Doroshkevich},
  \citenamefont {Zeldovich}, \citenamefont {Novikov},\ and\ \citenamefont
  {Sazhin}}]{Braginskii:1973vm}%
  \BibitemOpen
  \bibfield  {author} {\bibinfo {author} {\bibfnamefont {V.~B.}\ \bibnamefont
  {Braginskii}}, \bibinfo {author} {\bibfnamefont {L.~P.}\ \bibnamefont
  {Grishchuk}}, \bibinfo {author} {\bibfnamefont {A.~G.}\ \bibnamefont
  {Doroshkevich}}, \bibinfo {author} {\bibfnamefont {Y.~B.}\ \bibnamefont
  {Zeldovich}}, \bibinfo {author} {\bibfnamefont {I.~D.}\ \bibnamefont
  {Novikov}}, \ and\ \bibinfo {author} {\bibfnamefont {M.~V.}\ \bibnamefont
  {Sazhin}},\ }\href@noop {} {\bibfield  {journal} {\bibinfo  {journal} {Zh.
  Eksp. Teor. Fiz.}\ }\textbf {\bibinfo {volume} {65}},\ \bibinfo {pages}
  {1729} (\bibinfo {year} {1973})}\BibitemShut {NoStop}%
\bibitem [{\citenamefont {Pegoraro}\ \emph
  {et~al.}(1978{\natexlab{a}})\citenamefont {Pegoraro}, \citenamefont
  {Radicati}, \citenamefont {Bernard},\ and\ \citenamefont
  {Picasso}}]{Pegoraro:1978gv}%
  \BibitemOpen
  \bibfield  {author} {\bibinfo {author} {\bibfnamefont {F.}~\bibnamefont
  {Pegoraro}}, \bibinfo {author} {\bibfnamefont {L.~A.}\ \bibnamefont
  {Radicati}}, \bibinfo {author} {\bibfnamefont {P.}~\bibnamefont {Bernard}}, \
  and\ \bibinfo {author} {\bibfnamefont {E.}~\bibnamefont {Picasso}},\ }\href
  {\doibase 10.1016/0375-9601(78)90792-2} {\bibfield  {journal} {\bibinfo
  {journal} {Phys. Lett.}\ }\textbf {\bibinfo {volume} {A68}},\ \bibinfo
  {pages} {165} (\bibinfo {year} {1978}{\natexlab{a}})}\BibitemShut {NoStop}%
\bibitem [{\citenamefont {Pegoraro}\ \emph
  {et~al.}(1978{\natexlab{b}})\citenamefont {Pegoraro}, \citenamefont
  {Picasso},\ and\ \citenamefont {Radicati}}]{Pegoraro_1978}%
  \BibitemOpen
  \bibfield  {author} {\bibinfo {author} {\bibfnamefont {F.}~\bibnamefont
  {Pegoraro}}, \bibinfo {author} {\bibfnamefont {E.}~\bibnamefont {Picasso}}, \
  and\ \bibinfo {author} {\bibfnamefont {L.~A.}\ \bibnamefont {Radicati}},\
  }\href {\doibase 10.1088/0305-4470/11/10/013} {\bibfield  {journal} {\bibinfo
   {journal} {Journal of Physics A: Mathematical and General}\ }\textbf
  {\bibinfo {volume} {11}},\ \bibinfo {pages} {1949} (\bibinfo {year}
  {1978}{\natexlab{b}})}\BibitemShut {NoStop}%
\bibitem [{\citenamefont {Caves}(1979)}]{Caves:1979kq}%
  \BibitemOpen
  \bibfield  {author} {\bibinfo {author} {\bibfnamefont {C.~M.}\ \bibnamefont
  {Caves}},\ }\href {\doibase 10.1016/0370-2693(79)90227-2} {\bibfield
  {journal} {\bibinfo  {journal} {Phys. Lett. B}\ }\textbf {\bibinfo {volume}
  {80}},\ \bibinfo {pages} {323} (\bibinfo {year} {1979})}\BibitemShut
  {NoStop}%
\bibitem [{\citenamefont {Reece}\ \emph {et~al.}(1982)\citenamefont {Reece},
  \citenamefont {Reiner},\ and\ \citenamefont {Melissinos}}]{Reece:1982sc}%
  \BibitemOpen
  \bibfield  {author} {\bibinfo {author} {\bibfnamefont {C.~E.}\ \bibnamefont
  {Reece}}, \bibinfo {author} {\bibfnamefont {P.~J.}\ \bibnamefont {Reiner}}, \
  and\ \bibinfo {author} {\bibfnamefont {A.~C.}\ \bibnamefont {Melissinos}},\
  }\href@noop {} {\bibfield  {journal} {\bibinfo  {journal} {eConf}\ }\textbf
  {\bibinfo {volume} {C8206282}},\ \bibinfo {pages} {394} (\bibinfo {year}
  {1982})}\BibitemShut {NoStop}%
\bibitem [{\citenamefont {Reece}\ \emph {et~al.}(1984)\citenamefont {Reece},
  \citenamefont {Reiner},\ and\ \citenamefont {Melissinos}}]{Reece:1984gv}%
  \BibitemOpen
  \bibfield  {author} {\bibinfo {author} {\bibfnamefont {C.~E.}\ \bibnamefont
  {Reece}}, \bibinfo {author} {\bibfnamefont {P.~J.}\ \bibnamefont {Reiner}}, \
  and\ \bibinfo {author} {\bibfnamefont {A.~C.}\ \bibnamefont {Melissinos}},\
  }\href {\doibase 10.1016/0375-9601(84)90811-9} {\bibfield  {journal}
  {\bibinfo  {journal} {Phys. Lett. A}\ }\textbf {\bibinfo {volume} {104}},\
  \bibinfo {pages} {341} (\bibinfo {year} {1984})}\BibitemShut {NoStop}%
\bibitem [{\citenamefont {Bernard}\ \emph {et~al.}(2001)\citenamefont
  {Bernard}, \citenamefont {Gemme}, \citenamefont {Parodi},\ and\ \citenamefont
  {Picasso}}]{Bernard:2001kp}%
  \BibitemOpen
  \bibfield  {author} {\bibinfo {author} {\bibfnamefont {P.}~\bibnamefont
  {Bernard}}, \bibinfo {author} {\bibfnamefont {G.}~\bibnamefont {Gemme}},
  \bibinfo {author} {\bibfnamefont {R.}~\bibnamefont {Parodi}}, \ and\ \bibinfo
  {author} {\bibfnamefont {E.}~\bibnamefont {Picasso}},\ }\href {\doibase
  10.1063/1.1366636} {\bibfield  {journal} {\bibinfo  {journal} {Rev. Sci.
  Instrum.}\ }\textbf {\bibinfo {volume} {72}},\ \bibinfo {pages} {2428}
  (\bibinfo {year} {2001})},\ \Eprint {http://arxiv.org/abs/gr-qc/0103006}
  {arXiv:gr-qc/0103006 [gr-qc]} \BibitemShut {NoStop}%
\bibitem [{\citenamefont {Bernard}\ \emph {et~al.}(2002)\citenamefont
  {Bernard}, \citenamefont {Chincarini}, \citenamefont {Gemme}, \citenamefont
  {Parodi},\ and\ \citenamefont {Picasso}}]{Bernard:2002ci}%
  \BibitemOpen
  \bibfield  {author} {\bibinfo {author} {\bibfnamefont {P.}~\bibnamefont
  {Bernard}}, \bibinfo {author} {\bibfnamefont {A.}~\bibnamefont {Chincarini}},
  \bibinfo {author} {\bibfnamefont {G.}~\bibnamefont {Gemme}}, \bibinfo
  {author} {\bibfnamefont {R.}~\bibnamefont {Parodi}}, \ and\ \bibinfo {author}
  {\bibfnamefont {E.}~\bibnamefont {Picasso}},\ }\href@noop {} {\  (\bibinfo
  {year} {2002})},\ \Eprint {http://arxiv.org/abs/gr-qc/0203024}
  {arXiv:gr-qc/0203024 [gr-qc]} \BibitemShut {NoStop}%
\bibitem [{\citenamefont {Ballantini}\ \emph {et~al.}(2003)\citenamefont
  {Ballantini}, \citenamefont {Bernard}, \citenamefont {Chiaveri},
  \citenamefont {Chincarini}, \citenamefont {Gemme}, \citenamefont {Losito},
  \citenamefont {Parodi},\ and\ \citenamefont {Picasso}}]{Ballantini:2003nt}%
  \BibitemOpen
  \bibfield  {author} {\bibinfo {author} {\bibfnamefont {R.}~\bibnamefont
  {Ballantini}}, \bibinfo {author} {\bibfnamefont {P.}~\bibnamefont {Bernard}},
  \bibinfo {author} {\bibfnamefont {E.}~\bibnamefont {Chiaveri}}, \bibinfo
  {author} {\bibfnamefont {A.}~\bibnamefont {Chincarini}}, \bibinfo {author}
  {\bibfnamefont {G.}~\bibnamefont {Gemme}}, \bibinfo {author} {\bibfnamefont
  {R.}~\bibnamefont {Losito}}, \bibinfo {author} {\bibfnamefont
  {R.}~\bibnamefont {Parodi}}, \ and\ \bibinfo {author} {\bibfnamefont
  {E.}~\bibnamefont {Picasso}},\ }\href {\doibase 10.1088/0264-9381/20/15/316}
  {\bibfield  {journal} {\bibinfo  {journal} {Class. Quant. Grav.}\ }\textbf
  {\bibinfo {volume} {20}},\ \bibinfo {pages} {3505} (\bibinfo {year}
  {2003})}\BibitemShut {NoStop}%
\bibitem [{\citenamefont {Ballantini}\ \emph {et~al.}(2005)\citenamefont
  {Ballantini} \emph {et~al.}}]{Ballantini:2005am}%
  \BibitemOpen
  \bibfield  {author} {\bibinfo {author} {\bibfnamefont {R.}~\bibnamefont
  {Ballantini}} \emph {et~al.},\ }\href@noop {} {\  (\bibinfo {year} {2005})},\
  \Eprint {http://arxiv.org/abs/gr-qc/0502054} {arXiv:gr-qc/0502054}
  \BibitemShut {NoStop}%
\bibitem [{\citenamefont {Berlin}\ \emph {et~al.}(2023)\citenamefont {Berlin},
  \citenamefont {Blas}, \citenamefont {Tito~D'Agnolo}, \citenamefont {Ellis},
  \citenamefont {Harnik}, \citenamefont {Kahn}, \citenamefont
  {Sch\"utte-Engel},\ and\ \citenamefont {Wentzel}}]{Berlin:2023grv}%
  \BibitemOpen
  \bibfield  {author} {\bibinfo {author} {\bibfnamefont {A.}~\bibnamefont
  {Berlin}}, \bibinfo {author} {\bibfnamefont {D.}~\bibnamefont {Blas}},
  \bibinfo {author} {\bibfnamefont {R.}~\bibnamefont {Tito~D'Agnolo}}, \bibinfo
  {author} {\bibfnamefont {S.~A.~R.}\ \bibnamefont {Ellis}}, \bibinfo {author}
  {\bibfnamefont {R.}~\bibnamefont {Harnik}}, \bibinfo {author} {\bibfnamefont
  {Y.}~\bibnamefont {Kahn}}, \bibinfo {author} {\bibfnamefont {J.}~\bibnamefont
  {Sch\"utte-Engel}}, \ and\ \bibinfo {author} {\bibfnamefont {M.}~\bibnamefont
  {Wentzel}},\ }\href@noop {} {\  (\bibinfo {year} {2023})},\ \Eprint
  {http://arxiv.org/abs/2303.01518} {arXiv:2303.01518 [hep-ph]} \BibitemShut
  {NoStop}%
\bibitem [{\citenamefont {Weber}(1960)}]{PhysRev.117.306}%
  \BibitemOpen
  \bibfield  {author} {\bibinfo {author} {\bibfnamefont {J.}~\bibnamefont
  {Weber}},\ }\href {\doibase 10.1103/PhysRev.117.306} {\bibfield  {journal}
  {\bibinfo  {journal} {Phys. Rev.}\ }\textbf {\bibinfo {volume} {117}},\
  \bibinfo {pages} {306} (\bibinfo {year} {1960})}\BibitemShut {NoStop}%
\bibitem [{\citenamefont {Weber}(1966)}]{PhysRevLett.17.1228}%
  \BibitemOpen
  \bibfield  {author} {\bibinfo {author} {\bibfnamefont {J.}~\bibnamefont
  {Weber}},\ }\href {\doibase 10.1103/PhysRevLett.17.1228} {\bibfield
  {journal} {\bibinfo  {journal} {Phys. Rev. Lett.}\ }\textbf {\bibinfo
  {volume} {17}},\ \bibinfo {pages} {1228} (\bibinfo {year}
  {1966})}\BibitemShut {NoStop}%
\bibitem [{\citenamefont {Forward}(1971)}]{Forward:1971mel}%
  \BibitemOpen
  \bibfield  {author} {\bibinfo {author} {\bibfnamefont {R.~L.}\ \bibnamefont
  {Forward}},\ }\href {\doibase 10.1007/BF02450446} {\bibfield  {journal}
  {\bibinfo  {journal} {Gen. Rel. Grav.}\ }\textbf {\bibinfo {volume} {2}},\
  \bibinfo {pages} {149} (\bibinfo {year} {1971})}\BibitemShut {NoStop}%
\bibitem [{\citenamefont {Liccardo}\ \emph {et~al.}(2023)\citenamefont
  {Liccardo}, \citenamefont {Lenzi}, \citenamefont {Marinho}, \citenamefont
  {Aguiar}, \citenamefont {Frajuca}, \citenamefont {Bortoli},\ and\
  \citenamefont {Costa}}]{Liccardo:2023nzv}%
  \BibitemOpen
  \bibfield  {author} {\bibinfo {author} {\bibfnamefont {V.}~\bibnamefont
  {Liccardo}}, \bibinfo {author} {\bibfnamefont {C.~H.}\ \bibnamefont {Lenzi}},
  \bibinfo {author} {\bibfnamefont {R.~M.}\ \bibnamefont {Marinho}}, \bibinfo
  {author} {\bibfnamefont {O.~D.}\ \bibnamefont {Aguiar}}, \bibinfo {author}
  {\bibfnamefont {C.}~\bibnamefont {Frajuca}}, \bibinfo {author} {\bibfnamefont
  {F.~S.}\ \bibnamefont {Bortoli}}, \ and\ \bibinfo {author} {\bibfnamefont
  {C.~A.}\ \bibnamefont {Costa}},\ }\href@noop {} {\  (\bibinfo {year}
  {2023})},\ \Eprint {http://arxiv.org/abs/2302.01232} {arXiv:2302.01232
  [astro-ph.IM]} \BibitemShut {NoStop}%
\bibitem [{\citenamefont {Gottardi}\ \emph {et~al.}(2007)\citenamefont
  {Gottardi}, \citenamefont {de~Waard}, \citenamefont {Usenko}, \citenamefont
  {Frossati}, \citenamefont {Podt}, \citenamefont {Flokstra}, \citenamefont
  {Bassan}, \citenamefont {Fafone}, \citenamefont {Minenkov},\ and\
  \citenamefont {Rocchi}}]{Gottardi:2007zn}%
  \BibitemOpen
  \bibfield  {author} {\bibinfo {author} {\bibfnamefont {L.}~\bibnamefont
  {Gottardi}}, \bibinfo {author} {\bibfnamefont {A.}~\bibnamefont {de~Waard}},
  \bibinfo {author} {\bibfnamefont {A.}~\bibnamefont {Usenko}}, \bibinfo
  {author} {\bibfnamefont {G.}~\bibnamefont {Frossati}}, \bibinfo {author}
  {\bibfnamefont {M.}~\bibnamefont {Podt}}, \bibinfo {author} {\bibfnamefont
  {J.}~\bibnamefont {Flokstra}}, \bibinfo {author} {\bibfnamefont
  {M.}~\bibnamefont {Bassan}}, \bibinfo {author} {\bibfnamefont
  {V.}~\bibnamefont {Fafone}}, \bibinfo {author} {\bibfnamefont
  {Y.}~\bibnamefont {Minenkov}}, \ and\ \bibinfo {author} {\bibfnamefont
  {A.}~\bibnamefont {Rocchi}},\ }\href {\doibase 10.1103/PhysRevD.76.102005}
  {\bibfield  {journal} {\bibinfo  {journal} {Phys. Rev. D}\ }\textbf {\bibinfo
  {volume} {76}},\ \bibinfo {pages} {102005} (\bibinfo {year} {2007})},\
  \Eprint {http://arxiv.org/abs/0705.0122} {arXiv:0705.0122 [gr-qc]}
  \BibitemShut {NoStop}%
\bibitem [{\citenamefont {Cerdonio}\ \emph {et~al.}(1997)\citenamefont
  {Cerdonio} \emph {et~al.}}]{Cerdonio:1997hz}%
  \BibitemOpen
  \bibfield  {author} {\bibinfo {author} {\bibfnamefont {M.}~\bibnamefont
  {Cerdonio}} \emph {et~al.},\ }\href {\doibase 10.1088/0264-9381/14/6/016}
  {\bibfield  {journal} {\bibinfo  {journal} {Class. Quant. Grav.}\ }\textbf
  {\bibinfo {volume} {14}},\ \bibinfo {pages} {1491} (\bibinfo {year}
  {1997})}\BibitemShut {NoStop}%
\bibitem [{\citenamefont {Vinante}(2006)}]{Vinante:2006uk}%
  \BibitemOpen
  \bibfield  {author} {\bibinfo {author} {\bibfnamefont {A.}~\bibnamefont
  {Vinante}} (\bibinfo {collaboration} {AURIGA}),\ }\href {\doibase
  10.1088/0264-9381/23/8/S14} {\bibfield  {journal} {\bibinfo  {journal}
  {Class. Quant. Grav.}\ }\textbf {\bibinfo {volume} {23}},\ \bibinfo {pages}
  {S103} (\bibinfo {year} {2006})}\BibitemShut {NoStop}%
\bibitem [{\citenamefont {Harry}\ \emph {et~al.}(1996)\citenamefont {Harry},
  \citenamefont {Stevenson},\ and\ \citenamefont {Paik}}]{Harry:1996gh}%
  \BibitemOpen
  \bibfield  {author} {\bibinfo {author} {\bibfnamefont {G.~M.}\ \bibnamefont
  {Harry}}, \bibinfo {author} {\bibfnamefont {T.~R.}\ \bibnamefont
  {Stevenson}}, \ and\ \bibinfo {author} {\bibfnamefont {H.~J.}\ \bibnamefont
  {Paik}},\ }\href {\doibase 10.1103/PhysRevD.54.2409} {\bibfield  {journal}
  {\bibinfo  {journal} {Phys. Rev. D}\ }\textbf {\bibinfo {volume} {54}},\
  \bibinfo {pages} {2409} (\bibinfo {year} {1996})},\ \Eprint
  {http://arxiv.org/abs/gr-qc/9602018} {arXiv:gr-qc/9602018} \BibitemShut
  {NoStop}%
\bibitem [{\citenamefont {Aguiar}\ \emph {et~al.}(2009)\citenamefont {Aguiar},
  \citenamefont {Barroso}, \citenamefont {Marinho}, \citenamefont {Pimentel},\
  and\ \citenamefont {Tobar}}]{Aguiar:2009zzb}%
  \BibitemOpen
  \bibfield  {author} {\bibinfo {author} {\bibfnamefont {O.~D.}\ \bibnamefont
  {Aguiar}}, \bibinfo {author} {\bibfnamefont {J.~J.}\ \bibnamefont {Barroso}},
  \bibinfo {author} {\bibfnamefont {R.~M.}\ \bibnamefont {Marinho}}, \bibinfo
  {author} {\bibfnamefont {G.~L.}\ \bibnamefont {Pimentel}}, \ and\ \bibinfo
  {author} {\bibfnamefont {M.~E.}\ \bibnamefont {Tobar}},\ }\href {\doibase
  10.1142/S0218271809015849} {\bibfield  {journal} {\bibinfo  {journal} {Int.
  J. Mod. Phys. D}\ }\textbf {\bibinfo {volume} {18}},\ \bibinfo {pages} {2317}
  (\bibinfo {year} {2009})}\BibitemShut {NoStop}%
\bibitem [{\citenamefont {Arvanitaki}\ and\ \citenamefont
  {Geraci}(2013)}]{Arvanitaki:2012cn}%
  \BibitemOpen
  \bibfield  {author} {\bibinfo {author} {\bibfnamefont {A.}~\bibnamefont
  {Arvanitaki}}\ and\ \bibinfo {author} {\bibfnamefont {A.~A.}\ \bibnamefont
  {Geraci}},\ }\href {\doibase 10.1103/PhysRevLett.110.071105} {\bibfield
  {journal} {\bibinfo  {journal} {Phys. Rev. Lett.}\ }\textbf {\bibinfo
  {volume} {110}},\ \bibinfo {pages} {071105} (\bibinfo {year} {2013})},\
  \Eprint {http://arxiv.org/abs/1207.5320} {arXiv:1207.5320 [gr-qc]}
  \BibitemShut {NoStop}%
\bibitem [{\citenamefont {Aggarwal}\ \emph {et~al.}(2022)\citenamefont
  {Aggarwal}, \citenamefont {Winstone}, \citenamefont {Teo}, \citenamefont
  {Baryakhtar}, \citenamefont {Larson}, \citenamefont {Kalogera},\ and\
  \citenamefont {Geraci}}]{Aggarwal:2020umq}%
  \BibitemOpen
  \bibfield  {author} {\bibinfo {author} {\bibfnamefont {N.}~\bibnamefont
  {Aggarwal}}, \bibinfo {author} {\bibfnamefont {G.~P.}\ \bibnamefont
  {Winstone}}, \bibinfo {author} {\bibfnamefont {M.}~\bibnamefont {Teo}},
  \bibinfo {author} {\bibfnamefont {M.}~\bibnamefont {Baryakhtar}}, \bibinfo
  {author} {\bibfnamefont {S.~L.}\ \bibnamefont {Larson}}, \bibinfo {author}
  {\bibfnamefont {V.}~\bibnamefont {Kalogera}}, \ and\ \bibinfo {author}
  {\bibfnamefont {A.~A.}\ \bibnamefont {Geraci}},\ }\href {\doibase
  10.1103/PhysRevLett.128.111101} {\bibfield  {journal} {\bibinfo  {journal}
  {Phys. Rev. Lett.}\ }\textbf {\bibinfo {volume} {128}},\ \bibinfo {pages}
  {111101} (\bibinfo {year} {2022})},\ \Eprint
  {http://arxiv.org/abs/2010.13157} {arXiv:2010.13157 [gr-qc]} \BibitemShut
  {NoStop}%
\bibitem [{\citenamefont {Graham}\ \emph {et~al.}(2017)\citenamefont {Graham},
  \citenamefont {Hogan}, \citenamefont {Kasevich}, \citenamefont {Rajendran},\
  and\ \citenamefont {Romani}}]{Graham:2017pmn}%
  \BibitemOpen
  \bibfield  {author} {\bibinfo {author} {\bibfnamefont {P.~W.}\ \bibnamefont
  {Graham}}, \bibinfo {author} {\bibfnamefont {J.~M.}\ \bibnamefont {Hogan}},
  \bibinfo {author} {\bibfnamefont {M.~A.}\ \bibnamefont {Kasevich}}, \bibinfo
  {author} {\bibfnamefont {S.}~\bibnamefont {Rajendran}}, \ and\ \bibinfo
  {author} {\bibfnamefont {R.~W.}\ \bibnamefont {Romani}} (\bibinfo
  {collaboration} {MAGIS}),\ }\href@noop {} {\  (\bibinfo {year} {2017})},\
  \Eprint {http://arxiv.org/abs/1711.02225} {arXiv:1711.02225 [astro-ph.IM]}
  \BibitemShut {NoStop}%
\bibitem [{\citenamefont {El-Neaj}\ \emph {et~al.}(2020)\citenamefont {El-Neaj}
  \emph {et~al.}}]{AEDGE:2019nxb}%
  \BibitemOpen
  \bibfield  {author} {\bibinfo {author} {\bibfnamefont {Y.~A.}\ \bibnamefont
  {El-Neaj}} \emph {et~al.} (\bibinfo {collaboration} {AEDGE}),\ }\href
  {\doibase 10.1140/epjqt/s40507-020-0080-0} {\bibfield  {journal} {\bibinfo
  {journal} {EPJ Quant. Technol.}\ }\textbf {\bibinfo {volume} {7}},\ \bibinfo
  {pages} {6} (\bibinfo {year} {2020})},\ \Eprint
  {http://arxiv.org/abs/1908.00802} {arXiv:1908.00802 [gr-qc]} \BibitemShut
  {NoStop}%
\bibitem [{\citenamefont {Badurina}\ \emph {et~al.}(2020)\citenamefont
  {Badurina} \emph {et~al.}}]{Badurina:2019hst}%
  \BibitemOpen
  \bibfield  {author} {\bibinfo {author} {\bibfnamefont {L.}~\bibnamefont
  {Badurina}} \emph {et~al.},\ }\href {\doibase 10.1088/1475-7516/2020/05/011}
  {\bibfield  {journal} {\bibinfo  {journal} {JCAP}\ }\textbf {\bibinfo
  {volume} {05}},\ \bibinfo {pages} {011} (\bibinfo {year} {2020})},\ \Eprint
  {http://arxiv.org/abs/1911.11755} {arXiv:1911.11755 [astro-ph.CO]}
  \BibitemShut {NoStop}%
\bibitem [{\citenamefont {Loeb}\ and\ \citenamefont
  {Maoz}(2015)}]{Loeb:2015ffa}%
  \BibitemOpen
  \bibfield  {author} {\bibinfo {author} {\bibfnamefont {A.}~\bibnamefont
  {Loeb}}\ and\ \bibinfo {author} {\bibfnamefont {D.}~\bibnamefont {Maoz}},\
  }\href@noop {} {\  (\bibinfo {year} {2015})},\ \Eprint
  {http://arxiv.org/abs/1501.00996} {arXiv:1501.00996 [astro-ph.IM]}
  \BibitemShut {NoStop}%
\bibitem [{\citenamefont {Vutha}(2015)}]{Vutha:2015aza}%
  \BibitemOpen
  \bibfield  {author} {\bibinfo {author} {\bibfnamefont {A.~C.}\ \bibnamefont
  {Vutha}},\ }\href {\doibase 10.1088/1367-2630/17/6/063030} {\bibfield
  {journal} {\bibinfo  {journal} {New J. Phys.}\ }\textbf {\bibinfo {volume}
  {17}},\ \bibinfo {pages} {063030} (\bibinfo {year} {2015})},\ \Eprint
  {http://arxiv.org/abs/1501.01870} {arXiv:1501.01870 [physics.atom-ph]}
  \BibitemShut {NoStop}%
\bibitem [{\citenamefont {Kolkowitz}\ \emph {et~al.}(2016)\citenamefont
  {Kolkowitz}, \citenamefont {Pikovski}, \citenamefont {Langellier},
  \citenamefont {Lukin}, \citenamefont {Walsworth},\ and\ \citenamefont
  {Ye}}]{Kolkowitz:2016wyg}%
  \BibitemOpen
  \bibfield  {author} {\bibinfo {author} {\bibfnamefont {S.}~\bibnamefont
  {Kolkowitz}}, \bibinfo {author} {\bibfnamefont {I.}~\bibnamefont {Pikovski}},
  \bibinfo {author} {\bibfnamefont {N.}~\bibnamefont {Langellier}}, \bibinfo
  {author} {\bibfnamefont {M.~D.}\ \bibnamefont {Lukin}}, \bibinfo {author}
  {\bibfnamefont {R.~L.}\ \bibnamefont {Walsworth}}, \ and\ \bibinfo {author}
  {\bibfnamefont {J.}~\bibnamefont {Ye}},\ }\href {\doibase
  10.1103/PhysRevD.94.124043} {\bibfield  {journal} {\bibinfo  {journal} {Phys.
  Rev. D}\ }\textbf {\bibinfo {volume} {94}},\ \bibinfo {pages} {124043}
  (\bibinfo {year} {2016})},\ \Eprint {http://arxiv.org/abs/1606.01859}
  {arXiv:1606.01859 [physics.atom-ph]} \BibitemShut {NoStop}%
\bibitem [{\citenamefont {Bringmann}\ \emph {et~al.}(2023)\citenamefont
  {Bringmann}, \citenamefont {Domcke}, \citenamefont {Fuchs},\ and\
  \citenamefont {Kopp}}]{Bringmann:2023gba}%
  \BibitemOpen
  \bibfield  {author} {\bibinfo {author} {\bibfnamefont {T.}~\bibnamefont
  {Bringmann}}, \bibinfo {author} {\bibfnamefont {V.}~\bibnamefont {Domcke}},
  \bibinfo {author} {\bibfnamefont {E.}~\bibnamefont {Fuchs}}, \ and\ \bibinfo
  {author} {\bibfnamefont {J.}~\bibnamefont {Kopp}},\ }\href@noop {} {\
  (\bibinfo {year} {2023})},\ \Eprint {http://arxiv.org/abs/2304.10579}
  {arXiv:2304.10579 [hep-ph]} \BibitemShut {NoStop}%
\end{thebibliography}%

\acrodef{ADM}{Arnowitt-Deser-Misner}
\acrodef{AMR}{adaptive mesh-refinement}
\acrodef{BH}{black hole}
\acrodef{BBH}{binary black-hole}
\acrodef{BHNS}{black-hole neutron-star}
\acrodef{BNS}{binary neutron star}
\acrodef{CCSN}{core-collapse supernova}
\acrodefplural{CCSN}[CCSNe]{core-collapse supernovae}
\acrodef{CMA}{consistent multi-fluid advection}
\acrodef{CFL}{Courant-Friedrichs-Lewy}
\acrodef{DG}{discontinuous Galerkin}
\acrodef{HMNS}{hypermassive neutron star}
\acrodef{EM}{electromagnetic}
\acrodef{ET}{Einstein Telescope}
\acrodef{EOB}{effective-one-body}
\acrodef{EOS}{equation of state}
\acrodef{FF}{fitting factor}
\acrodef{GR}{general-relativistic}
\acrodef{GRLES}{general-relativistic large-eddy simulation}
\acrodef{GRHD}{general-relativistic hydrodynamics}
\acrodef{GRMHD}{general-relativistic magnetohydrodynamics}
\acrodef{GW}{gravitational wave}
\acrodef{ILES}{implicit large-eddy simulations}
\acrodef{LIA}{linear interaction analysis}
\acrodef{LES}{large-eddy simulation}
\acrodefplural{LES}[LES]{large-eddy simulations}
\acrodef{MRI}{magnetorotational instability}
\acrodef{NR}{numerical relativity}
\acrodef{NS}{neutron star}
\acrodef{PNS}{protoneutron star}
\acrodef{RMNS}{remnant massive neutron star}
\acrodef{SASI}{standing accretion shock instability}
\acrodef{SGRB}{short $\gamma$-ray burst}
\acrodef{SPH}{smoothed particle hydrodynamics}
\acrodef{SN}{supernova}
\acrodefplural{SN}[SNe]{supernovae}
\acrodef{SNR}{signal-to-noise ratio}
\acrodef{ZAMS}{zero age main sequence}

\end{document}